\documentclass[10pt,letterpaper]{article}
\usepackage[top=0.85in,left=2.75in,footskip=0.75in]{geometry}

\usepackage{amsmath,amssymb}

\usepackage{changepage}

\usepackage[utf8x]{inputenc}

\usepackage{textcomp,marvosym}

\usepackage{cite}

\usepackage{nameref,hyperref}

\usepackage{microtype}
\DisableLigatures[f]{encoding = *, family = * }

\usepackage[table]{xcolor}

\usepackage{array}

\newcolumntype{+}{!{\vrule width 2pt}}

\newlength\savedwidth

\raggedright
\usepackage[aboveskip=1pt,labelfont=bf,labelsep=period,justification=raggedright,singlelinecheck=off]{caption}

\makeatletter
\renewcommand{\@biblabel}[1]{\quad#1.}
\makeatother

\usepackage{lastpage,fancyhdr,graphicx}
\usepackage{epstopdf}
\fancyheadoffset[L]{2.25in}
\fancyfootoffset[L]{2.25in}
\newcommand{\maragogiistypical}{{section 1}}
\newcommand{\SMC}{{section 2}}
\newcommand{\householdnetworks}{{section 3}}
\newcommand{\dataprocessing}{{section 4}}
\newcommand{\networkssocialdynamics}{{section 5}}
\newcommand{\COMORBUSS}{{section 5}}
\newcommand{\calibrationinitialization}{{section 6}}
\newcommand{\airborne}{{section 7}}
\newcommand{\robustness}{{section 8}}

\newcommand{\DateInitial}{May 9th, 2020}%

\newcommand{\DateFinal}{July 25th, 2020}%

\begin{document}
\vspace*{0.2in}

\begin{flushleft}
{\Large
\textbf\newline{Quantifying protocols for safe school activities} 
}
\newline
\\
Juliano Genari\textsuperscript{1\Yinyang},
Guilherme Tegoni Goedert\textsuperscript{2,3,4\Yinyang},
S\'ergio H. A. Lira\textsuperscript{5},
Krerley Oliveira\textsuperscript{5},
Adriano Barbosa\textsuperscript{5},
Allysson Lima\textsuperscript{6},
Jos\'e Augusto Silva\textsuperscript{5},
Hugo Oliveira\textsuperscript{5},
Maur\'icio Maciel\textsuperscript{5},
Ismael Ledoino\textsuperscript{7},
Lucas Resende\textsuperscript{8},
Edmilson Roque dos Santos\textsuperscript{1},
Dan Marchesin\textsuperscript{8},
Claudio J. Struchiner\textsuperscript{9},
Tiago Pereira\textsuperscript{1*},
\\
\bigskip
\textbf{1} Instituto de Ci\^encias Matem\'aticas e Computa\c{c}\~ao, Universidade de S\~ao Paulo,  S\~ao Paulo, S\~ao Paulo, Brazil
\\
\textbf{2} Dipartimento di Fisica, Universit\`a degli Studi di Roma Tor Vergata, Rome, Lazio, Italy
\\
\textbf{3} Aachen Institute for Advanced Study in Computational Engineering Science, RWTH AACHEN University, Aachen, North Rhine-Westphalia, Germany
\\
\textbf{4} Computation-based Science and Technology Research Center, The Cyprus Institute, Nicosia, District of Nicosia, Cyprus
\\
\textbf{5} Universidade Federal de Alagoas, Maceió, Alagoas, Brazil
\\
\textbf{6} Universidade Federal de Grande Dourados, Dourados, Mato Grosso do Sul, Brazil
\\
\textbf{7} Laborat\'orio Nacional de Computa\c{c}\~ao Cient\'ifica, Petrópolis, Rio de Janeiro, Brazil
\\
\textbf{8} Instituto de Matem\'atica Pura e Aplicada, Rio de Janeiro, Rio de Janeiro, Brazil
\\
\textbf{9} Escola de Matemática Aplicada, Funda\c{c}\~ao Get\'ulio Vargas, Rio de Janeiro, Rio de Janeiro, Brazil
\\
\bigskip

%
%
\Yinyang These authors contributed equally to this work.

\ddag These authors also contributed equally to this work.

\textcurrency Current Address: Dept/Program/Center, Institution Name, City, State, Country 

\dag Deceased

\textpilcrow Membership list can be found in the Acknowledgments section.

* tiago@icmc.usp.br

\end{flushleft}
\section*{Abstract}
By the peak of COVID-19 restrictions on April 8, 2020, up to 1.5 billion students across 188 countries were by the suspension of physical attendance in schools. Schools were among the first services to reopen as vaccination campaigns advanced. With the emergence of new variants and infection waves, the question now is to find safe protocols for the continuation of school activities. We need to understand how reliable these protocols are under different levels of vaccination coverage,  as many countries have a meager fraction of their population vaccinated, including Uganda where the coverage is about 8\%. We investigate the impact of face-to-face classes under different protocols and quantify the surplus number of infected individuals in a city. Using the infection transmission when schools were closed as a baseline, we assess the impact of  physical school attendance in classrooms with poor air circulation. We find that (i) resuming school activities with people only wearing low-quality masks leads to a near fivefold city-wide increase in the number of cases even if all staff is vaccinated, (ii) resuming activities with students wearing good-quality masks and staff wearing N95s leads to about a threefold increase, (iii) combining high-quality masks and active monitoring, activities may be carried out safely even with low vaccination coverage. These results highlight the effectiveness of good mask-wearing. Compared to ICU costs, high-quality masks are inexpensive and can help curb the spreading. Classes can be carried out safely, provided the correct set of measures are implemented.

\section*{Author summary}
The World Bank-UNESCO-UNICEF report \cite{WB_Unesco} estimates that learning losses from the COVID-19 pandemic could cost this generation \$17 trillion dollars in lifetime earnings. Despite the surging pressure to keep schools open, many countries lack guidelines for safe school activities. Using the empirical transmission level for closed schools as a baseline, we quantify the impact of distinct non-pharmaceutical interventions (NPIs) on infection rates and different values of vaccine coverage. Strikingly, we show that classes can be kept safe, provided the correct wearing of good quality masks together with to the proper combination of other NPIs. In such scenarios, the increase in infections can be kept below 20\% compared to suspending classes.


\section*{Introduction}
The educational system plays a fundamental role in the socio-intellectual development and mental health of children and adolescents. During the COVID-19 pandemic, the impact of school closures on society has been massive. UNESCO reported that, as of April 8, 2020, up to 188 countries closed schools nationwide.  In developing countries, such as Brazil, nutritional wellbeing of children was put in jeopardy as families rely on the provision of school meals. And yet, in Brazil alone, schools remained closed full-time for 191 days in 2020 affecting 44.3 million children. However, given the frequent contact during a school day, the prevalence of mild symptoms in children and the role of school as a source of contacts bridging family nuclei, there is understandable concern that face-to-face classes could be driving uncontrolled spreading of the virus.

In view of the negative physical and mental consequences for students, together with the educational deficit imposed by school closings, the ECDC agency points out that measures of transmission mitigation are necessary for students to have a safe socialization and learning environment \cite{ECDC_schools}. Thus a major concern is the assessment of mitigation protocols \cite{Claudio_questions} to understand the impact of each measure within the school community. 

Living in a household with a child who goes to school physically increases the risk of becoming infected by up to 38\%. Similarly, teachers working in school are 1.8 times more likely to be infected than those working from home   \cite{Lessler_2021_in_person_schooling} and resuming face-to-face classes has been directly related to outbreaks \cite{ECDC_schools}. Mitigation measures such as separating student groups, quarantining exposed students and professionals, wearing masks, maintaining adequate air ventilation, vaccinating risk groups, and monitoring case emergence, can all decrease the number of new cases \cite{Gurdasani_2021_lancet,Munday_2021_estimation_UK, Lessler_2021_in_person_schooling}. 

Often, mitigation measures are put in place simultaneously, which makes it difficult to disentangle their individual impact on transmission from temporal case report datasets. The lack of infrastructure, personnel and lab equipment may also limit the use of these measures in developing countries, especially when they are based on resource intensive practices such as testing and subsequent contact tracing of cases. Thus, it becomes crucial to identify effective mitigation practices \textit{a priori}.

Our aim is to assess quantitatively the effects of vaccination \cite{Tiago_Claudio_PNAS} and NPIs protocols and find effective protocols for school activities. This is done via the COMORBUSS computer model, an agent-based model developed by our group, focusing on the social dynamics and individualized agent biology. To achieve this end, we collected data from a Brazilian city with 33 thousand people and income profile close to the median for the country. The dataset contains information on COVID-19 hospitalizations, detailed city infrastructure and anonymous social data representing household clusters. We used these data to calibrate COMORBUSS and run hundreds of simulations with individual and combined mitigation measures. We take the 77 days during which the schools were closed citywide (and whose prevalence of COVID-19 was about 11.50 ± 1.02\%) as a baseline rate for comparison.

Our results suggest that keeping schools open with students wearing poor quality masks and no further mitigation measures can increase the number of infections fivefold across the whole community. When staff and student wear good quality masks the number of infections decrease by 300\%.  Similarly, vaccination of professionals accounts for a sizable decrease in cases but is not enough by itself. The relative increase in cases reduces with the successive adoption of other non-pharmaceutical interventions (NPIs), the most important of which being the adoption of case monitoring protocols. The relative effectiveness rank between intervention remains preserved under scenarios of larger cities and more infectious variants, which evidence the robustness of our findings.

Our study shows that classes can be kept open safely, provided that the correct combination of measures be adopted. Relying on a single measure is mostly not effective nor stable, but simple measures can go a long way when properly combined and implemented.

\section*{Materials and methods}
\subsection*{Data collection}
The city of Maragogi in Northeastern Brazil has 33,000 inhabitants \cite{IBGE_maragogi} and is a representative of at least $40\%$ of Brazilian cities in terms of income and demographics. {Moreover, its demography is also typical worldwide, being} located above the $ 50 \%$ quantile in a sample of 28,372 {North American cities and 41,000 global cities}, using the \texttt{simplemaps} database \cite{us_cities,world_cities}, see {\bf { \nameref{S1_SI} \maragogiistypical}} for further details. 

\begin{figure*}[hbt!]
	\centering
	\includegraphics[width=1.0\linewidth]{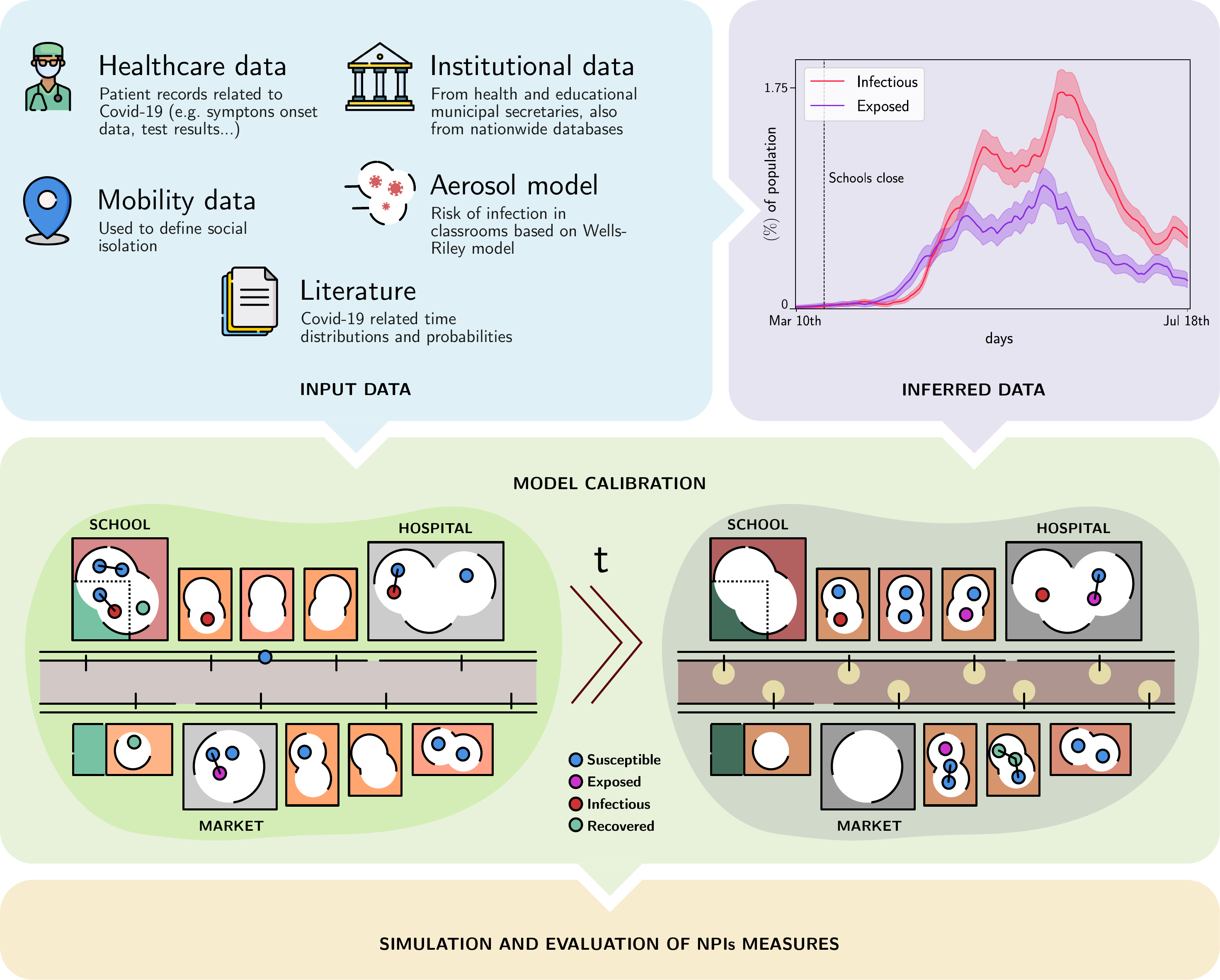}%
	\caption{\textbf{Pipeline overview description.} Data is collected as patients attend health institutions. Health professionals register patients' personal, epidemiological and geo-location data to the Clinical Monitoring System (CMS) that is blended with socio-economical and household data. Using these data we estimate the number of Exposed (blue), Infectious (red) and Recovered (green) individuals. All the pre-processed data is used to calibrate our Stochastic Agent Based model, COMORBUSS. From bottom left to right: a schematic representation of the social dynamics of COMORBUSS, producing contacts between individuals in different social contexts. The colored circles represent the state of individuals, and  lines represent relevant physical contacts capable of producing contagions. Once calibrated the model is used to estimate the effectiveness of NPIs.}
	\label{Fig1}
\end{figure*}

Through a partnership with the city of Marogogi, established since March 2020, we developed a Clinical Monitoring System to track and trial all severe acute respiratory syndrome patients, see description in {\bf { \nameref{S1_SI} \SMC}}. We also geo-localized patients and integrated this information with public data to obtain the household socio-economical data and family clusters {\bf { \nameref{S1_SI} \householdnetworks}}. The data integration is illustrated in the upper left panel of Fig \ref{Fig1}. For our study, we used the data from  \DateInitial~ to \DateFinal, consisting of 18 confirmed deaths and 119 hospitalizations. In this period 1722 tests were performed, namely 52 RT-PRC tests and 1670 antibody tests (in majority COVID-19 IgG/IgM, see {\bf { \nameref{S1_SI} \dataprocessing}} for further details).

\subsubsection*{Services} We mapped the services that were allowed to be open during the period under government regulations and interviewed a sample of businesses for an estimate of daily occupation. The bulk of such services are food stores, building supply stores, restaurants and other minor retailer services as described in { {\bf \nameref{S1_SI} \COMORBUSS}}.

\subsubsection*{Street markets} We estimated the usage of important open air services such as street markets by images collected via drones. We processed the images using the marking tool of the Drone Deploy mapping software \cite{drone_deploy} in order to evaluate the mean size and duration of cluster of people less than 2 meters apart during opening hours, as well as the average time spent by individuals in the street market. In { {\bf \nameref{S1_SI} \maragogiistypical}}, we also show that cities with demographics similar to Marogogi have analogous street market behavior.

\subsubsection*{Health services}
During the period considered, the triage of all COVID-19 related cases was performed in a field hospital. We interviewed the staff of the health secretary to obtain data on the appointment mean time, and the mean number of contacts a patient has with doctors and other patients. This also provided data on the mean number of contacts among staff, see { {\bf \nameref{S1_SI} \networkssocialdynamics}}.

\subsection*{Inference of states from data} 
We estimate the epidemiological SEIR curve from the attendance data of our Clinical Monitoring System. The SEIR curve corresponds to the trajectory of the population over the period of observation in the states: susceptible, exposed, infectious and recovered. The challenge is to transform the information of an individual reported in the attendance data into these states of the entire city population over time, correcting for sub-notification.

Under the hypothesis that all severe cases (hospitalization and death) are reported in our Clinical Monitoring System, for each reported individual we estimated the number of unreported infected individuals using a negative binomial (NB) distribution, and consequently, the total number of cases in the city over the period of observation. We modelled the total number of cases by $T = NB(p_h, 119) + 119$, where $p_h \approx 3.304\%$ is the estimated hospitalization probability for the city. We assume that these unreported individuals present their first symptoms at the same time as reported individuals.

Having all the individuals carrying the virus, we estimated how they progress across the SEIR compartments based on the severity of the case and the distribution of permanence of each state \cite{Wolfel_2020}. We rerun the statistical model 400 times to obtain SEIR curve samples for the city, see {\bf { \nameref{S1_SI} \dataprocessing}} for further details. We denote by $\hat{\nu}$ the (empirical) distribution induced by these samples, e.g., the measure given by the uniform distribution over the 400 obtained samples.

\subsection*{Agent based modeling}
We developed an agent based model, called
COMORBUSS (COmmunitary Malady Observer of Reproduction and Behavior via Universal Stochastic Simulations), gathering all biological and social aspects of the disease spreading as a contact process. This allows us to pinpoint the impact of a specific service on spreading as well as track the resultant infection tree.  To avoid overspecialized simulations of a single city, COMORBUSS stochastically produces for every simulation a realization of the transmission trajectory for the city in the class defined by the desired demographic and infrastructure data. For instance, each simulation has its own household network while satisfying the same distribution.

\subsubsection*{Modeling disease}
Each agent is characterized by its age, which determines the agent's susceptibility, probability of developing symptoms and probability of dying from the disease. When a susceptible agent encounters an infectious one (pre-symptomatic, asymptomatic, mildly or severely symptomatic), it has a probability of becoming exposed. After an incubation period, this agent becomes pre-symptomatic, and after an activation period, its state is converted to either asymptomatic, mildly or severely symptomatic. The distribution of these states is estimated empirically from actual	statistics \cite{Linton_2020_incubation_period,Verity_2020}. After a random period, agents are converted to recovered (or dead), see {\bf  \nameref{S1_SI} \COMORBUSS}.

\subsubsection*{Vaccine modeling and effect}
\begin{figure*}[hbt!]
	\centering
	\includegraphics[width=1\linewidth]{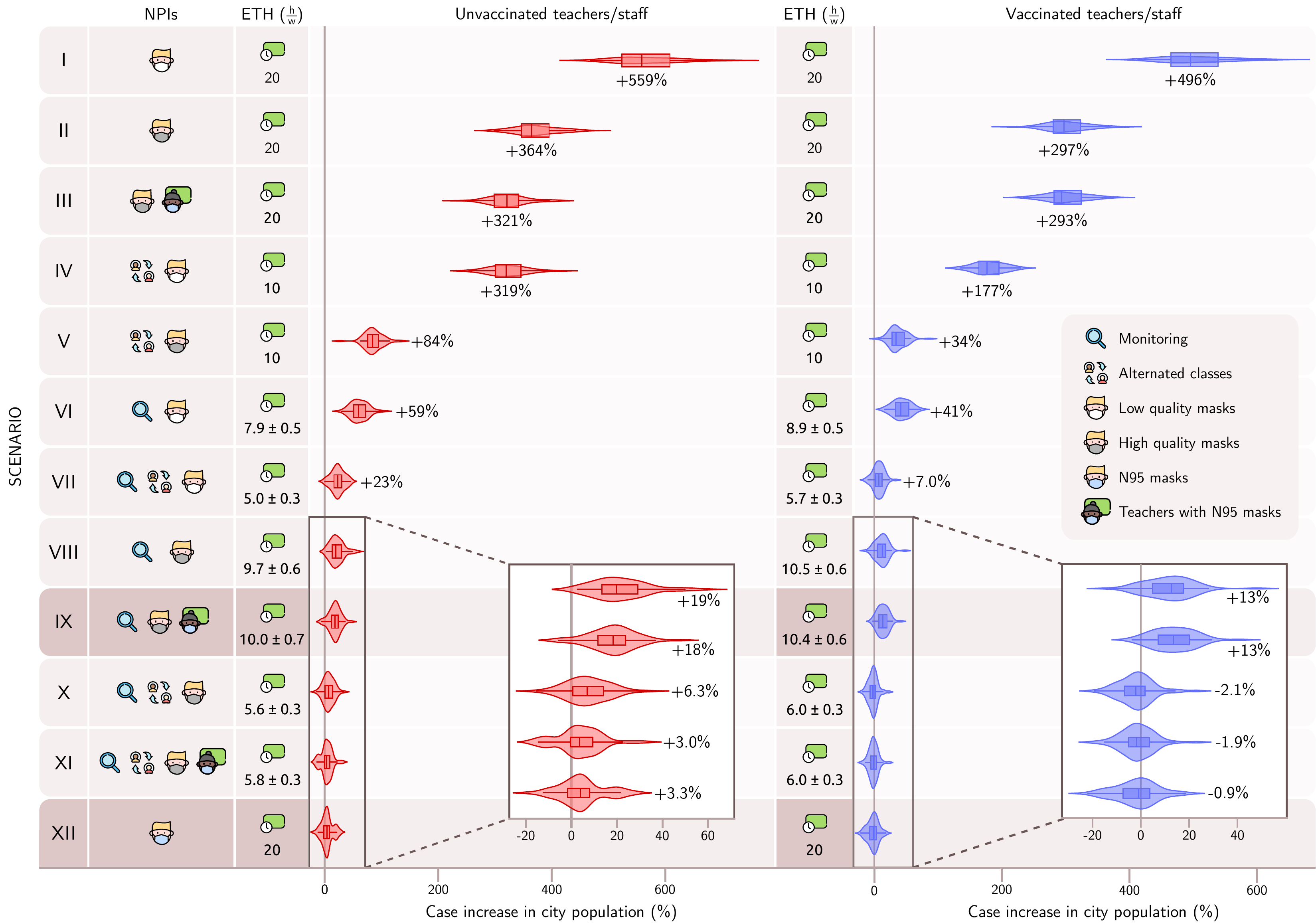}%
	\caption{\textbf{Combination of NPIs measures in comparison to the baseline model settings.} Left panel: Case increase under different scenarios with unvaccinated teachers and staff. Right panel: Case increase under different scenarios with vaccinated teachers and staff. The effective teaching hours in hours/week $\frac{h}{w}$ and case increase in school population with respect to baseline are displayed for each NPIs combination.  In case the active monitoring is also applied, the mean and standard deviation over 60 realizations for the effective teaching hours are shown. The proportional increase in the number of cases is displayed as violin plots (median, lower and upper quartiles), with kernel density estimates for distributions.}
	\label{Fig2}
\end{figure*}

As our aim is to evaluate transmission patterns under different mitigation strategies, we are naturally interested in vaccines that can also affect disease dynamics either by blocking virus infection or transmission. Unfortunately, studies of vaccine efficacy so far did not address these outcomes directly and we lack data for modeling these mechanisms. We argue, based on the results shown in Fig \ref{Fig2}, that vaccines are important mechanisms for individual protection, and, in turn, they might also have an impact at the population level by decreasing transmission of the virus to susceptible individuals.  However, this is a secondary effect: what really defines the trend for relative reduction of cases is the chosen combination of NPIs. This is seen by comparing the worst and best-case scenarios regarding vaccination (left and right in Fig \ref{Fig2}, respectively). We simulate a best-case scenario by initializing all school workers as perfectly vaccinated, such that their vaccination blocks all infection. We observe that the general picture remains unchanged within error bars, preserving almost the same ordering among scenarios. It is still unknown whether current vaccines can block infections. With all these and other unknowns in mind, we find it very reassuring that the two limiting cases we have studied are structurally similar and lead to sound conclusions regarding the effectiveness of school activity protocols.

\begin{figure}[hbt!]
	\vspace*{0.5cm} 
	\centering
	\includegraphics[width=0.8\linewidth]{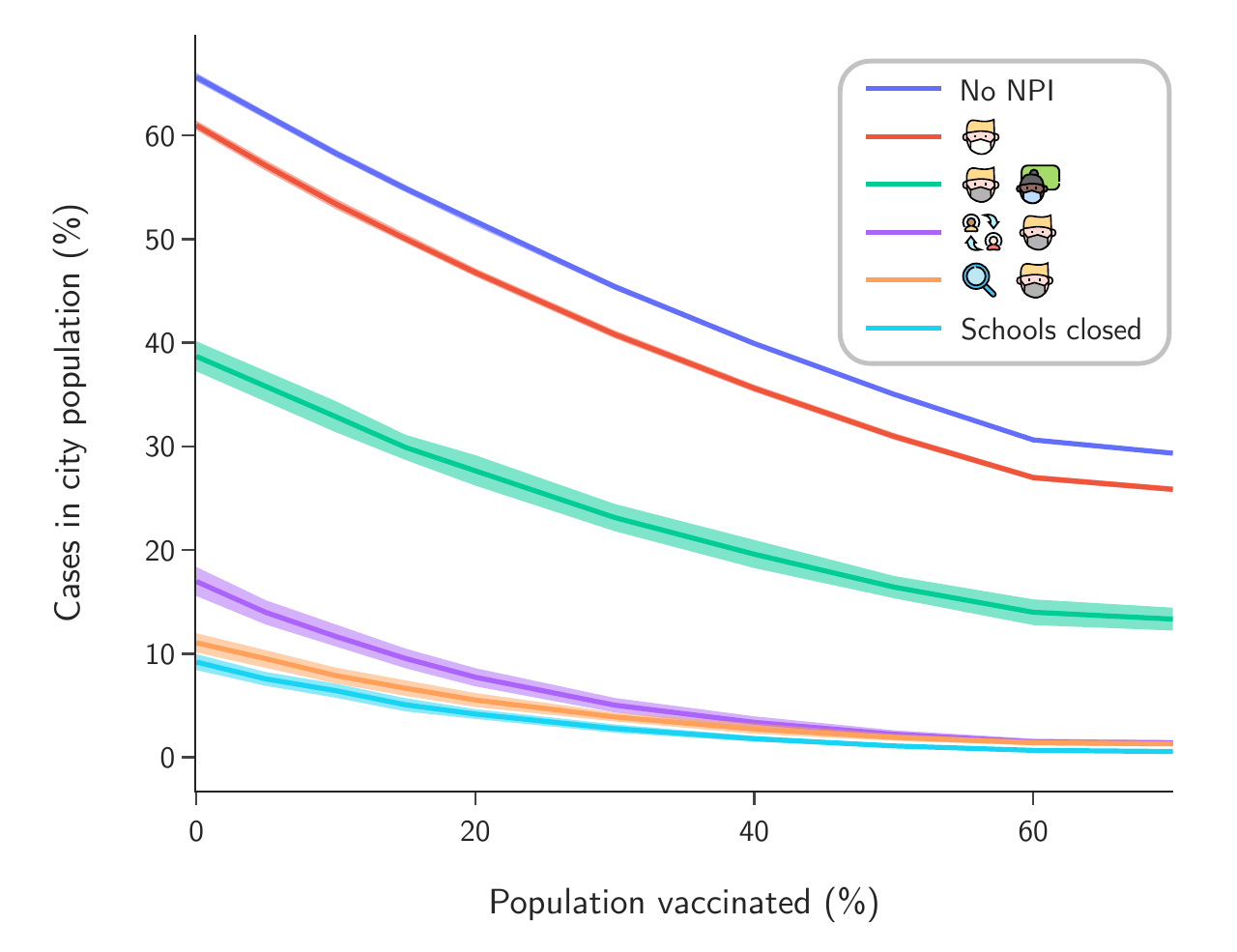}%
	\caption{\textbf{Population fraction infected at the end of the simulation period (77 days) under varying vaccination coverage.}}
	\label{Fig3}
\end{figure}

Secondly, we investigate the effects of NPI adoption under different scenarios of partial vaccination for the general population (see Fig \ref{Fig3}). Our main interest in this analysis is to evaluate the viability of the proposed measures for countries with different vaccination coverage, both in the well covered European continent and in the under-vaccinated African continent. We observe that the correct choice in NPIs can effectively protect the community even for low vaccination coverage, while poor adoption of NPIs can lead to high infection rates even for high vaccination coverage. Since we are dealing with larger segments of the population instead of just the school sub population, these simulations were performed with a more realistic vaccination model which only partially protects each agent with a biological efficacy of $98\%$ for infection, resulting in an effective vaccine efficacy of $90\%$ for the scenario where no NPIs are adopted.  Although it tends to be more realistic, this model is highly complex to adjust and interpret because the measured vaccine efficacy is closely related to the running epidemiological scenario which responds to the adopted NPIs \cite{Kaslow, Struchiner2007, Claudio_challenges}.

\subsubsection*{Modeling Services}

The city infrastructure is modeled by creating individual instances for each service (schools, hospitals, markets, restaurants, shops etc.) and by assigning agents to work/visit that location if they belong to an appropriate age group (a child may not work at a shop, and an adult may not attend class). Worker agents are relocated to that service location during their shifts, while the visits of client agents are simulated stochastically. An hourly visitation rate of a service by an agent is empirically	estimated, taking into account the service's opening hours and average visitation frequency of real clients; for details see { \bf \nameref{S1_SI} \networkssocialdynamics}. Additionally, agents may be allocated as guests to special services, which implies that their standard location is changed from their homes to that service instance. In this way we distinguish between hospitalizations, hotel quarantines and nursing home patients.

A novel point of our model is the creation of contact networks contextualized by social activity. The ratio of encounters between workers and clients as well as the clustering properties of a contact network is naturally dependent on the observed social context. For example, in restaurants there is a clustering of clients belonging to the same table, and contact between different tables is mediated by the contact of a shared waiter. Contact networks in schools, hospitals, stores etc., are all considerably different from each other. COMORBUSS updates random contact networks every hour for all the agents in the service instances, while respecting the characteristic architecture of the contact network of that type of service and distinguishing between the social roles of agents. Details and examples may be found in { \bf \nameref{S1_SI} \networkssocialdynamics}.

\subsection*{Model calibration and closed schools as baseline}

We aggregate socio-geographical data, as well as epidemiological data to COMORBUSS from May 9th 2020 to July 25th 2020, and leave the infection probability $p$ and the mean number of contacts $c$ in the City Hall to be calibrated using the empirical measure $\hat{\nu}$ obtained from the inference of states from data (see Section C). For a given $y = (q, d) \in [0,1] \times \mathbb{R}_{+}$ we denote by $\hat{\mu}^y$ the empirical measure given by 400 independent realizations of COMORBUSS with $p$ and $c$ chosen as $(p, c)=y$. We construct an estimate $\hat{x}$ for $x=(p,c)$ by minimizing over all $y$ the $L_1$-Wasserstein distance between $\hat{\nu}$ and $\hat{\mu}^y$, see {\bf   \nameref{S1_SI} \calibrationinitialization}.

We initialize the community according to its demographics and household distribution, see {\bf  \nameref{S1_SI} \calibrationinitialization}. The disease state of agents is proportional to the average inferred epidemiological data for day May 9th 2020. The calibrated model is in excellent agreement with the estimated data and we use it as a baseline.  This scenario resulted in an average of $3007 \pm 249$ new infections in the population, in which $25\%$ of those infections occurred in the school population, a measure that will serve as a baseline for keeping schools open in study cases.

\subsection*{Poorly ventilated classrooms}
In poorly ventilated classrooms, the main transmission mechanism is by aerosols emitted by an infected agent. The aerosols can remain suspended in the air, thereby reaching agents far from the original emitters \cite{MORAWSKA_2020_world_reality,Poydenot_2021_risk_assessment}. To model this exhaled air without reference to the microscopic pathogen concentration, we follow the exposition in \cite{Miller_Skagit_2021,Bazant_PNAS}, describing the evolution of concentration of quanta in a closed space. \textit{Quanta, introduced by Wells, measure the expected} rate of disease transmission, interpreted as infection quanta transference between pairs of infected and susceptible agent \cite{Riley_1978_airborne}. 

\begin{figure}[hbt!]
	\centering
	\includegraphics[width=0.8\linewidth]{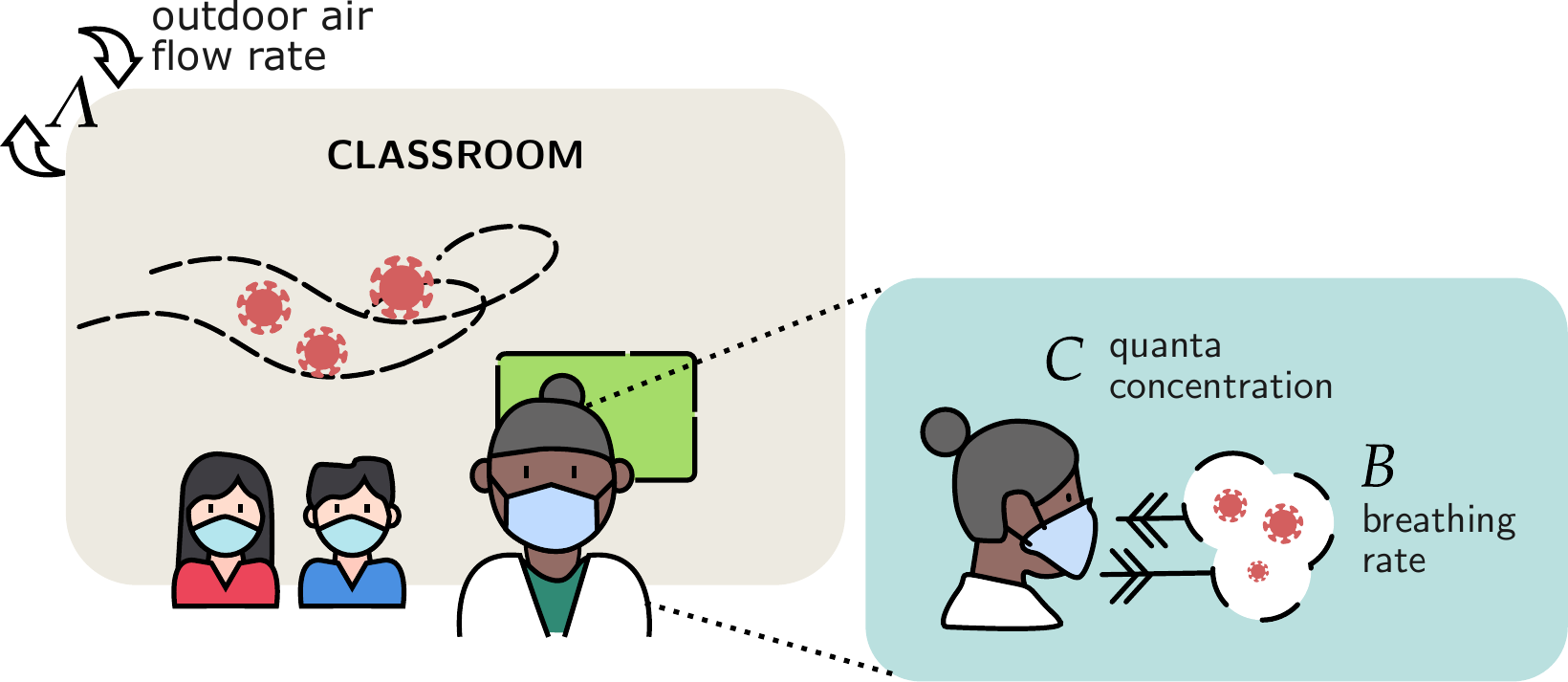}%
	\caption{\textbf{Airborne transmission model inside school environment.} The classroom is an enclosed space in which airborne transmission has a high chance of occurrence. Contaminated particles are spread over the classroom, allowing long range infections. The fresh air rate flow $\Lambda$ quantifies the classroom ventilation. The quanta concentration $C$ varies in the environment depending on the breathing activity. }
	\label{Fig4}
\end{figure}

In our model, we denote by $C$ ($quanta/m^3$) the total concentration of quanta inside a classroom of volume $V$. Classrooms contain a total of $N$ agents, with $S$ susceptible	individuals, $I_s$ infected students and $I_t$ infected teachers. All breathe uniformly at a rate $B = 0.5~m^3/h$. Since mask wearing can decrease the amount of aerosols emitted to the air, we denote for each agent the penetration mask factor $p_{m}^i \in (0, 1)$, with $i=s,t$: see Fig \ref{Fig4}.

Each person exchanges quanta with the air depending on breathing activity. We introduce the concentration of quanta expelled by students $C_s = 40$ ($quanta/m^3$) and teachers $C_t = 72$ ($quanta/m^3$) \cite{Bazant_PNAS}(corresponding to voiced counting \cite{MORAWSKA_2009_droplets_size}).
Under a well-mixed room assumption, the total concentration of quanta $C$ ($quanta/m^3$) inside the classroom satisfies the mass equation:
\begin{align}\label{eq:mass_balance_equation}
V \frac{d C}{d t} = -(\Lambda V + N B) C + B (C_s I_s p_{m}^s  + C_t I_t p_{m}^t).
\end{align}
Note that our setting relies on the fact that the airborne particles remain airborne before being extracted by the outdoor air flow $\Lambda$ (typically  reported as air changes per hour or ACH) or inhaled by an agent. We investigate the poor ventilation limit $\Lambda = 0$ and fresh air flow ($\Lambda > 0$), see { {\bf \nameref{S1_SI} \airborne}}.

The amount of quanta inhaled by the $i$th agent inside the class over a time $t$ is the inhaled dose  $D_i(t) =  B p_m^i \int_0^{t} C(t) d t$. We evaluate this integral over the solution to Equation~\eqref{eq:mass_balance_equation}. Using the inhaled dose of each agent, we plug it into the Wells-Riley model to calculate the probability of a susceptible individual being infected \cite{Miller_Skagit_2021,Poydenot_2021_risk_assessment}, which consists in  estimating the risk of infection in indoor environments via
\begin{align*}
p_{\rm indoor}^i(t) = 1 - e^{-r_i D_i(t)},
\end{align*}
where $r_i$ is the relative susceptibility (an age-based measure \cite{Zhang_2020_susceptibility}) for the agent $i$. We set the relative susceptibility of children (aged 0 years to 14 yr), adults (aged 15 yr to 64 yr) and the elderly (over 65 yr) to $r_i$ = 0.23, 0.68, 1, respectively. To determine the source of infection of a particular exposed individual, we pick a random individual uniformly from all of the infectious individuals in the enclosed space, see { {\bf \nameref{S1_SI} \airborne}}.

\section*{Results and Discussion}

\subsection*{NPIs and vaccination}
\begin{figure}[hbt!]
	\centering
	\includegraphics[width=0.7\columnwidth]{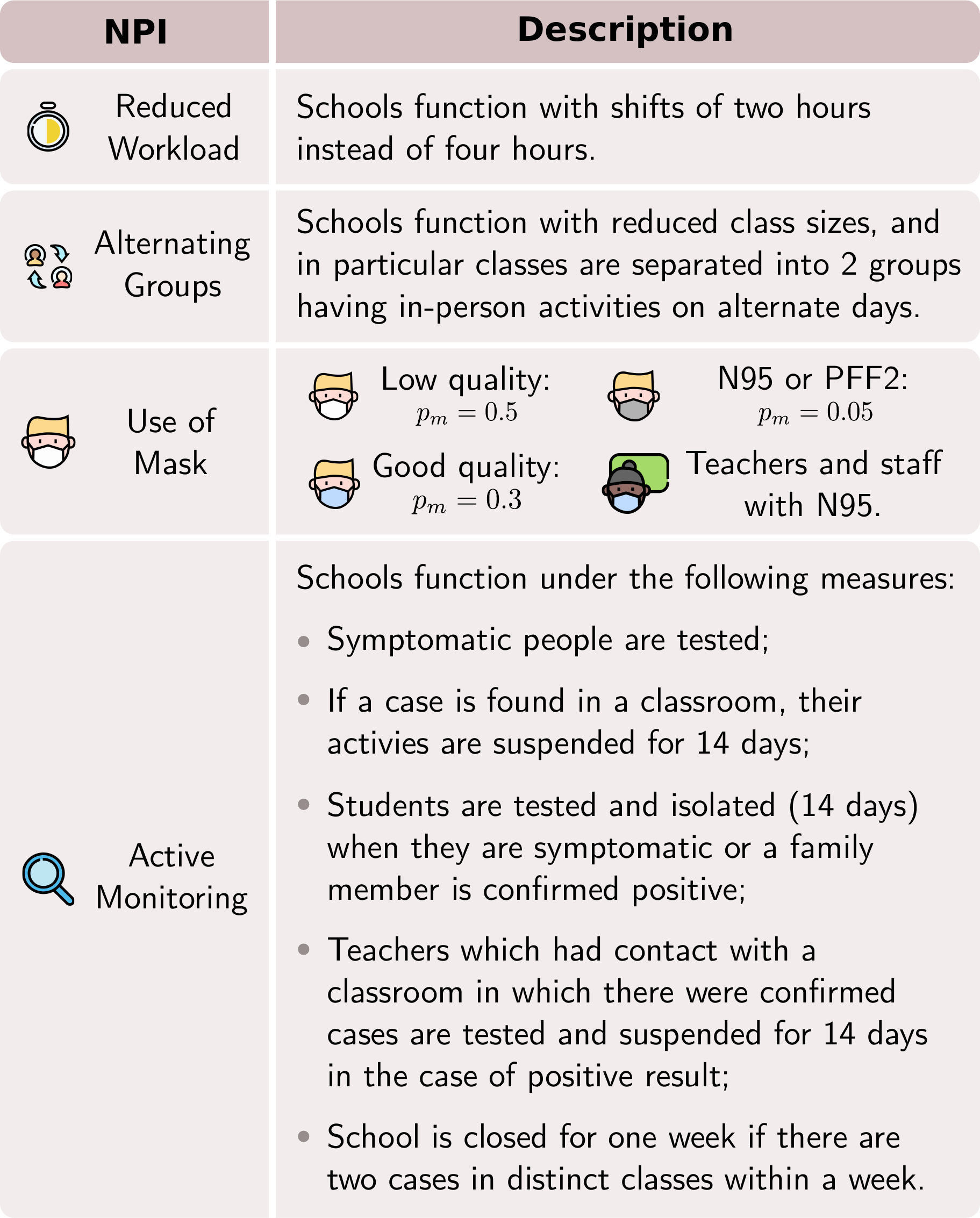}%
	\caption{\textbf{NPIs description}. The icons distinguish the non-pharmaceutical interventions evaluated in this study. In scenarios involving masks, the mask penetration factor $p_m$ is uniform for all individuals, except for teachers wearing PFF2 masks.}
	\label{Fig5}
\end{figure}

Across 27 schools, the total school population is 8,528, with 7,557 students. We quantified the effects of five NPIs on the school population, which consists of teachers, school staff and students. Each NPI is described in Fig \ref{Fig5}. Although there is still controversy in the literature about the efficiency of surgical masks for filtering particles \cite{Cheng_science_mask_2021} and side-effects \cite{Kisielinski_2021_mask_sideffect}, we assign mask quality via their permeability factors $p_m$, as indicated in Fig \ref{Fig5}.

We simulate school activities with different NPI and compute the percentage increase of cases with respect to the baseline. The results are presented in Fig \ref{Fig2} along with the effective teaching hours. Conducting classes in full shift and wearing only poor quality masks leads to a $559 \%$ increase in infections. We note that the wearing of N95 masks by teachers and staff is particularly effective at reducing the number of cases compared to other scenarios, and we highlight this NPI in Fig \ref{Fig2} (darker color). Active monitoring curbs spreading, at the expense of the effective number of teaching hours.

We assume in the simulation that vaccinated teachers and staff are initialized with protective neutralizing antibodies against COVID-19. This blocks any possible infection chain starting from these individuals. The right panel in Fig \ref{Fig2} displays the effectiveness of NPI combinations with vaccinated employees. If employees are not vaccinated, case rates increase in all scenarios. The case increases in the highlighted (darker color) scenarios are reduced for both unvaccinated and vaccinated employees, indicating that they are a potential source of infection for the school population.

We also analyze the robustness of our results when considering a larger city, using as example the regional capital of Curitiba with almost 2 million inhabitants. We observe how bad protocols lead to sharp increase in infections while good ones successfully avoid this phenomena. Most remarkably, the relative effectiveness rank between intervention is preserved, even if the case increase relative to the baseline is less pronounced, see further details in { {\bf \nameref{S1_SI} \robustness}}. This not only shows the stability of the protocols but also indicates that smaller cities are more vulnerable and in need of appropriate protocols.

We also consider the effectiveness of NPI scenarios under different levels of vaccination coverage, see Fig \ref{Fig3}. Our motivation is to asses the viability and safety of public health decisions even in countries with low coverage, such as African countries. In fact, even with low vaccination coverage, we find that a good choice of NPIs in schools also protects the wider community better. At the same time, poorly chosen or non-existent NPIs may leave the communities highly exposed, regardless of vaccination coverage. We therefore stress the importance of appropriate NPIs and protocols, whether or not the underlying country enjoys good vaccine coverage. We recall that the cities are modelled with only essential services operating, including schools. Lessons drawn here extend to other services and social contexts to avoid the worsening of outbreaks.

\subsection*{Sensitivity analysis: mask penetration and ventilation}

\begin{figure*}[hbt!]
	\centering
	\includegraphics[width=0.8\textwidth]{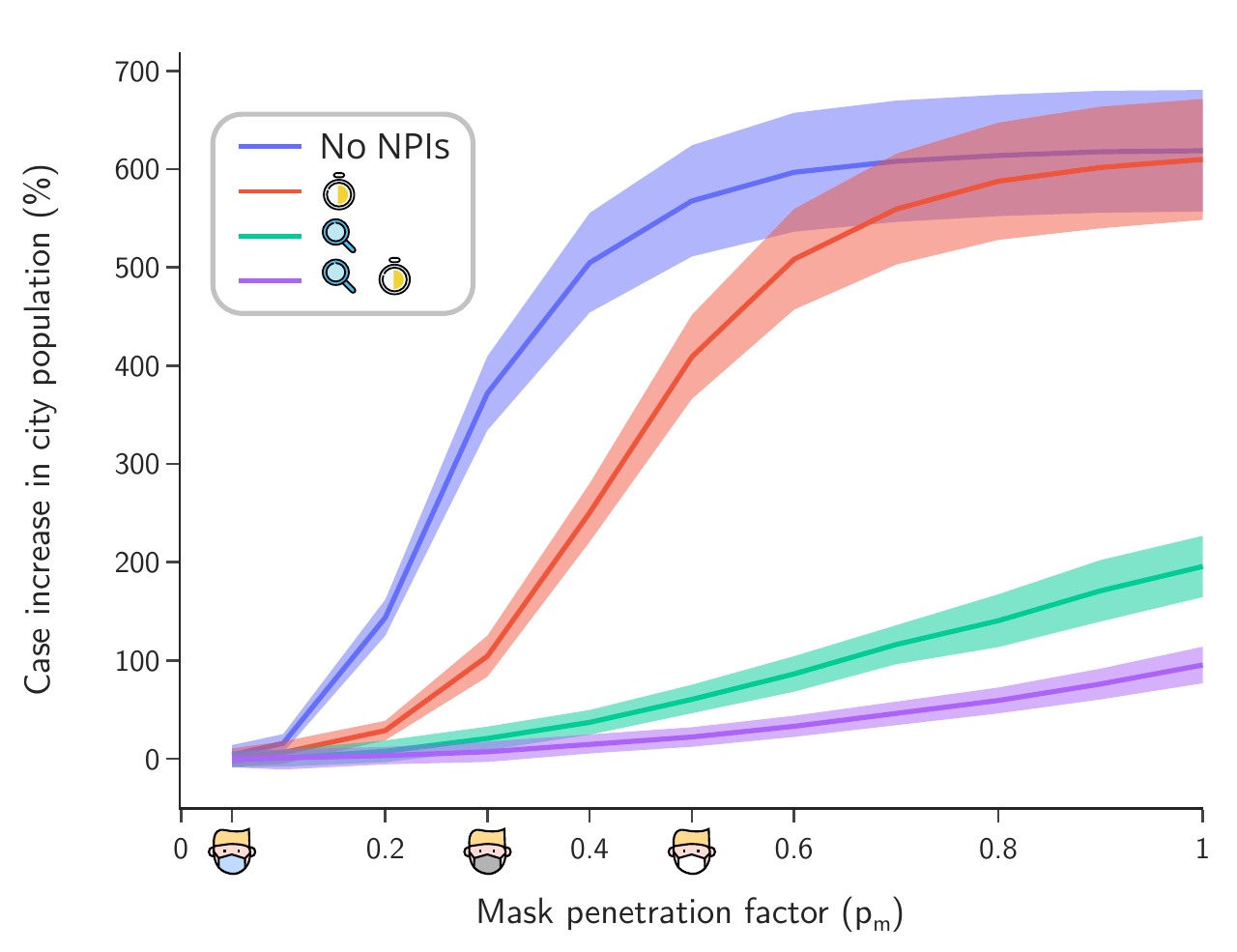}
	\caption{\textbf{Sensitivity analysis across mask penetration factor $p_m$.} Cases increase in school population (solid lines) versus the mask penetration (mean values over 60 realizations for each $p_m$ value).}
	\label{Fig6}
\end{figure*}

We quantify the relevance of the mask penetration factor $p_m$ and ventilation air flow rate $\Lambda$ for the increase of COVID-19 cases in the cities. Assuming that all pupils wear masks with the same $p_m$, Fig \ref{Fig6} shows the impact of the penetration factor on the number of cases if schools are kept open. Results are sensitive to the penetration factor of the masks, as seen by comparing the first (poor quality or practices, $p_m = 0.5$) and second (high quality masks, $p_m = 0.3$) simulation scenarios, showing a decrease of almost $200$\% in cases regardless of the vaccination status of employees. We also observe that the use of N95 masks by employees increases the effective teaching hours in the scenarios with active monitoring.

\begin{figure*}[hbt!]
	\centering
	\includegraphics[width=0.8\textwidth]{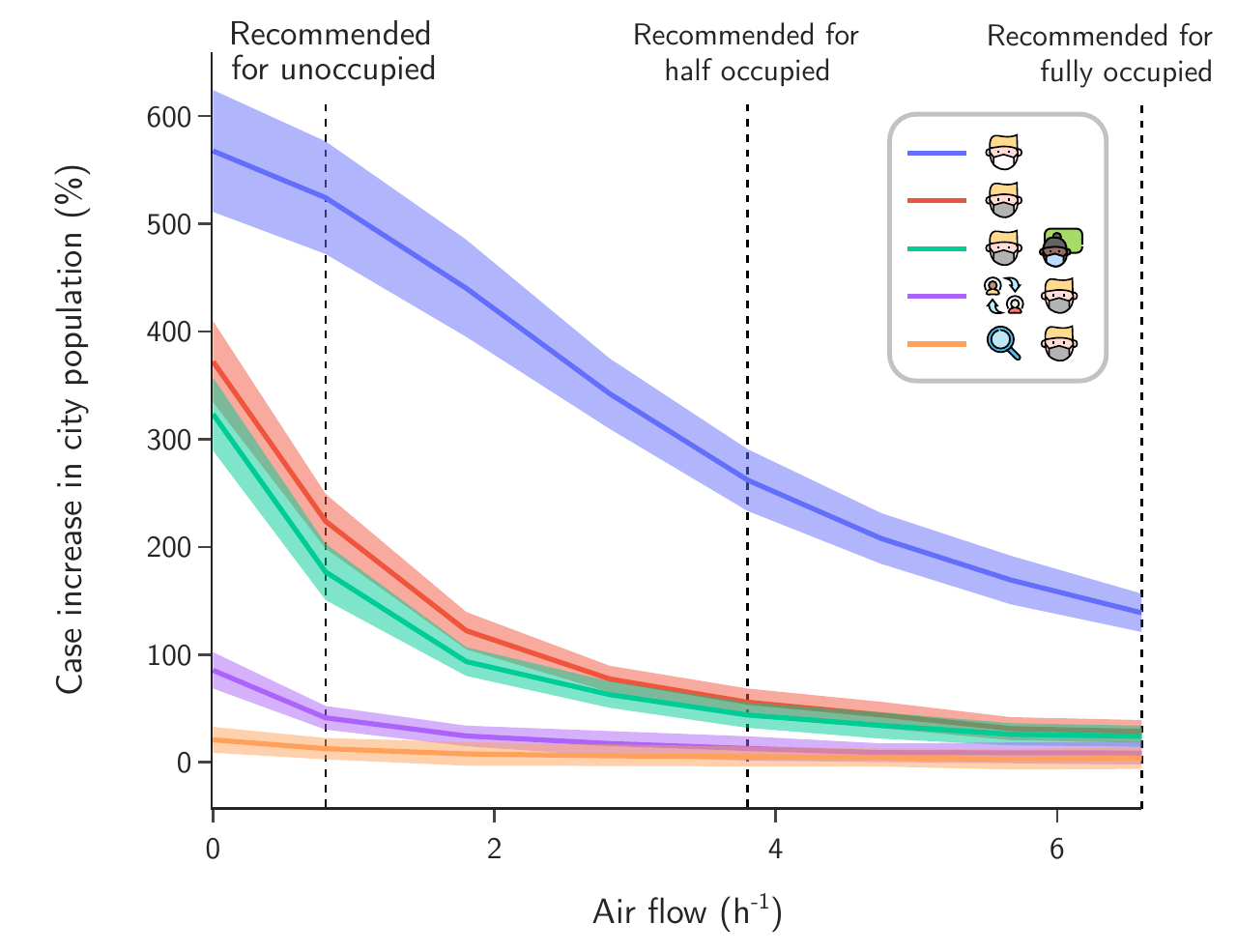}
	\caption{\textbf{Sensitivity analysis across ventilation $\Lambda$.} Cases increase in school population (mean and standard deviation) as a function of classroom ventilation rate. Dashed lines indicate the recommended ventilation rates: $\Lambda_1 = 0.8 ~h^{-1}$ (unoccupied room), $\Lambda_2 = 3.8 ~h^{-1}$ (half occupied room) and $\Lambda_3 = 6.6 ~h^{-1}$ (fully occupied room), following ASHRAE standard for an average classroom in Maragogi.}
	\label{Fig7}
\end{figure*}

Fig \ref{Fig7} shows the sensitivity analysis when the ventilation rate is varied inside classrooms. Based on recomendations by the American Society of Heating, Refrigerating and Air Conditioning Engineers (ASHRAE)\footnote{ASHRAE 62.1-2019 (ASHRAE 62.1) - Ventilation for Acceptable Indoor Air Quality.}, we calculated the minimal ventilation rate of $\Lambda_1 = 0.8 ~h^{-1}$ for unoccupied classrooms using their average dimensions in Maragogi. Ventilation rates for half full and full classrooms are $\Lambda_2 = 3.8 ~h^{-1}$ and $\Lambda_3 = 6.6 ~h^{-1}$, respectively; for further details see {\bf { \nameref{S1_SI} \airborne}}.

\subsection*{Scenarios with more infectious variants}
In investigating the effectiveness of school safety protocols during infection waves caused by new, more infectious variants, we are drawn to the limiting worst case scenarios. As such, we assume that the new variant completely avoids the acquired immunity from vaccination or previous infections. New variants are modeled by an increase in the population susceptibility, therefore encompassing both our contact and aerosol transmission models. Susceptibility is increased by the multiplying factor over all age groups as a limiting case.

\begin{figure}[hbt!]
	\centering
	\includegraphics[width=0.8\linewidth]{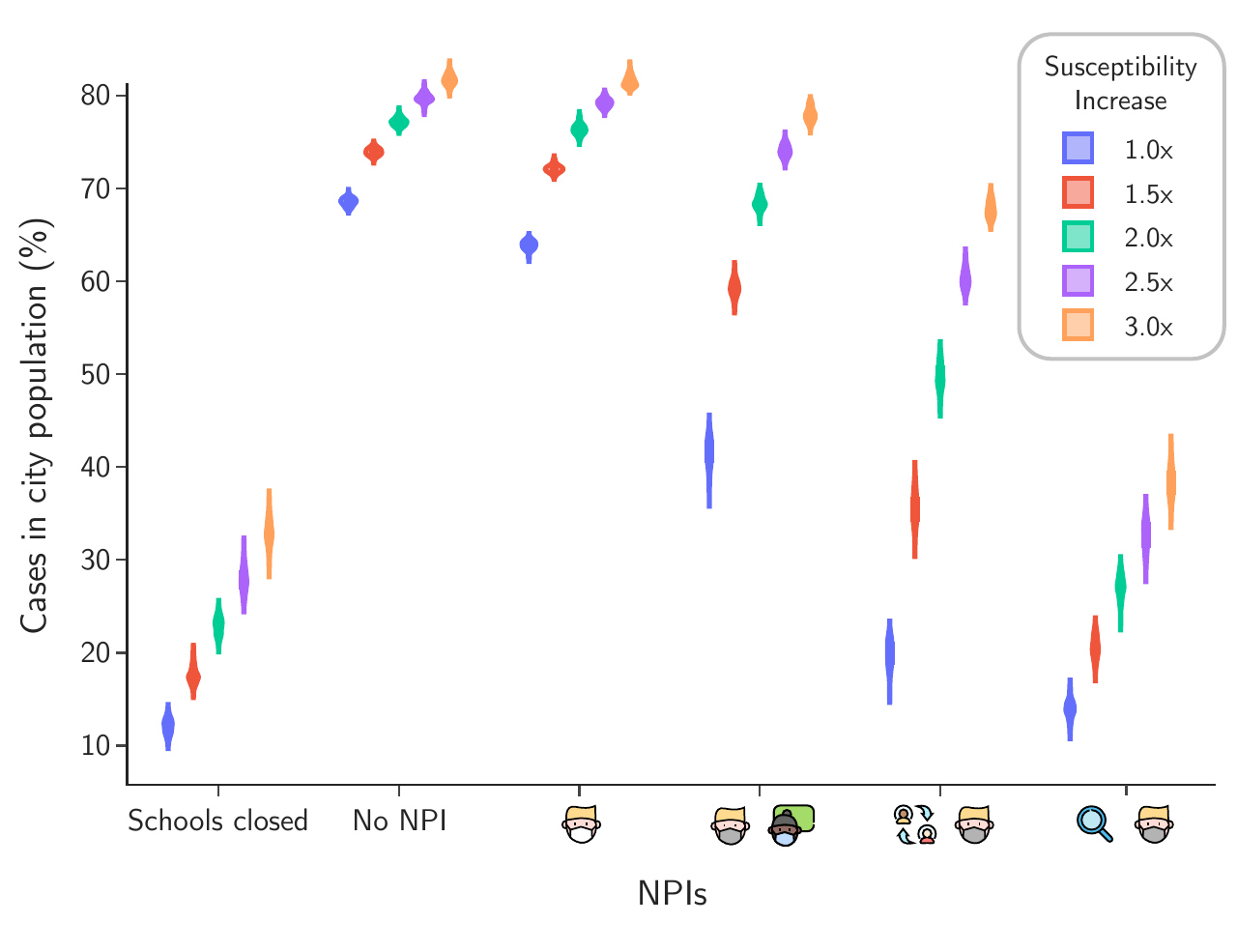}
	\caption{\textbf{Population infected in case of increase in susceptibility.} For each intervention scenario, we show the distribution in the percentile of the population infected provided the susceptibility of the population is increased uniformly by a multiplying factor.}
	\label{Fig8}
\end{figure}

The results are depicted in Fig \ref{Fig8}. As expected, the total population infected increases monotonously with the increase in susceptibility, with poor protocols for school activities leading to extreme infection rates across the community. Most importantly, not only good protocols still lead to remarkable decrease in infection rates but the relative rank of effectiveness between protocols is preserved regardless of how much susceptibility is increased. This shows the stability of good protocols and makes the point that their adoption should always be a top priority even when facing new potentially variants.

\section*{Conclusion}

The airborne transmission mechanism of COVID-19 is the main cause of infections in school environments in classrooms with poor air circulation. Since many classrooms are equipped with air conditioning or heating, most have poor air circulation. Therefore, reducing the class size does not necessarily curb spreading because an infected person can emit aerosols that stay in the air and infect students far away in the same classroom.

Vaccination of employees is an essential measure. Still, in the absence of other measures such as monitoring and quarantines, the number of cases in the cities is likely to increase by $177$\% if only the use of low quality masks and alternated classes are implemented.

The penetration factors provided by manufactures and used in our simulations are idealized. In practice, the fit of a mask and the practices of users result in lower filtration efficacy. Indeed, after testing a model of contagion based on a study of Canadian classrooms \cite{Hou_2021_school_canada}, we compared the ensuing results with our own aerosol model under the same class conditions but varying penetration factors, in order to estimate its value in these classrooms. We were alarmed to find that the effective penetration factor for the Canadian classrooms in that study was only 0.5, despite the assumption of high quality masks. It would therefore be of great benefit to instruct the general population in proper mask use. Otherwise, the potential effectiveness of sanitary protocols will be compromised as the achieve penetration factor increases (Fig \ref{Fig6}).

All these findings can be explained by three facts: teachers are more susceptible than children, they expel more virulent particles since they are constantly speaking loudly and they are the most effective bridges of transmission between isolated classes. Therefore, high quality masks not only protect the individual teacher, but also suppress community infection.

Our most striking result is that one must adopt the appropriate NPIs and behavioral protocols in order to safely continue school activities during a pandemic, regardless of vaccination coverage. Good protocols can protect countries even with poor vaccine coverage. Conversely, bad protocols may seriously aggravate the underlying public health crises even in countries with very high vaccination coverage. This is in great part due to the long duration of social contacts in schools, easily leading to breakthrough infections without proper protocols. This is particularly relevant given that in many countries children are not routinely vaccinated for COVID-19, or when preparing for the emergence of new variants with potentially low cross immunity.

There is no single solution to a pandemic, but we draw hope in showing that the proper combination of NPIs, vaccination and behaviors permit the safe continuation of activities as fundamental and important as teaching.

\section*{Supporting information}


\paragraph*{S1 File.}
\label{S1_SI}
{\bf Supplementary Information.} Detailed description of data collection, data analysis, COMORBUSS software, calibration, and sensibility analysis.

\section*{Acknowledgments}
This work was supported by  the Center for Research in Mathematics Applied to Industry (FAPESP grants 2013/07375-0), by the Royal Society London, by the Brazilian National Council for Scientific and Technological Development (CNPq; grants 301778/2017-5, 302836/2018-7, 304301/2019-1, 306090/2019-0, 403679/2020-6) and by the Serrapilheira Institute (Grant No. Serra-1709-16124). GTG has received funding from the European Union Horizon 2020 research and innovation programme under the Marie Skłodowska-Curie grant agreement No 765048. ERS was supported by FAPESP grant 2018/10349-4. KO and SHAL acknowledge the project promat-maragogi and CJS acknowledges the financial support of CNPq and FAPERJ, DM received funding from the grants CNPq-306566/2019-2 and FAPERJ - E-26/202.764/2017. The graphics used to prepare the Figures were extracted from FontAwesome and FreePick.


%
%
%
%
%
%
%


\begin{thebibliography}{10}

\bibitem{WB_Unesco}
{The State of the Global Education Crisis: A Path to Recovery}; 2021.
\newblock \url{https://www.unicef.org/reports/state-global-education-crisis}.

\bibitem{ECDC_schools}
{COVID-19 in children and the role of school settings in transmission - second
  update}; 2021.
\newblock
  \url{https://www.ecdc.europa.eu/sites/default/files/documents/COVID-19-in-children-and-the-role-of-school-settings-in-transmission-second-update.pdf}.

\bibitem{Claudio_questions}
Thompson RN, et~al CJS.
\newblock Key questions for modelling COVID-19 exit strategies.
\newblock Proceedings of the Royal Society B. 2020;287.
\newblock doi:{https://doi.org/10.1098/rspb.2020.1405}.

\bibitem{Lessler_2021_in_person_schooling}
Lessler J, Grabowski MK, Grantz KH, Badillo-Goicoechea E, Metcalf CJE,
  Lupton-Smith C, et~al.
\newblock {Household COVID-19 risk and in-person schooling}.
\newblock Science. 2021;372(6546):1092--1097.
\newblock doi:{10.1126/science.abh2939}.

\bibitem{Gurdasani_2021_lancet}
Gurdasani D, Alwan NA, Greenhalgh T, Hyde Z, Johnson L, McKee M, et~al.
\newblock {School reopening without robust COVID-19 mitigation risks
  accelerating the pandemic}.
\newblock The Lancet. 2021;397(10280):1177--1178.
\newblock doi:{10.1126/science.abh2939}.

\bibitem{Munday_2021_estimation_UK}
Munday JD, I~Jarvis C, Gimma A, Wong KL, Zandvoort Kv, Group CCW, et~al.
\newblock {Estimating the impact of reopening schools on the reproduction
  number of SARS-CoV-2 in England, using weekly contact survey data}.
\newblock medRxiv. 2021;~.
\newblock doi:{10.1101/2021.03.06.21252964}.

\bibitem{Tiago_Claudio_PNAS}
Silva PJS, Sagastizábal C, Nonato LG, Struchiner CJ, Pereira T.
\newblock Optimized delay of the second COVID-19 vaccine dose reduces ICU
  admissions.
\newblock Proceedings of the National Academy of Sciences of the United States
  of America. 2021;118(35).
\newblock doi:{https://doi.org/10.1073/pnas.2104640118}.

\bibitem{IBGE_maragogi}
{IBGE panorama --- Maragogi-AL }; 2020.
\newblock \url{https://cidades.ibge.gov.br/brasil/al/maragogi/panorama}.

\bibitem{us_cities}
{United States Cities Database }; 2021.
\newblock \url{https://simplemaps.com/data/us-cities}.

\bibitem{world_cities}
{World Cities Database}; 2021.
\newblock \url{https://simplemaps.com/data/world-cities}.

\bibitem{drone_deploy}
Drone Deploy mapping software; 2020.
\newblock \url{https://www.dronedeploy.com/}.

\bibitem{Wolfel_2020}
W{\"o}lfel R, Corman VM, Guggemos W, Seilmaier M, Zange S, M{\"u}ller MA,
  et~al.
\newblock Virological assessment of hospitalized patients with COVID-2019.
\newblock Nature. 2020;581(7809):465--469.
\newblock doi:{10.1038/s41586-020-2196-x}.

\bibitem{Linton_2020_incubation_period}
Linton NM, Kobayashi T, Yang Y, Hayashi K, Akhmetzhanov AR, Jung Sm, et~al.
\newblock Incubation Period and Other Epidemiological Characteristics of 2019
  Novel Coronavirus Infections with Right Truncation: A Statistical Analysis of
  Publicly Available Case Data.
\newblock Journal of Clinical Medicine. 2020;9(2).
\newblock doi:{10.3390/jcm9020538}.

\bibitem{Verity_2020}
Verity R, Okell LC, Dorigatti I, Winskill P, Whittaker C, Imai N, et~al.
\newblock Estimates of the severity of coronavirus disease 2019: a model-based
  analysis.
\newblock The Lancet Infectious Diseases. 2020;20(6):669--677.
\newblock doi:{10.1016/S1473-3099(20)30243-7}.

\bibitem{Kaslow}
Kaslow DC.
\newblock Force of infection: a determinant of vaccine efficacy?
\newblock npj Vaccines. 2021;6.
\newblock doi:{doi.org/10.1038/s41541-021-00316-5}.

\bibitem{Struchiner2007}
STRUCHINER CJ, HALLORAN ME.
\newblock Randomization and baseline transmission in vaccine field trials.
\newblock Epidemiology and Infection. 2007;135(2):181–194.
\newblock doi:{10.1017/S0950268806006716}.

\bibitem{Claudio_challenges}
Madewell ZJ, Dean NE, Berlin JA, Coplan PM, Davis KJ, Struchiner CJ, et~al.
\newblock Challenges of evaluating and modelling vaccination in emerging
  infectious diseases.
\newblock Epidemics. 2021;37:100506.
\newblock doi:{https://doi.org/10.1016/j.epidem.2021.100506}.

\bibitem{MORAWSKA_2020_world_reality}
Morawska L, Cao J.
\newblock {Airborne transmission of SARS-CoV-2: The world should face the
  reality}.
\newblock Environment International. 2020;139:105730.
\newblock doi:{https://doi.org/10.1016/j.envint.2020.105730}.

\bibitem{Poydenot_2021_risk_assessment}
Poydenot F, Abdourahamane I, Caplain E, Der S, Haiech J, Jallon A, et~al.
\newblock Risk assessment for long and short range airborne transmission of
  SARS-CoV-2, indoors and outdoors, using carbon dioxide measurements.
\newblock medRxiv. 2021;~.
\newblock doi:{10.1101/2021.05.04.21256352}.

\bibitem{Miller_Skagit_2021}
Miller SL, Nazaroff WW, Jimenez JL, Boerstra A, Buonanno G, Dancer SJ, et~al.
\newblock {Transmission of SARS-CoV-2 by inhalation of respiratory aerosol in
  the Skagit Valley Chorale superspreading event}.
\newblock Indoor Air. 2021;31(2):314--323.
\newblock doi:{https://doi.org/10.1111/ina.12751}.

\bibitem{Bazant_PNAS}
Bazant MZ, Bush JWM.
\newblock {A guideline to limit indoor airborne transmission of COVID-19}.
\newblock Proceedings of the National Academy of Sciences. 2021;118(17).
\newblock doi:{10.1073/pnas.2018995118}.

\bibitem{Riley_1978_airborne}
RILEY EC, MURPHY G, RILEY RL.
\newblock {AIRBORNE SPREAD OF MEASLES IN A SUBURBAN ELEMENTARY SCHOOL}.
\newblock American Journal of Epidemiology. 1978;107(5):421--432.
\newblock doi:{10.1093/oxfordjournals.aje.a112560}.

\bibitem{MORAWSKA_2009_droplets_size}
Morawska L, Johnson GR, Ristovski ZD, Hargreaves M, Mengersen K, Corbett S,
  et~al.
\newblock {Size distribution and sites of origin of droplets expelled from the
  human respiratory tract during expiratory activities}.
\newblock Journal of Aerosol Science. 2009;40(3):256--269.
\newblock doi:{https://doi.org/10.1016/j.jaerosci.2008.11.002}.

\bibitem{Zhang_2020_susceptibility}
Zhang J, Litvinova M, Liang Y, Wang Y, Wang W, Zhao S, et~al.
\newblock Changes in contact patterns shape the dynamics of the COVID-19
  outbreak in China.
\newblock Science. 2020;368(6498):1481--1486.
\newblock doi:{10.1126/science.abb8001}.

\bibitem{Cheng_science_mask_2021}
Cheng Y, Ma N, Witt C, Rapp S, Wild PS, Andreae MO, et~al.
\newblock {Face masks effectively limit the probability of SARS-CoV-2
  transmission}.
\newblock Science. 2021;372(6549):1439--1443.
\newblock doi:{10.1126/science.abg6296}.

\bibitem{Kisielinski_2021_mask_sideffect}
Kisielinski K, Giboni P, Prescher A, Klosterhalfen B, Graessel D, Funken S,
  et~al.
\newblock {Is a Mask That Covers the Mouth and Nose Free from Undesirable Side
  Effects in Everyday Use and Free of Potential Hazards?}
\newblock International Journal of Environmental Research and Public Health.
  2021;18(8).
\newblock doi:{10.3390/ijerph18084344}.

\bibitem{Hou_2021_school_canada}
Hou D, Katal A, Wang LL.
\newblock Bayesian Calibration of Using CO2 Sensors to Assess Ventilation
  Conditions and Associated COVID-19 Airborne Aerosol Transmission Risk in
  Schools.
\newblock medRxiv. 2021;~.
\newblock doi:{10.1101/2021.01.29.21250791}.

\end{thebibliography}

\end{document}


\maketitle

\begin{flushleft}
	{\Large
		\textbf\newline{Quantifying protocols for safe school activities} 
	}
	\newline
	\\
	Juliano Genari\textsuperscript{1$\dagger$},
	Guilherme Tegoni Goedert\textsuperscript{2,3,4$\dagger$},
	S\'ergio H. A. Lira\textsuperscript{5},
	Krerley Oliveira\textsuperscript{5},
	Adriano Barbosa\textsuperscript{5},
	Allysson Lima\textsuperscript{6},
	Jos\'e Augusto Silva\textsuperscript{5},
	Hugo Oliveira\textsuperscript{5},
	Maur\'icio Maciel\textsuperscript{5},
	Ismael Ledoino\textsuperscript{7},
	Lucas Resende\textsuperscript{8},
	Edmilson Roque dos Santos\textsuperscript{1},
	Dan Marchesin\textsuperscript{8},
	Tiago Pereira\textsuperscript{1},
	Claudio J. Struchiner\textsuperscript{9*},
	\\
	\bigskip
	\textbf{1} Instituto de Ci\^encias Matem\'aticas e Computa\c{c}\~ao, Universidade de S\~ao Paulo,  S\~ao Paulo, S\~ao Paulo, Brazil
	\\
	\textbf{2} Dipartimento di Fisica, Universit\`a degli Studi di Roma Tor Vergata, Rome, Lazio, Italy
	\\
	\textbf{3} Aachen Institute for Advanced Study in Computational Engineering Science, RWTH AACHEN University, Aachen, North Rhine-Westphalia, Germany
	\\
	\textbf{4} Computation-based Science and Technology Research Center, The Cyprus Institute, Nicosia, District of Nicosia, Cyprus
	\\
	\textbf{5} Universidade Federal de Alagoas, Maceió, Alagoas, Brazil
	\\
	\textbf{6} Universidade Federal de Grande Dourados, Dourados, Mato Grosso do Sul, Brazil
	\\
	\textbf{7} Laborat\'orio Nacional de Computa\c{c}\~ao Cient\'ifica, Petrópolis, Rio de Janeiro, Brazil
	\\
	\textbf{8} Instituto de Matem\'atica Pura e Aplicada, Rio de Janeiro, Rio de Janeiro, Brazil
	\\
	\textbf{9} Escola de Matemática Aplicada, Funda\c{c}\~ao Get\'ulio Vargas, Rio de Janeiro, Rio de Janeiro, Brazil
	\\
	\bigskip
	
	%
	%
	$\dagger$ These authors contributed equally to this work.
	* Corresponding author
\end{flushleft}	

\tableofcontents
\printnomenclature

\chapter{Maragogi-AL put into context}
\label{sec:maragogi_is_typical}

\begin{figure}[h]
    \centering
    \includegraphics[width = .8\textwidth]{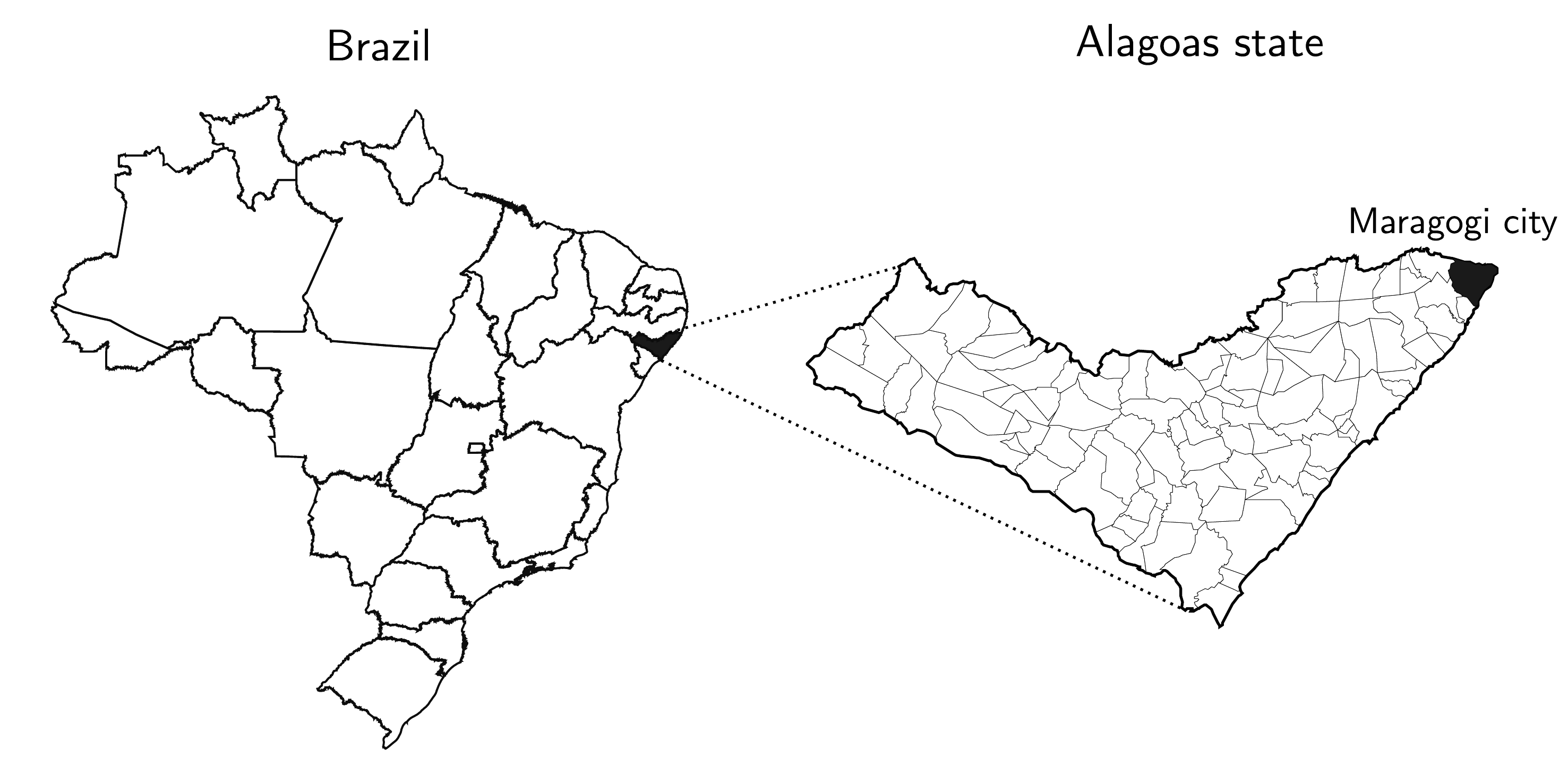}
    \caption{\textbf{Maragogi location in Brazil.} Left panel depicts 27 administrative divisions of Brazil, where Alagoas state is highlighted in black. Right panel displays the city of Maragogi (in black) inside Alagoas state. }
    \label{fig:maragogi_on_map}
\end{figure}

Maragogi is located in the northeast of Alagoas state approximately 137 km from the capital Maceió, see Figure \ref{fig:maragogi_on_map}.

\textbf{Demographics.} The national 2010 survey \cite{IBGE_maragogi} estimated that Maragogi had 28749 inhabitants, see Table \ref{table:py_IBGE_2010}. Note in 2010 the population was mostly composed by young people (0 - 40 yo) and when compared to the current estimate, we observe a significant shift towards the mid age (29 - 69 yo). 

\begin{table}[ht]
\centering
\begin{tabular}{c|c||ccccccccc||r}
 & & 0-9 & 10-19 & 20-29 & 30-39 & 40-49 & 50-59 & 60-69 & 70-79 & 80+ & Total\\
\hline
2010 & & 6016 & 6694 & 5220 & 4160 & 2861 & 1850 & 1177 &  539 &  232 & 28749\\
& (\%) & 20.93 & 23.28 & 18.16 & 14.47 & 9.95 &  6.44 & 4.09 & 1.87 & 0.81 & 100 \\
\hline 
2019 &  & 5542	&6276	&5967	&4704	&4102	& 2954	&1933	&1005	&219 & 32702\\
& (\%) & 16.95	&19.20	&18.25 &14.39	&12.54 & 9.04	&5.91	&3.07	&0.67	& 100 \\
\end{tabular}
\caption{\textbf{Age pyramid of Maragogi.} The age pyramid shown in the first row corresponds to the national 2010 survey \cite{IBGE_maragogi} . In the second row, the age pyramid for 2019 is constructed using two databases and corrected due to biases in the data (such as duplicate registers for same individuals). }
\label{table:py_IBGE_2010}
\end{table}

The national 2019 survey estimated the population size in 32702 and 33351 in 2021 \cite{IBGE_maragogi}. To construct the age pyramid of Maragogi in 2019 we merged two databases. For the interval 0-79 y we used the Programa da Saúde da Família (PSF) ---  public health assistance program, see Section \ref{sec:household_networks} --- summing over a total of 34598 inhabitants. For the interval 80 y - 100+ we imported the individuals of each 5 years interval from the age pyramid of Maragogi estimated in the national 2019 survey \cite{IBGE_2019_pop_est}, and the total number . We constructed the age pyramid of Table \ref{table:py_IBGE_2010} multiplying by the factor $32702/34598$ which corresponds to fraction between the total population size estimated in 2019 and the population size from PSF data. 

\begin{figure}[h!]
    \centering
    \includegraphics[width = .75\textwidth]{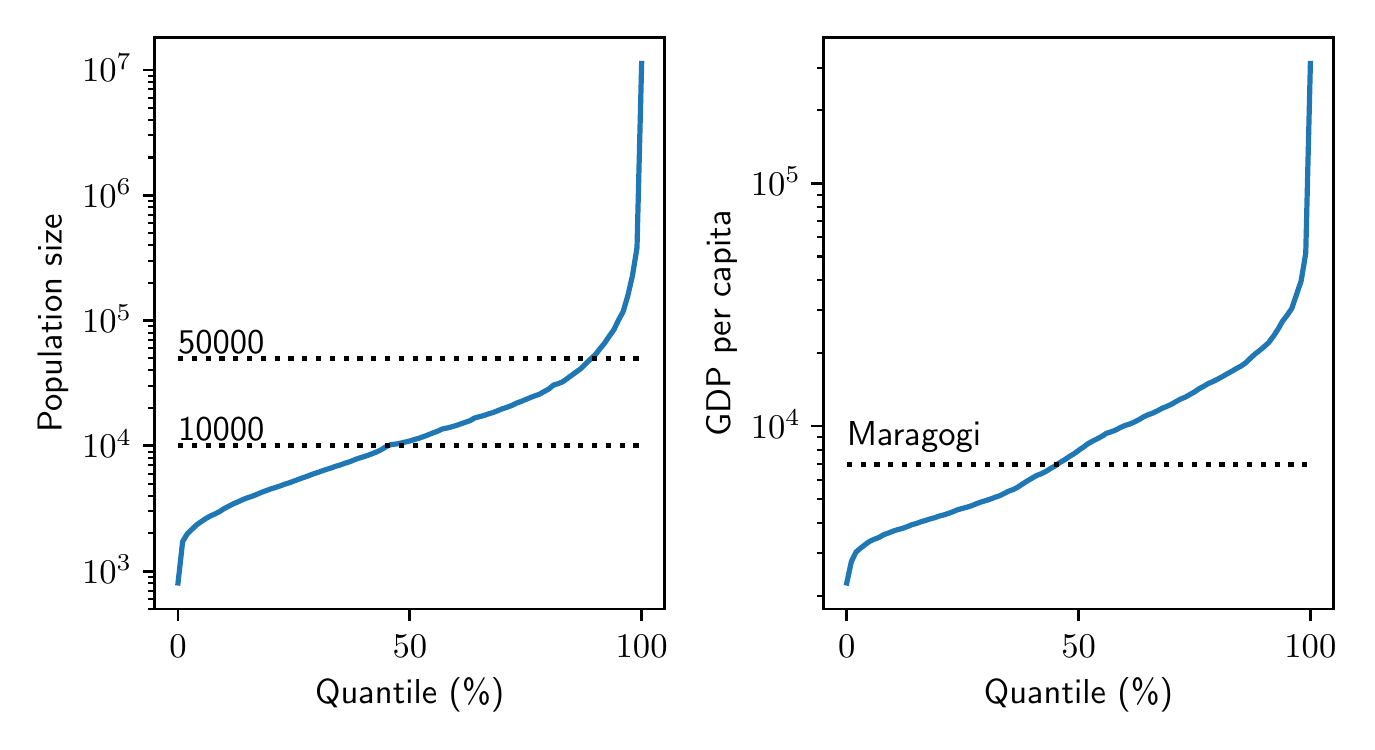}
    \caption{\textbf{Maragogi-AL in comparison with Brazilian cities}. Cumulative histogram of the total population of Brazilian cities (left panel), GDP \textit{per capita} (right panel) as a function of the proportion of municipalities \cite{IBGE_2010_census}. The GDP \textit{per capita} is conditioned on the group of Brazilian cities between 10000 and 50000 inhabitants.}
    \label{fig:comparative_probs_with_brazil}
\end{figure}

Comparing with other Brazilian cities, the left panel of Figure \ref{fig:comparative_probs_with_brazil} displays the population size range between 10000 and 50000 inhabitants, which is the range correspondingly to $44\%$ of Brazilian cities, and encompasses the city of Maragogi located inside it.

This range of cities between 10000 and 50000 inhabitants encompasses mostly cities sharing common features in terms of social and epidemiological synergy: small population size, low occupation density, and disease vectors such as public transport are not significant. Moreover, there is a small portion of vertical urbanization.

\begin{figure}[h!]
    \centering
    \includegraphics[width = .75\textwidth]{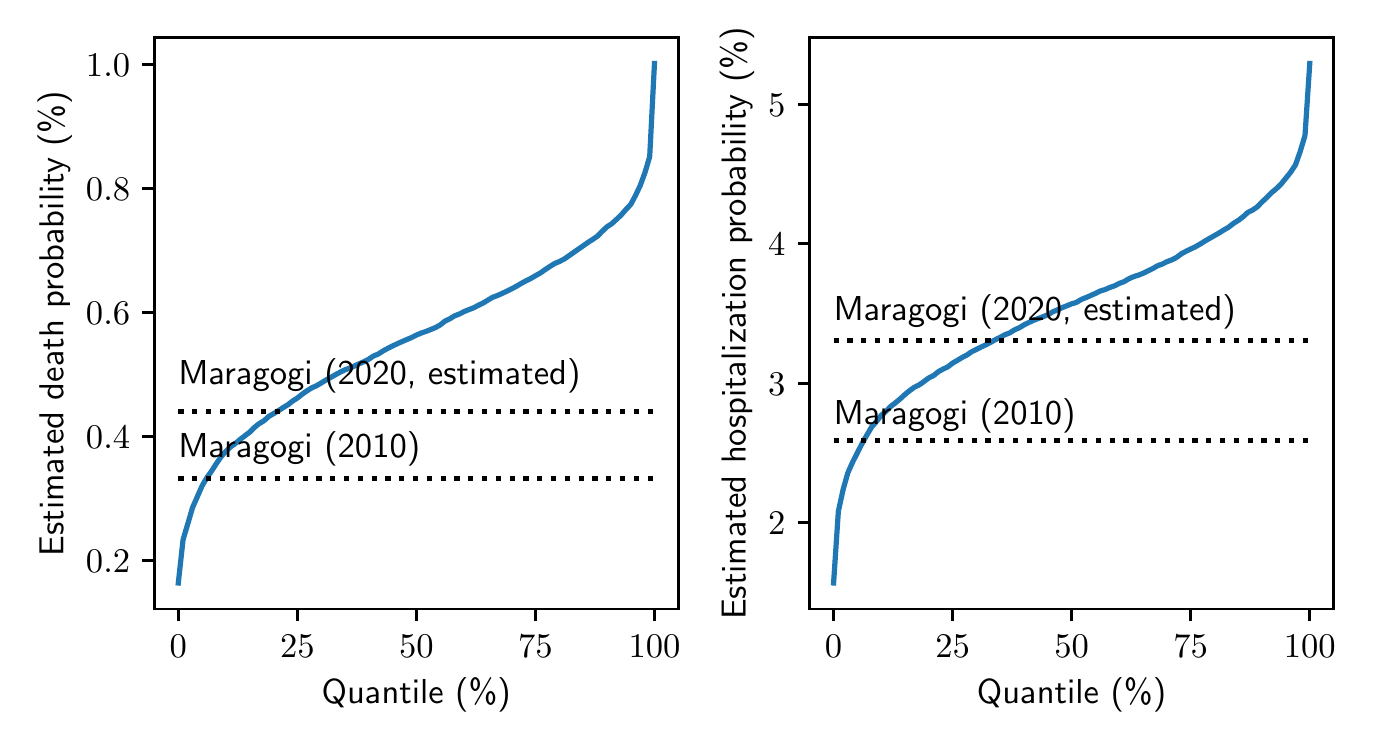}
    \caption{\textbf{Age based probabilities of death and hospitalization for COVID-19 calculated for Brazilian cities.} Expected death (left panel) and hospitalization (right panel) probabilities for Brazilian cities in the range 10000 and 50000 inhabitants. For hospitalization is assumed any individual developing COVID-19 severe symptoms, see Table \ref{table:probs}.}
    \label{fig:death_hospi_prob}
\end{figure}

Table \ref{table:probs} contains the probabilities of symptomatic cases, severe cases and deaths aggregated by age group. Crossing those proportions with Maragogi's age pyramid (Table \ref{table:py_IBGE_2010}) we obtain an expected hospitalized/infected ratio of $p_h = 3.304\%$ and a death/infected ratio of $p_d = 0.441\%$ overall. Figure \ref{fig:death_hospi_prob} displays the age based probabilities of death and hospitalization for COVID-19 (computed using the statistics in Table \ref{table:probs}) calculated for Brazilian cities.

\begin{table}[ht]
\centering
\begin{tabular}{c||ccccccccc||}
Age & 0-9 & 10-19 & 20-29 & 30-39 & 40-49 & 50-59 & 60-69 & 70-79 & 80+ \\
\hline
$p_{sym}$ & 0.5	& 0.55	& 0.6	&0.65	&0.70	& 0.75	& 0.80	&0.85	&0.9 \\
$p_{hosp}$ & 0.0001	& 0.0001	& 0.011 & 0.034	& 0.043 & 0.082	& 0.118	&0.166	&0.184	\\
$p_{death}$ & 0.00002	& 0.00006	& 0.0003 & 0.0008	& 0.0015 & 0.006	& 0.022	& 0.051	& 0.093\\
\end{tabular}
\caption{Age based probabilities for COVID-19.}
\label{table:probs}
\end{table}

\begin{figure}[h!]
    \centering
    \includegraphics[width = .75\textwidth]{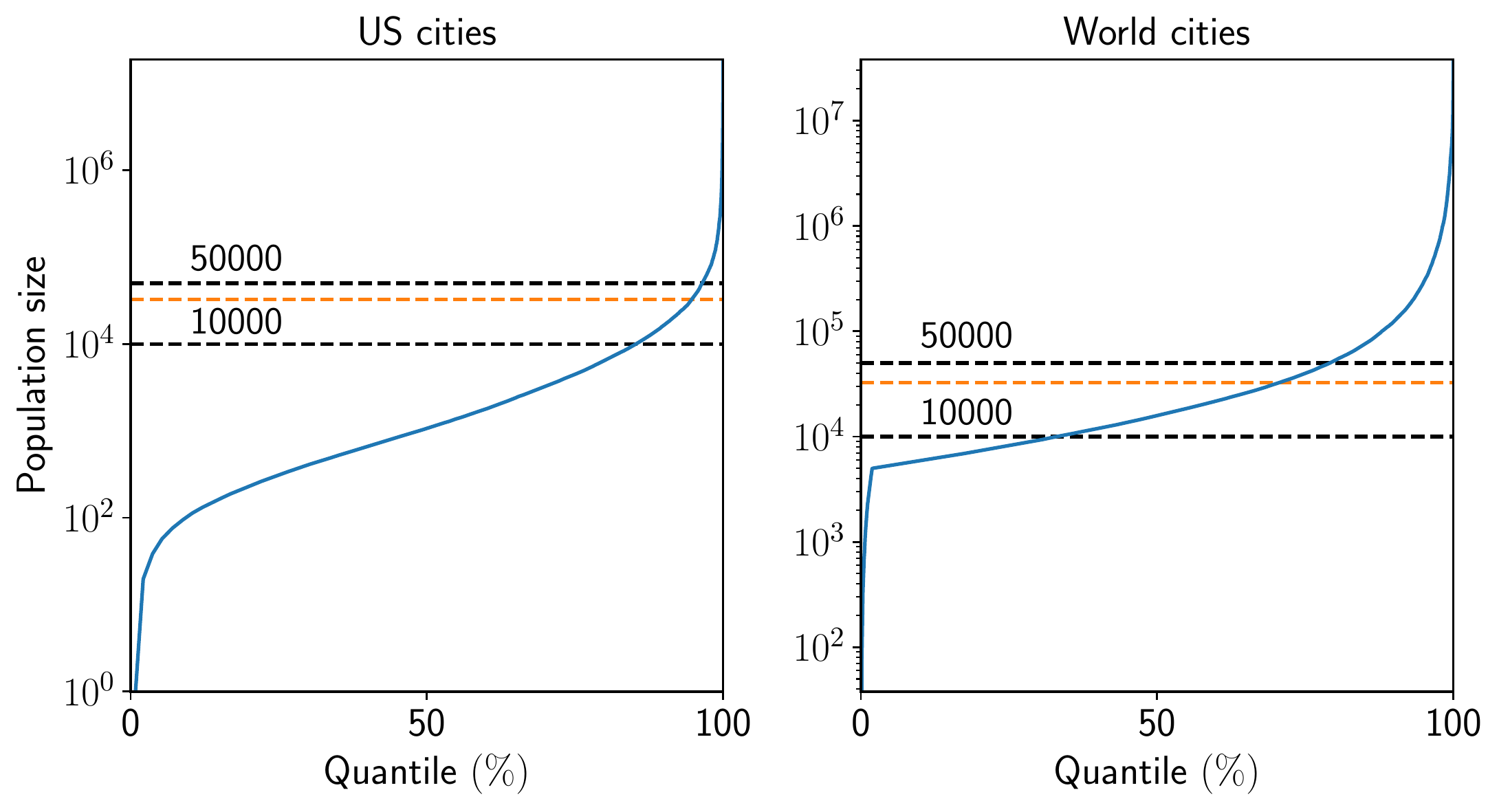}
    \caption{\textbf{Maragogi-AL in comparison with cities in United States and World}. Left panel displays the cumulative histogram of the total population of US cities \cite{us_cities} and right panel displays in World cities  \cite{world_cities} as a function of the proportion of cities. Dashed lines in black correspond to 10000 and 50000 inhabitants for reference while the orange shows the Maragogi population.}
    \label{fig:comparative_us_world}
\end{figure}

To put the city of Maragogi into context worldwide, Figure \ref{fig:comparative_us_world} shows the cumulative histogram of the total population from the \texttt{simplemaps} database containing 28372 and 41000 cities corresponding to US and World cities, respectively \cite{us_cities,world_cities}. We observe the city of Maragogi is above the center in both cases, which suggests that is a small urban area with an
average size population worldwide \cite{OECD}. 

\begin{figure}
    \centering
    \includegraphics[width =0.75\textwidth]{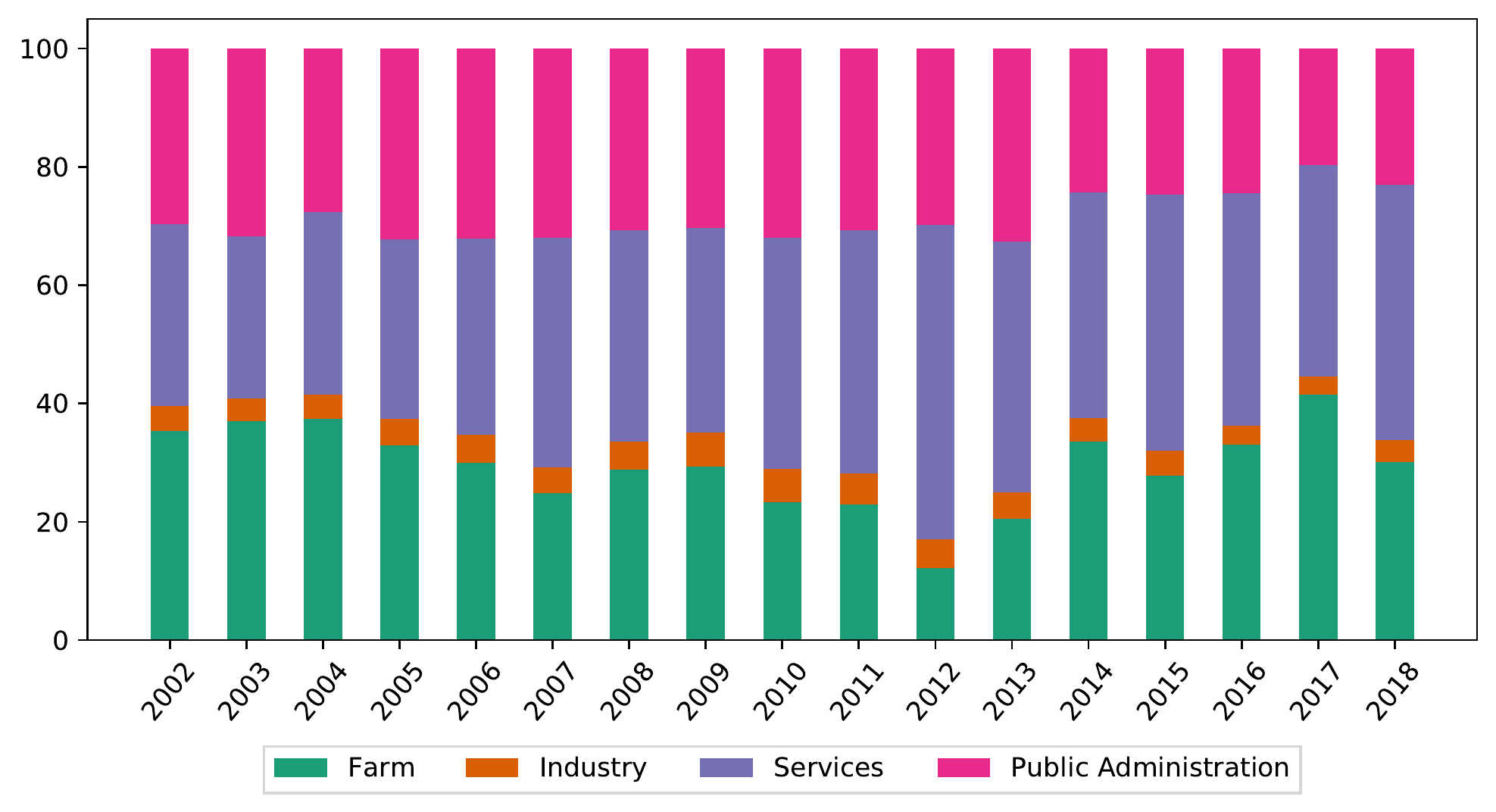}
    \caption{\textbf{Economic value added from 2002 to 2018}. Each economic sector contribution in Maragogi with respect to the total. Administration public includes Defense, Education, Public health and Social security.}
    \label{fig:maragogi_activities}
\end{figure}

\textbf{Economic aspects.} If we narrow our analysis to this 10000 and 50000 inhabitants range, center panel in Figure \ref{fig:comparative_probs_with_brazil} shows that Maragogi had GDP \textit{per capita} close to the median in 2010. To illustrate the distribution of socioeconomic activities in the city, see Figure \ref{fig:maragogi_activities}, which shows the Economic value added in the last years. 

The service sector is represented by a network of hotels and establishments providing accommodations for travelers. We discarded this hospitality service sector from our analysis because most accommodation establishments were closed during the period of our analysis (Strategic plan May 2020 - from City Hall information). 


The farming activity splits into crops ($44.7\%$), pastures ($33.6\%$), woods and forests ($7.9\%$) in 2017. The Instituto Nacional de Colonização e Reforma Agrária - INCRA has registered in terms of the Cadastro Ambiente Rural (CAR) 363 farming organizations, which 89$\%$ correspond to smallholder farming organizations. Inside this category of smallholder farming organizations, approximately 6$\%$ consist of rural settlements\footnote{
Rural settlement is defined as a portion of land that rural workers undertake to live on the plot and exploit it for their livelihood, using exclusively family labor \cite{INCRA_assentamento_2021}.}, where 1475 families practice agricultural activities \cite{INCRA_2021, CAR_2021}. This familiar agricultural activity results in commercialization of products weekly in street market (under initiative of the City Hall), see Section \ref{sec:street_market}. 
    
\begin{figure}[h!]
    \centering
    \includegraphics[width = .9\textwidth]{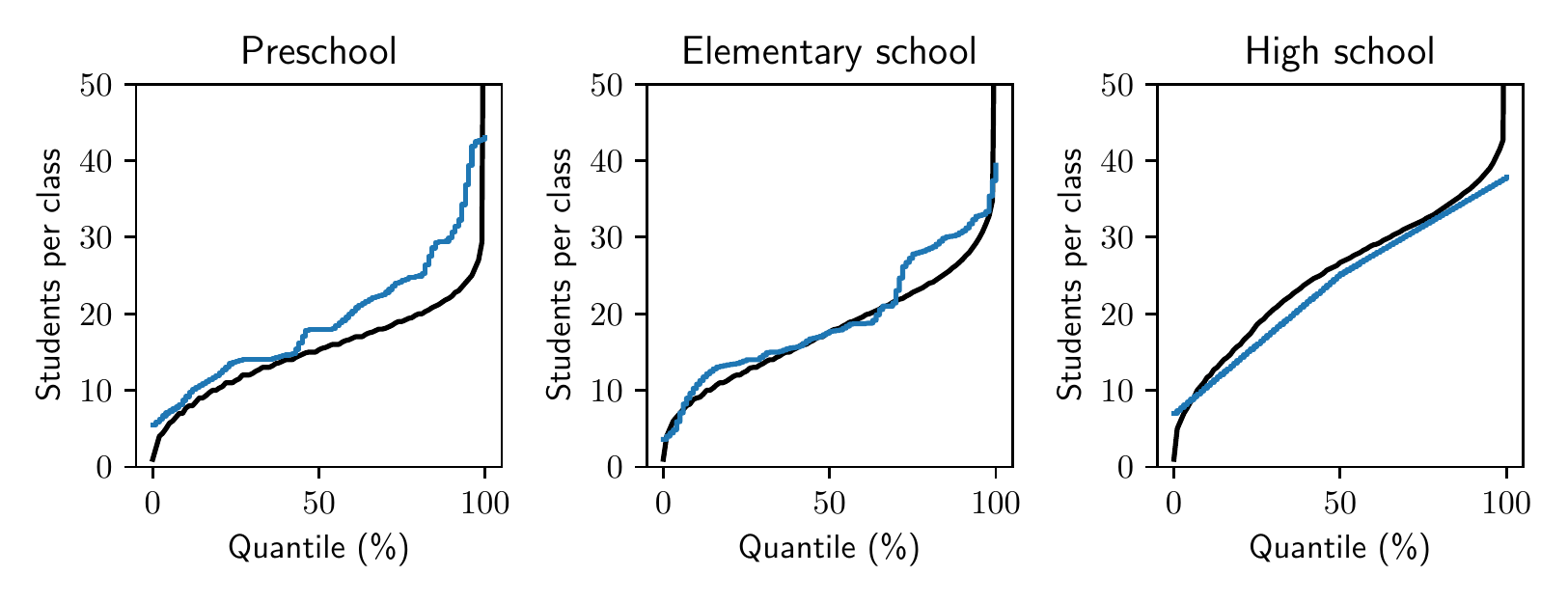}
    \caption{\textbf{Student per class distribution inside schools over cities within the range of 10000 and 50000.} The solid black line corresponds to the occupation density distribution over Brazilian cities within the range of 10000 and 50000 inhabitants and Maragogi's is represented in blue. Left, center and right panels display the distribution for Early Childhood Education, Elementary school and High school, respectively.}
    \label{fig:compare_classes_brazil}
\end{figure}

\textbf{Education.} We filtered the data for schools belonging to the municipalities in the range of interest. The data is composed by educational institution and school level (kindergarten, elementary and high school) of INEP 2020 \cite{INEP_students_per_class}, see Figure \ref{fig:compare_classes_brazil}. Figure \ref{fig:compare_classes_brazil} shows the occupation density of schools in Maragogi into context over Brazilian cities within the range of interest. The distributions are similar in all
levels of education, in particular  overall Early Childhood Education (ECE), Elementary and High schools in the range of interest the average of students per class corresponds to 15.25, 17.87 and 25.30, respectively, compared to 19.54, 19.49 and 23.37 of Maragogi schools.

\section{Street market}
\label{sec:street_market}

Agricultural street markets exist throughout Brazil, ranging from small to large cities and the commercialized products consists of vegetables and typical products from the market's location surroundings. This commerce event also allows producers commercializing their agricultural production, and there is no intervention of third parties but a direct channel of trade between producer and consumer \cite{de2019importancia}.

These street markets dating back from 1548 \cite{RM} have significant economical importance for small and medium cities \cite{araujo2018feiras}. For farmers, fairs represent gains ranging from 1 to 3 monthly minimum wages \cite{gastal2016construccao, ribeiro2005programa}. In fact, in Minas Novas, which is located in the state of Minas Gerais and has the same indicators of Maragogi, the $40\%$ of farmers had fairs as the only source of income and for $64\%$ fairs accounted for over half of farmers income \cite{ribeiro2005programa}. Mobility constrains imposed by COVID-19 prevention measures impacted farmers revenues \cite{claudinoimpactos, canela2021impacto}.

\textbf{Street markets in Maragogi.} The street market has commercialization of products derived from farmers and families from the rural settlement (Strategic Plan). It opens weekly and according to INCRA, fruit growing is responsible for $57\%$ of the total production of the rural settlement. 

\begin{figure}[h!]
    \centering
    \includegraphics[width = .5\textwidth]{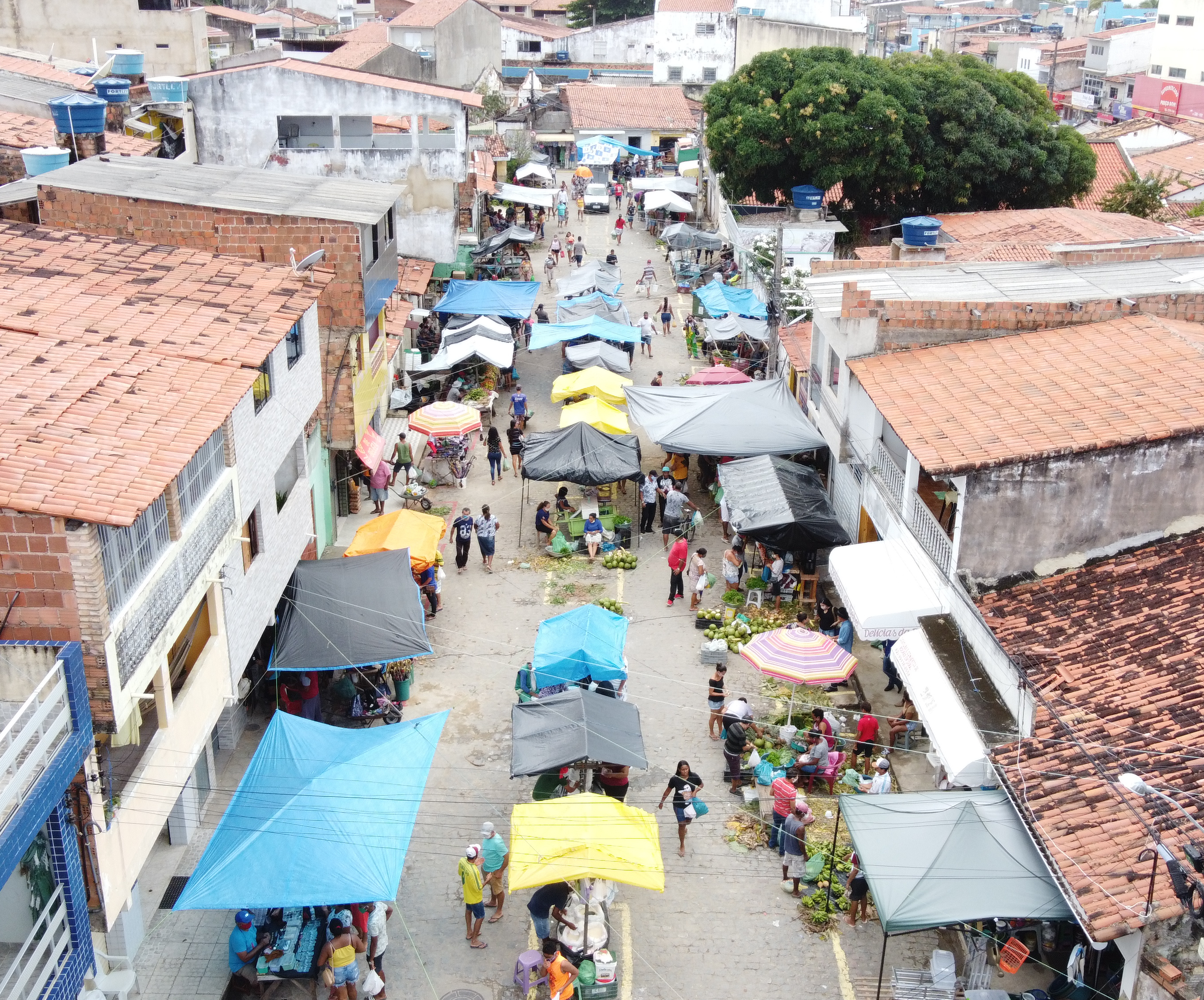}
    \caption{\textbf{Street market in Maragogi.}  Photo captured by drone (model Drone DJI Mavic Mini) in August 8th, 2020. }
    \label{fig:feira_maragogi}
\end{figure}

\textbf{Drone data and local video recording.} Early in the outbreak in Maragogi, the City Hall provided several measures to avoid that the street market could be a source of high number of COVID-19 cases. This motivated our collaboration along with the municipal administration \cite{feira_piaui}. Based on the drone video recording we analyzed the street market.

\begin{figure}[h!]
    \centering
    \includegraphics[width = 1.0\textwidth]{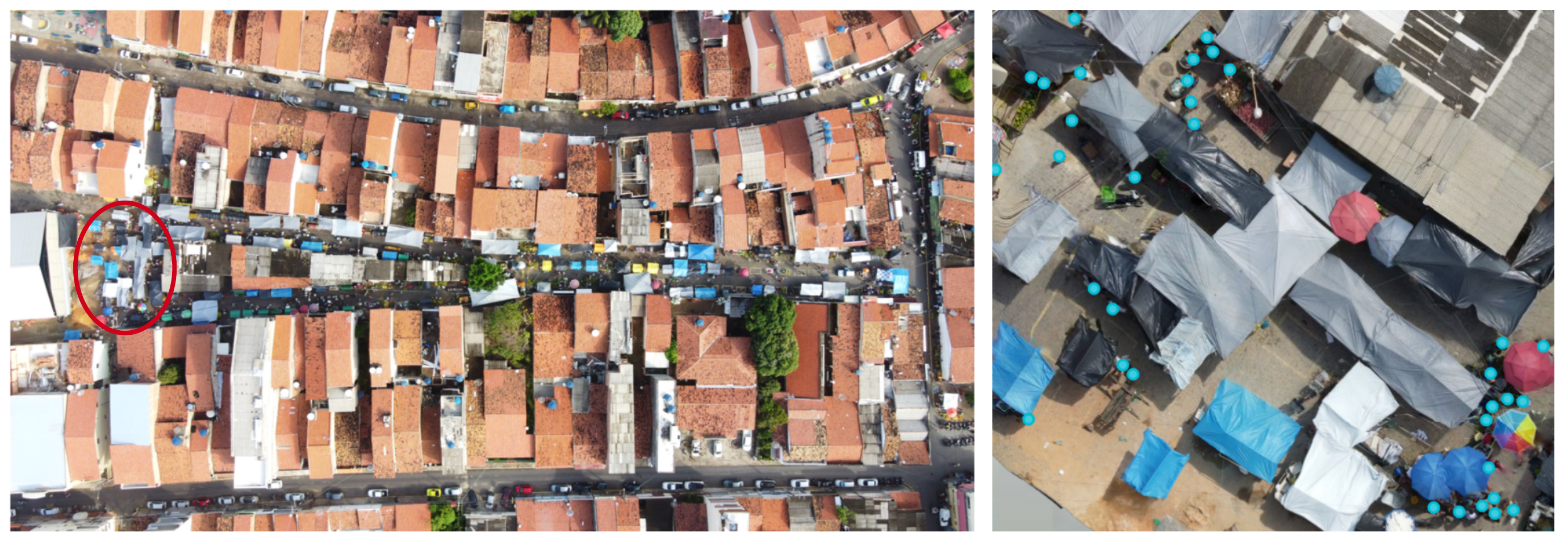}
    \caption{\textbf{Aerial recording of the street market in Maragogi.} Left panel displays the aerial photography of the full extension of the street market. The red circles highlights the sector which one of the samples from the video recording were used to estimate the occupation density, shown in the right panel. The marking tool of the Drone Deploy mapping software \cite{drone_deploy} identifies moving targets using blue dots. Photo captured by drone (model Drone DJI Mavic Mini) in August 22th, 2020.}
    \label{fig:feira_counting_analysis}
\end{figure}

We estimate that the passable area corresponds to 1600 $m^2$ out of 2500 $m^2$ total internal area, and the average distance between 2 people were 48 cm (under high crowding) and 180 cm (low crowding). In August 8, 2020 at 8:00 am, we video recorded the street market and performed counting analysis using the Drone Deploy mapping software \cite{drone_deploy}, see Figure \ref{fig:feira_counting_analysis} for an example. In this analysis we aimed to identify clusters of people, and in the occasion, we estimate there were 1 cluster of 5 persons, 3 clusters of 4, 11 clusters of 3 and 23 clusters of 2 persons.

We estimate there were around 120 tents, under which on average half of them containing 2 persons and 3 persons on the other half, see Table \ref{table:counting_analysis} for the counting for August 22.  
\begin{table}[ht]
\centering
\begin{tabular}{c||c}
Hour & People count \\
\hline
6:00 & 261 \\
6:30 & 270 \\
7:00 & 341 \\
8:00 & 336 \\
8:30 & 316 \\
9:00 & 262 \\
9:30 & 202 \\
\end{tabular}
\caption{\textbf{Data counting from street market.} The counting analysis were performed exclusively from visible persons. The actual number is larger due to many visitors are under tents. Video recording performed on August 22, 2020. }
\label{table:counting_analysis}
\end{table}

\chapter{The Clinical Monitoring System (CMS)}
\label{sec:SMC}
\section{Collecting real data from local health service}

The Clinical Monitoring System (CMS) software is written in Python and Django framework using MySQL database,  and is available through the internet as a web application for each one of the health units of the city. Health staff (nurses and medical doctors) were able to collect in real time the following data: 

\begin{itemize}
	\item \emph{Personal information}: Name of each person, name of their mother, date of birth, national health id (CNS), National Personal Number (CPF) and borough information.  
	\item \emph{Medical information}: Medical information about the evolution of the patient status through the year (date of each symptom, type of test for COVID-19 used, date and result of the test etc). Date of admission and discharge to an intensive care unit (ICU) bed or a clinical bed.  
	\item \emph{Family information}: Relatives that had visited the health system and declared primary contacts.
\end{itemize}

We emphasize that all officially confirmed cases, hospitalizations and deaths by covid-19 in the city of Maragogi were registered in the CMS database. Therefore, this database was essential for estimating the real incidence of cases in the city.

There were several other features available at CMS used by the health authority of the city that were not used in this study, such as:

\begin{itemize} 
	\item  \emph{Complementary Personal information}:  Geo-located address of the house that the person lives,  telephone number, gender, height, weight, ethnics info, occupation, tobacco use, number of people living at the same house, recent visited places.
	
	\item  \emph{Complementary Medical information}: Previous vaccination against influenza, comorbidities,  recent use of oxygen assistance, current oxygen saturation, temperature, hearth rate, arterial pressure, recent contact with confirmed cases, medical diagnosis, medical prescription. 
	
	\item \emph{Data analysis and visualization} Automatic reports of number of new confirmed and recovered cases per week, patients per day, total admissions for ICU and Clinical Beds per day, forecast of new bed demand for the next 4 days based on new patients records, risk classification (suspected cases) of patient visits based on symptoms, geographic map visualization of the spreading of the infection based on medical records, automatic epidemiological reports for the city.
	
    \item	\emph{Other features:} User management, panel of the consumables needed on each health unit, several types of reports based on patients, visits and addresses. 

\end{itemize}

When a new suspected or confirmed case is detected,  a team of trained staff proceed with  periodic  followup calls to check the current patient's health status and update the medical record of the patient, filling missing information, if any.

This software began its use in May, 5th and replaced  excel spreadsheets used by the health professionals. Along April/2020, first month of the pandemics in Maragogi, the first 33 confirmed cases of COVID-19 were reported in this spreadsheet. These 33 cases constituted the initial data loading of the database of CMS.  By December, 21th of 2020, the database of CMS registered a total of 1972 patients and 2607 medical appointments.

To inform new COVID-19 cases in Brasil, the Ministry of Health of Brazil developed a software, \emph{ESUS-VE} - \emph{Epidemiological Surveillance  Unique Health System}, which accounts for mandatory registration of attendances and laboratory tests. The first version was released in March/2020, specifically to cover the demand caused by COVID-19. The goal of this software was to provide a unified database of COVID-19 cases within Brazil.  Along the pandemics course, this software changed several times, as well as the basic information required to inform a new COVID-19 case. 
 
ESUS-VE is just one of the five mandatory softwares provided by the Ministry of Health of Brazil that need to be used during the COVID-19 pandemics by the municipalities. The other softwares are \emph{Gerenciador de Ambiente Laboratorial} (laboratory testing), \emph{Cadastro Nacional de Saúde} (personal database), \emph{Sistema de Informação de Vigilância da Gripe}  (severe cases), \emph{Sistema de informação de Mortalidade} (deaths). These software do not have any integration whatsoever,  creating a great administrative challenge to keep all these databases coherent and updated. The personal data of a given single case could be typed in 4 different systems before its resolution in several different sectors of the administration.   This lack of efficiency and  integration was reflected in the willingness of the health staff in adopt new software tools.  In order to overcome this difficulty  and since ESUS-VE  has no application programming interface (API), we use automated bots to automatically insert the data from CMS into ESUS-VE. 



\begin{figure}[ht]
    \centering
     \includegraphics[width=0.8\linewidth]{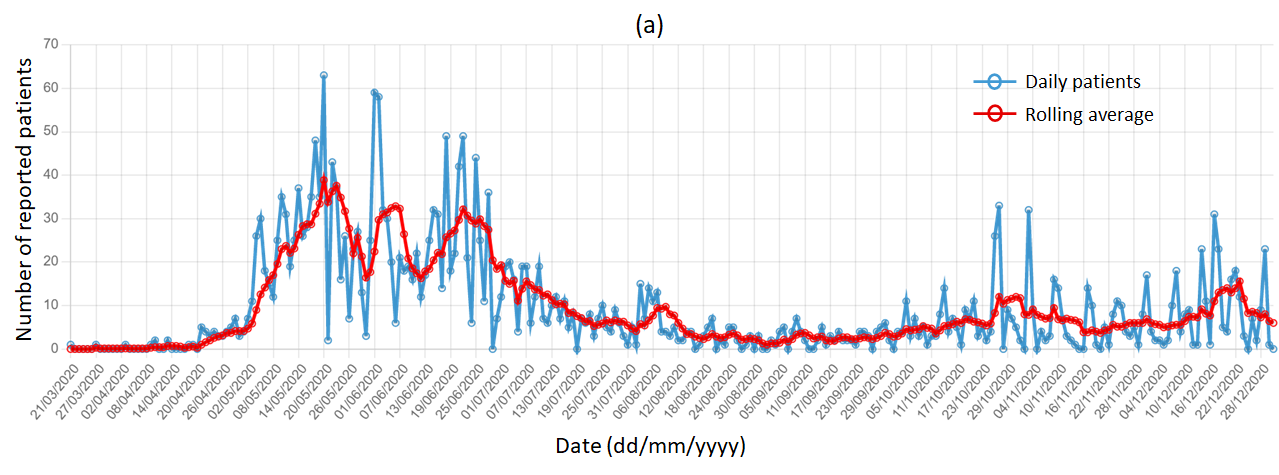}
      \includegraphics[width=0.8\linewidth]{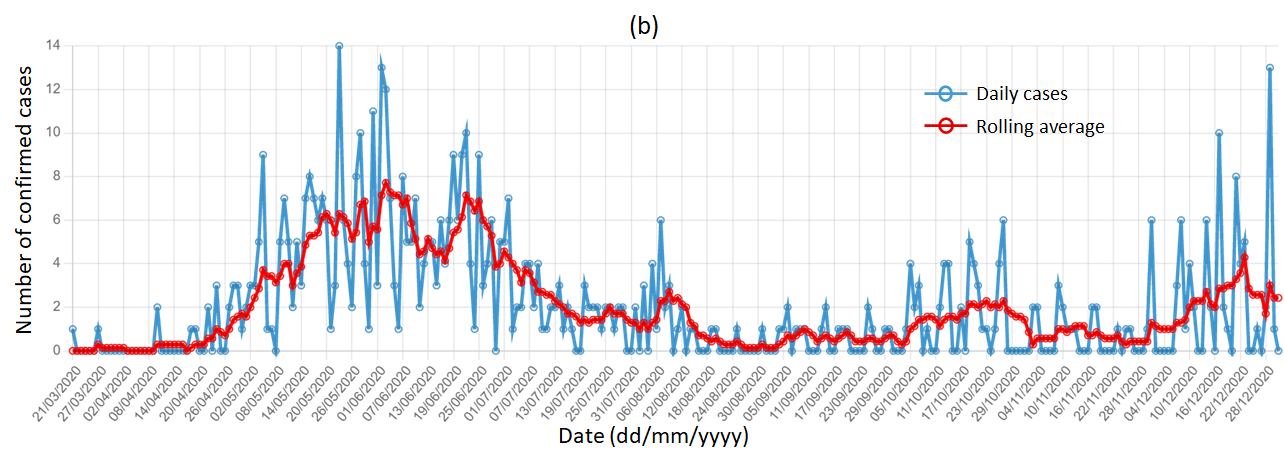}
    \caption{\textbf{Real data automatically plotted by the CMS dashboard.} (a) Number of patients with COVID-like symptoms as a function of the date of attendance in Maragogi. (b) Number of confirmed COVID-19 cases in Maragogi as a function of the date of symptoms onset. Both panels are promptly available to municipal health managers.}
    \label{fig:dashboard_smc_a}
\end{figure}


\chapter{Household network}
\label{sec:household_networks}







We use three databases were to reconstruct a social network of household contacts: \emph{Programa Saúde da Família} (\textbf{PSF}), \emph{Programa Bolsa Família} (\textbf{BOLSA}) and \emph{Sistema de Monitoramento da COVID-19} (\textbf{SMC}). These city owned databases correspond respectively to a public health assistance program, a social assistance program and a software for registering covid-19 health attendance. The data from the first two databases were previously collected from non-structured sources such as PDF files and processed. By combining data from the other two sources we managed to capture the household family size distribution for at least $2/3$ of the city's population.

\par Each of the following tables were constructed containing one column that specifies for each person (row in the table) which household group it belongs, hence by grouping rows in the table by this \emph{group-column} value we can obtain the network structure.

The \emph{Programa da Saúde da Família} is the largest database containing $26721$ rows but is the poorest in detail with only $4$ columns, namely:

\begin{itemize}
	\item \emph{nome}: Token representing the name of each person;
	\item \emph{cns}: Token representing the health program id, (\emph{cadastro nacional de saúde - cns}) of each person (will be used later for merging the tables);
	\item \emph{idade}: The age of each person;
	\item \emph{codigo\_familiar\_psf}: A token representing the group (family code) of each person;
\end{itemize}

Since the data was available in PDF format, we made a JavaScript script to download files from each family, extracting the text using the python library called \emph{pdfplumber}. We noticed that the fields were well defined between some specific sets of words, so we used string matching techniques to filter the information from each field and structure the text.

The \emph{Programa Bolsa Família} database is the second largest database, containing $18682$ rows and $11$ columns concerning rich data about the beneficiary of the social program, with $99\%$ of the data collected in 2016 or later. For this reason we chose this table to be the fundamental source of data for the construction of the network, as it will be detailed in section II. The columns contain data relating to:

\begin{itemize}
	\item \emph{Personal information}: Tokens representing the name of each person, name of their parents, \emph{cns} id and \emph{cpf} id which stands for \emph{cadastro de pessoa física}, an individual id used in Brazil that can uniquely represent each person throughout the databases. The age of each person.
	\item \emph{Work information}: Various columns detailing work information.
	\item \emph{Family information}: A token representing the group (family code) of each person in the database as well a field (column) describing the family-role of the beneficiary. The address of the house that the each person lives (is the same for each person in the same group).
\end{itemize}

The \emph{Sistema de Monitoramento COVID-19} database is the one containing detailed information about the health status of people in the city, concerning the actual pandemic. It is by far the smallest database with only $1602$ rows and $14$ columns with data relating to:

\begin{itemize}
	\item \emph{Personal information}: Tokens representing the name of each person, name of their mother, \emph{cns} id and \emph{cpf}. The age of each person. 
	\item \emph{Medical information}: Various columns detailing medical information about the evolution of the patient status through the year (date of first symptoms, date that the patient tested positive or recovered) etc.
	\item \emph{Family information}: A token representing the group (family code) of each person in the database. The address (neighborhood) of the house that the each person lives.
\end{itemize}

The idea of the integration of the three databases presumes the following propositions:

\begin{itemize}
	\item BOLSA database offers us a reasonable idea of the distribution of families across the population.
	\item Medical information from all patients, whenever possible, must be used.
\end{itemize}


An important fact to notice is that the three databases don't share a common key e.g. \emph{cpf} or \emph{cns} in order to merge the data without repetition. Hence, we propose the following approach to merge the databases (let $B$\nomenclature{$B$}{Bolsa Família database}, $S$\nomenclature{$S$}{Sistema de Monitoramento da Covid-19 database}, $P$\nomenclature{$P$}{Programa Saúde da Família database} denote BOLSA, CMS and PSF databases respectively):

\begin{enumerate}
	\item Let $I_1 = B \cap S$. Intersection is found by using the key \emph{cpf}.
	\item If $C = S - I_1$, let $I_2 = C \cap P$. Intersection is found by using the key \emph{cns}.
	\item The merged database $M$ is obtained by the disjoint union of the following tables: $I_1 \cup I_2 \cup (B - I_1) \cup (S - I_2 \cup I_1)$.
\end{enumerate}

\nomenclature{$I_1$}{Intersection of tables $B$ and $S$}
\nomenclature{$C$}{Complement of table $I_1$ with respect to table $S$}
\nomenclature{$I_2$}{Intersection of tables $C$ and $P$}

We finish this process with a final table $M$\nomenclature{$M$}{Merged database} with $19973$ rows and several columns, containing roughly $2/3$ of the city's population, regarding household contact, economic and health data of its citizens. That table is used assign each person to a group, based on the \emph{grouping column} (family code) of each source database.

\begin{figure}[H]
\centering
  \includegraphics[width=0.7\linewidth]{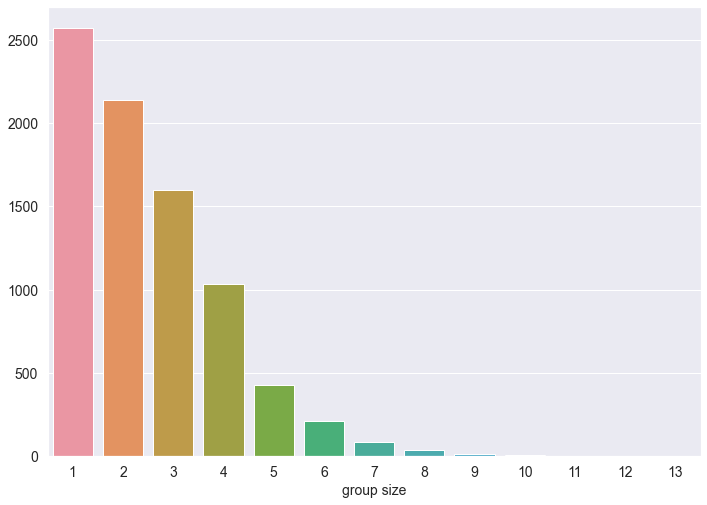}
  \caption{Group size distribution of the M database.}
  \label{fig:boat1}
\end{figure}

We further expand the database $M$ by incorporating persons registered only in the $P$ table that have a relative (same group person) in the $I_2$ table, as described above, resulting in a final table with $20350$ rows.

\begin{figure}[H]
  \includegraphics[width=\linewidth]{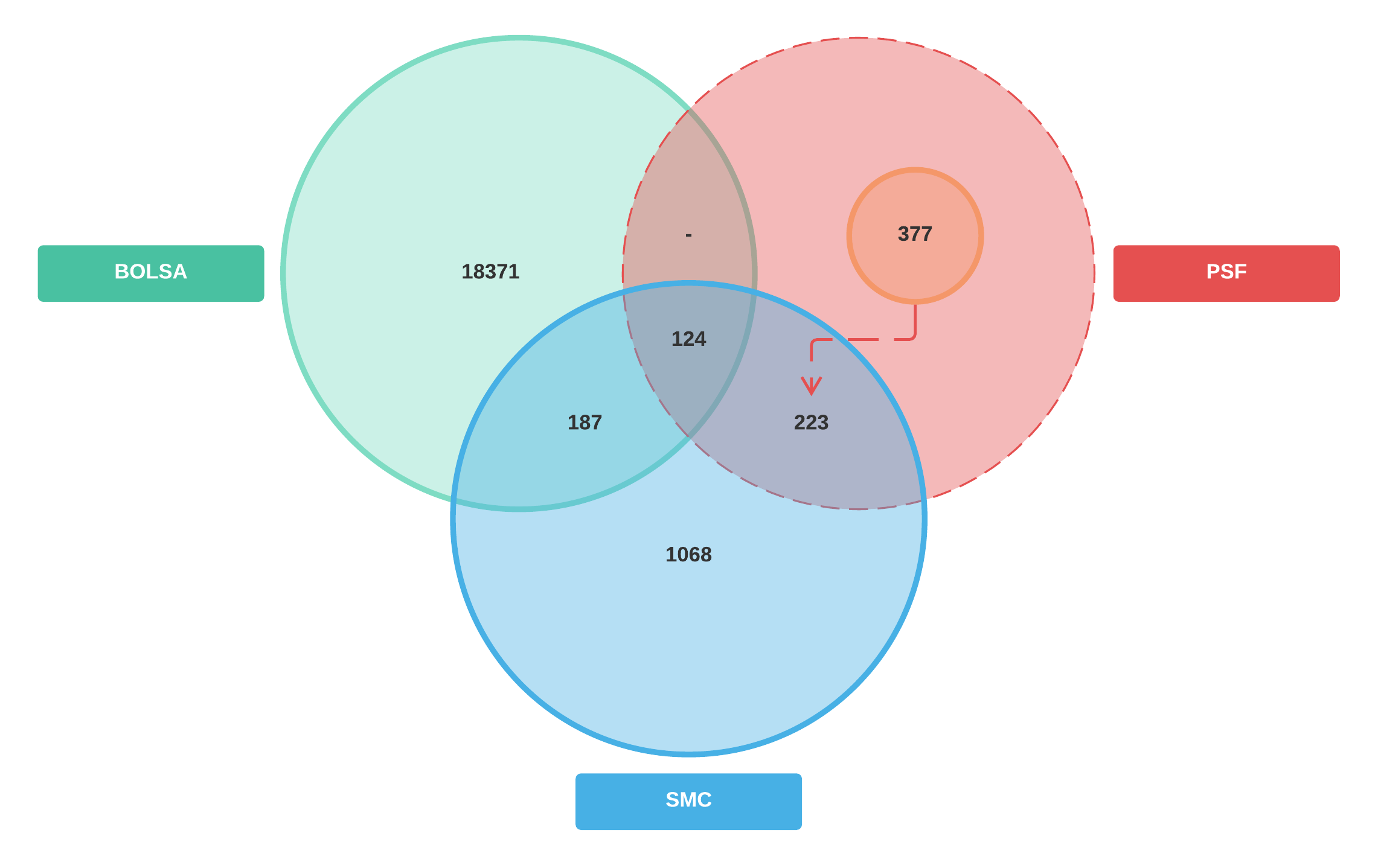}
  \caption{\textbf{Database diagram.}  PSF - Programa Saúde da Família, BOLSA-Programa Bolsa Família and SMC-Sistema de Monitoramento Clínico.}
  \label{fig:database_diagram}
\end{figure}

The merged database $M$ is used to construct the network of contacts $G$\nomenclature{$G$}{Constructed network of contacts} trough the following steps:

\begin{enumerate}
    \item Create, for each row in $M$ a node in $G$ with its respective attributes (columns of $M$).
    \item Group each node in $G$ by its \emph{grouping column} (family code). If a node has more than 1 valid family code we choose to group it by the following priority: firstly family code from $B$, secondly from $P$ and thirdly from $S$.
    \item If vertices $u$ and $v$ are in the same group as constructed in the previous step, create an edge $(u, v)$ in the network. That process ultimately yields a network composed of only fully connected components (cliques) which represents the household contacts.
    \item We try to further connect cliques in the network by checking parenthood relationships based on the mother/father names in $B$ section of database (those nodes who have a valid $B$ family code attribute). We first select nodes that have unique names and then map each name to its node label. For each node $u$ in the network we find nodes $f$ and $m$ that have the name attribute equals to father name and mother name attribute of node $u$, respectively. If $f$ and $m$ both have the same family code we create edges $(u, m)$ and $(u, f)$. Notice that it is possible to find only node $f$ or only node $m$ (exclusively). In such case, we just create the edge between child and parent node.
\end{enumerate}

The caution taken in the last step of checking uniqueness of names and if nodes $m$ and $f$ belongs to a same family is due to the fact that common names might pose a problem on creating edges in such way, as they would form clusters that are little related to the real parenthood relationships.

Edges created by step $3$ are labeled \emph{INTRAFAMILIAR edges} (concerning contacts within families), whereas edges created by step $4$ are labeled \emph{INTERFAMILIAR edges} (concerning contacts between distinct, but related, families).

\begin{figure}[H]
    \begin{center}
    \includegraphics[scale=0.25]{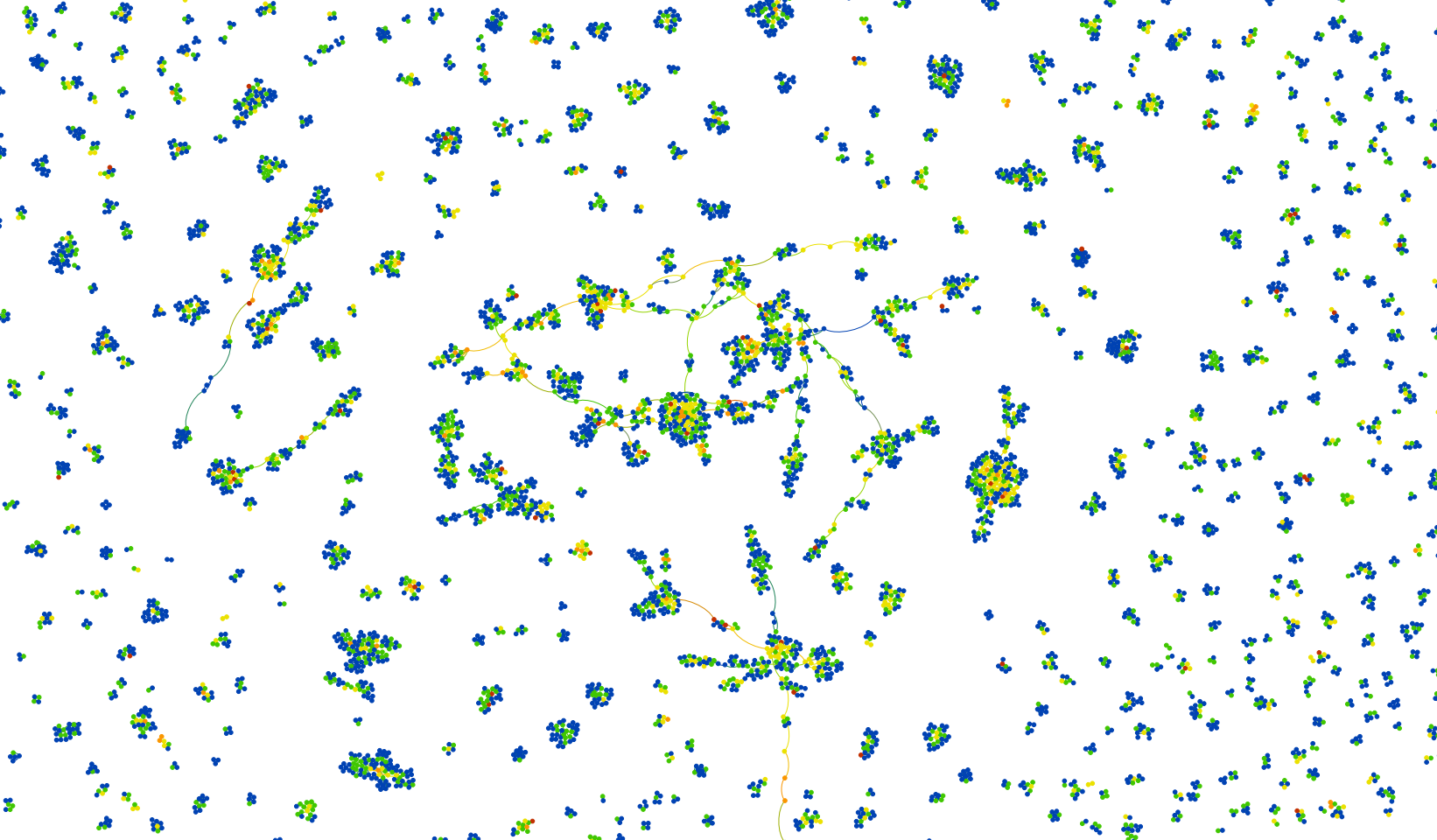}
    \caption{\textbf{Local picture of the network.} One can see the cliques (dense clusters) interconnected by edges. Nodes are colored by age.}
    \end{center}
    \label{fig:localNet}
\end{figure}

The resulting network has $27235$ edges, $24596$ being \emph{INTRAFAMILIAR}. The average degree of the network considering only such edges is about $2.4$ as we can see in the histogram below:

\begin{figure}[H]
    \begin{center}
    \includegraphics[scale=0.6]{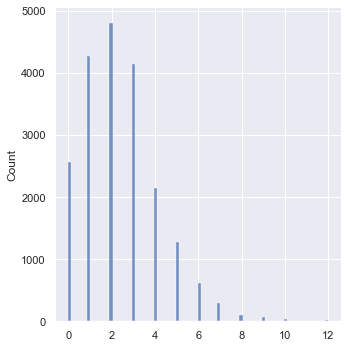}
    \caption{Node degree distribution, considering only edges within cliques.}
    \end{center}
    \label{fig:globalNet}
\end{figure}

\chapter{SARS-Cov2 data processing}
\label{sec:data_processing}

Our reconstruction is based on the SMC database, see \ref{sec:SMC}. From the anonymous database, we accessed attendances of each tested patient, in case of hospitalization, the hospitalization date, and in case of death, the death date. Each attendance entry is composed of attendance date, symptom onset date, test type (rapid or RT-PCR) and test result (positive or negative), see Listing \ref{patientdata}.

\begin{lstlisting}[
    language=C,
    frame=single,
    caption={Patient data example. This patient had two appointments, the first with a negative result and the last, one month later, with a positive result. The patient was hospitalized one day after the second appointment, but did not die.},
    captionpos=b,
    label=patientdata,
    basicstyle=\small\ttfamily,
    frame=none, 
    backgroundcolor=\color{lbcolor},  
    showstringspaces=false
]
"5": { //anonymized patient id
        "attendances": {
            "1174": { //attendance unique id
                "result": "negative",
                "test_type": "rapid",
                "attendance_date": "2020-05-11",
                "symptom_onset_date": "2020-05-07"
            },
            "1375": {
                "result": "positive",
                "test_type": "TR-PCR",
                "attendance_date": "2020-06-17",
                "symptom_onset_date": null //unfilled attendance date
            }
        },
        "hospitalization_date": "2020-06-18",
        "death_date": null,
    }
\end{lstlisting}

Most quantities required for the reconstruction, such as number of hospitalizations, deaths and attendances, evolve over time. We chose to reconstruct the curve until July 25.

To account for false negatives and false positives we also needed information about tests specificity and sensitivity. Overall, only 52 of the 1722 tests realized until July 25 were RT-PCR tests.

Since different rapid test brands were used during the year. The utilization dates in Table \ref{table:testsintime} were informed by Maragogi's health professionals and the accuracies were taken from a \cite{anvisatests}. The RT-PCR test was assumed to have $100\%$ specificity and sensitivity. 

\begin{table}[ht]
\centering
\begin{tabular}{llcc}
Test brand & Utilization                & Specificity & Sensitivity \\ \hline \hline
Wondfo     & Apr 11 - Jun 25        & $99,57\%$   & $86,43\%$   \\
{\scriptsize One Step COVID-2019 Test}     &        &   &   \\\hline
MedTeste    & May 01 - Jun 22          & $99,3\%$    & $97,4\%$    \\
{\scriptsize MedTeste Coronavírus (COVID-19) IgG/IgM}     &        &   &   \\\hline
Advagen    & Jun 23 - Aug 31       & $96\%$      & $85\%$      \\
{\scriptsize COVID-19 IgG/IgM LF}     &        &   &   \\\hline
Lungene    & Sep 01 - Oct 28 & $96,48\%$   & $91,06\%$  \\
{\scriptsize COVID-19 IgG/IgM Rapid Test. Cassette}     &        &   &   \\\hline
\end{tabular}
\caption{Usage and accuracy of rapid tests.}
\label{table:testsintime}
\end{table}

\begin{figure}[h]
    \centering
    \includegraphics[width=\linewidth]{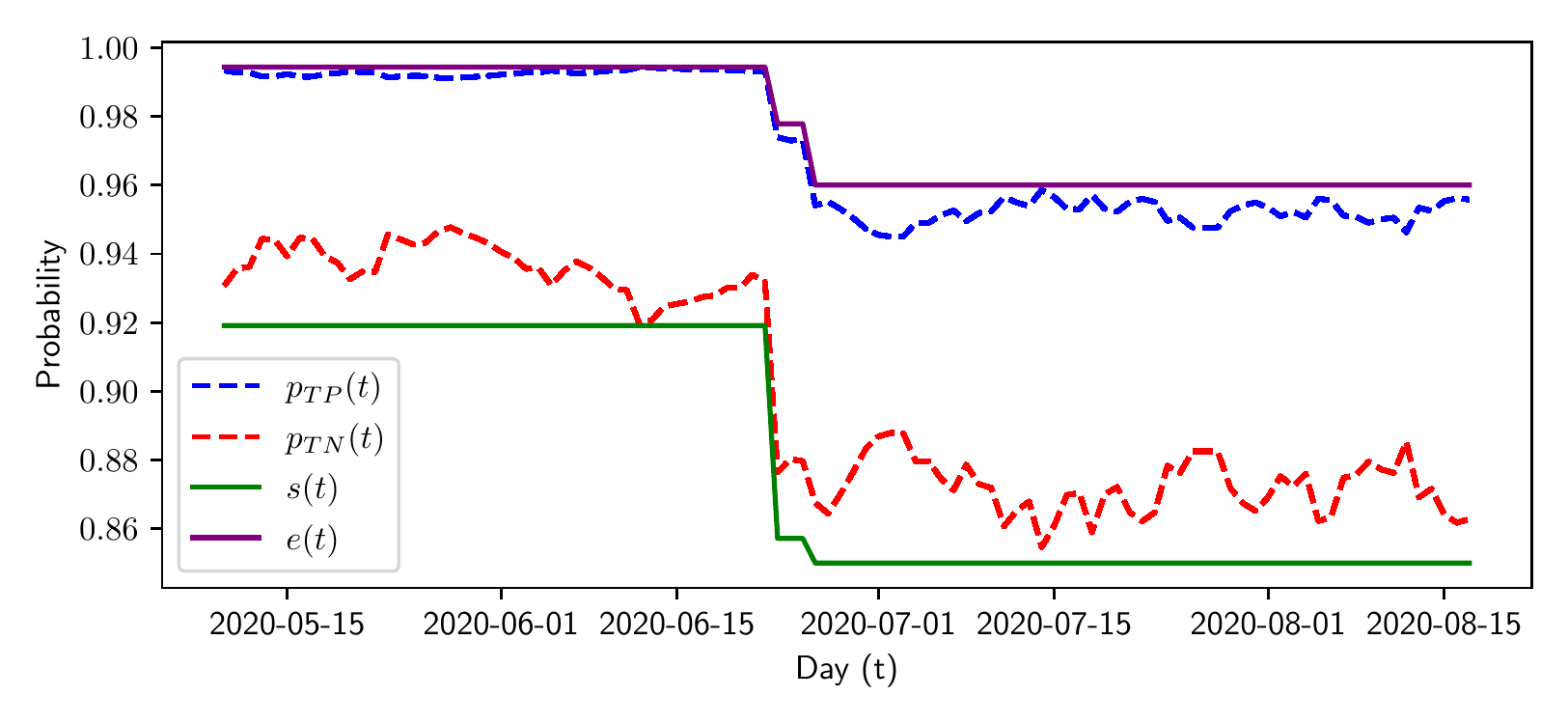}
    \caption{\textbf{Daily specificity $e(t)$, sensitivity $s(t)$ and constructed probabilities $p_{TP}(t)$ and $p_{TN}(t)$ in Equation \eqref{eq:prob_TP_TN}}. Note the dashed curves also rely on the sampling incomplete test types, so it changes over each realization of the reconstructed curve. Since the standard deviation are minimal, we chose to plot only the mean curve. Until April 28, only negative results were reported by rapid tests and the moving average has a window of 21 days. }
    \label{fig:spesentptn}
\end{figure}

Using data from Table \ref{table:testsintime} and assuming when more than one rapid test is available they are equally used, we obtain the overall daily specificity and sensitivity of rapid tests (see Figure \ref{fig:spesentptn}). From Table \ref{table:probs} in Section \ref{sec:maragogi_is_typical}, we use the resulting expected probabilities of hospitalized/infected ratio $p_h = 3.304\%$ and death/infected ratio of $p_d = 0.441\%$ overall.





Finally, the distributions of infection times were given by \cite{kerr2020covasim}, namely:
\begin{itemize}
    \item Incubation period (length of time between exposition and viral shedding): log-normal with mean $4.6$ days and deviation $4.8$;
    \item Symptom onset period (length of time after viral shedding has begun and before an individual has symptoms, when one has symptoms): log-normal with mean $1$ day and deviation $1$;
    \item Recovery period (length of time after incubation while the individual is infectious): log-normal with mean $8$ days and deviation $2$ for non hospitalized patients or with mean $14$ days and deviation $2.4$ for hospitalized patients.
\end{itemize}

\section{The reconstruction algorithm}

The reconstruction of the susceptible, exposed, infectious and recovered curves was performed by taking the mean over 400 curves generated stochastically. Each generated curve is also saved to be used in the calibration.

To build these curves we need to know, for each infected person, when one enters and leaves each compartment. For instance, we know the attendance date of the patient in Listing \ref{patientdata} for both attendances. We also know the hospitalization date and the symptom onset date for the first attendance. But the date when the patient was exposed to the virus, when it became infectious or recovered is unknown. This missing information will be reconstructed using previously known distributions, as listed in the last section, or re-sampling from the data. 

The reconstruction has tree main steps: test data correction, individual timeline reconstruction and cases estimate.

\subsection{Test data correction}

The test data correction step relies on two minor steps: sampling incomplete test type and inference of true positives ($TP$)\nomenclature{$TP$}{True positive} and true negatives ($TN$)\nomenclature{$TN$}{True negative}. 

\subsubsection{Sampling incomplete test type} First of all, we treated incomplete data. For each incomplete test type field (104 out of 1722) with date $t$\nomenclature{$t$}{Sample date}, we sampled its type (either rapid or RT-PCR) using all tests with known test type from the same date $t$. 

\subsubsection{Inference of true positives and true negatives}

The next step is to arbitrate if the test results are correct or not (for rapid tests, since RT-PCR tests are always assumed to be correct). 
    
Let $TP$ be the percentage of true positives, $TN$ of true negatives, $FP$\nomenclature{$FP$}{False positive} of false positives and $FN$\nomenclature{$FN$}{False negative} of false negatives. By definition, specificity (e)\nomenclature{$e$}{specificity of the test} and sensitivity (s)\nomenclature{$s$}{sensitivity of the test} are given by
\begin{equation}
    \textit{e} = \frac{TN}{TN+FP}\text{ and }\textit{s} = \frac{TP}{TP+FN},
    \label{eq:spesen}
\end{equation}
but we want to evaluate the probability of true positives ($p_{TP}$)\nomenclature{$p_{TP}$}{probability of true positives} and true negatives ($p_{TN}$)\nomenclature{$p_{TN}$}{probability of true negative}, i.e.,
\begin{equation}
    p_{TP} = \frac{TP}{TP+FP}\text{ and }p_{TN} = \frac{TN}{TN+FN}.
\end{equation}

We aim at writing both above equations in terms of known quantities: specificity and sensitivity are known from the technical notes \cite{anvisatests} and $p = TP+FP$ comes from the total number of positive tests along the period. From Equation (\ref{eq:spesen}) we have
\begin{equation}
    TN\left( 1-\frac{1}{e} \right) + FP = 0\text{ and }TP\left( 1-\frac{1}{s} \right) + FN = 0
\end{equation}
and from $TP+FP+TN+FN=1$,
\begin{equation}
    TP+FP = p\text{ and }TN+FN = 1-p.
\end{equation}

Thus,
\begin{equation}
    TP - TN\left( 1-\frac{1}{e} \right) = p\text{ and }TN - TP\left( 1-\frac{1}{s} \right) = 1-p.
    \label{eq:tptn}
\end{equation}

Solving (\ref{eq:tptn}), we have
\begin{equation}
    TP = \frac{ p + (1-p)\left( 1-\frac{1}{e} \right) }{ 1 - \left( 1-\frac{1}{s} \right)\left( 1-\frac{1}{e} \right) }\text{ and }TN = \frac{ (1-p) + p\left( 1-\frac{1}{s} \right) }{ 1 - \left( 1-\frac{1}{s} \right)\left( 1-\frac{1}{e} \right) }.
\end{equation}

Therefore,
\begin{equation}\label{eq:prob_TP_TN}
    p_{TP} = \frac{ 1 + \frac{1-p}{p}\left( 1-\frac{1}{e} \right) }{ 1 - \left( 1-\frac{1}{s} \right)\left( 1-\frac{1}{e} \right) }\text{ and }p_{TN} = \frac{ 1 + \frac{p}{1-p} \left( 1-\frac{1}{s} \right) }{ 1 - \left( 1-\frac{1}{s} \right)\left( 1-\frac{1}{e} \right) }.
\end{equation}

These quantities evolve with time since the proportion of positive tests varies over time. So, for a given day $t$ we let $p(t)$ be the ratio $p$ computed using a window of 21 days centered on $t$ (which matches the disease cycle used on the calibration). Also, let $e(t)$ and $s(t)$ be the mean specificity and sensitivity of the rapid tests available at day $t$.

Using the curves $p_{TP}$ and $p_{TN}$, one can determine if a given rapid test was positive or negative. We run that decision stochastically for each attendance with a rapid test. From now on, when we refer to positive tests, we are talking about the tests we judged as positive.

\subsection{Individual timeline reconstruction}

Each individual has one or more attendances. From the first attendance with a positive test result, if exists, we took the symptom onset date and the attendance date.

Let $i$ be an agent, $\tau_{E}^i$ its exposition date\nomenclature{$\tau_{E}^i$}{exposition date of an agent $i$}, $\tau_{I}^i$ the day it becomes infectious\nomenclature{$\tau_{I}^i$}{the day agent $i$ becomes infectious}, $\tau_{sym}^i$ the symptom onset day\nomenclature{$\tau_{sym}^i$}{the symptom onset day of an agent $i$} and $\tau_{R}^i$ the recovery (or death) date\nomenclature{$\tau_{R}^i$}{the recovery (or death) date of an agent $i$}, the individual disease timeline is the tuple $(\tau_E^i, \tau_I^i, \tau_{sym}^i, \tau_R^i)$. The value $\tau_{sym}^i$ is the only one we know and any other value can be stochastically constructed using the distributions in \cite{kerr2020covasim}, namely: 
\begin{align*}
 \tau_I^i - \tau_E^i &\sim \mbox{lognormal}(4.6, 4.8)\\
 \tau_{sym}^i - \tau_I^i &\sim \mbox{lognormal}(1, 1)\\
\end{align*}
 and for non hospitalized patients
\begin{align*}
     \tau_{R}^i - \tau_I^i \sim \mbox{lognormal}(8, 2)
\end{align*}
or hospitalized patients
\begin{align*}
 \tau_{R}^i - \tau_I^i \sim \mbox{lognormal}(14, 2.4).   
\end{align*}
Some attendances have no information about the symptom onset date (around $23\%$ of the positive cases). Again, from the filled data we derived the distribution of the time between symptom onset and medical attendance, only over positives cases, and then sampled the onset date of unfilled entries.

\subsection{Number of Cases estimation}

The ratios
\begin{equation}
    \frac{\textit{number of hospitalizations}}{\textit{number of cases}} \text{ and } \frac{\textit{number of deaths}}{\textit{number of cases}}
\end{equation}
should approximate the inferred ratios $p_h = 0.03304$ and $p_d = 0.00441$, respectively. Let $NB(q, n)$ be the negative binomial distribution with success probability $q$, which counts the number of Bernoulli failures should occur until $n$ successes. In the period till July 25th, a total of 18 individuals ended dying and 119 were hospitalized. So, we can model the number of cases as
\begin{equation}
    T_h = NB(p_h, 119) + 119 \text{ or as } T_d = NB(p_d, 18) + 18.
\end{equation}

Using the number of hospitalizations we have $\mathbb{E}(T_h) = \frac{119}{p_h} \approx 3601$ with a $90\%$ confidence interval of $[2966, 4033]$. Using the number of deaths we have $\mathbb{E}(T_d) = \frac{18}{p_d} \approx 4086$ with a $90\%$ confidence interval of $[2623,5760]$. Both confidence intervals agree, although the confidence interval estimated using deaths is larger. Since it has a narrower confidence interval, we use $T = T_h$ to estimate the total number of cases.

It is also interesting to notice that the data seems consistent, the ratio between recorded deaths and recorded hospitalizations is $\frac{18}{119} \approx 15.1\%$ and the ratio $\frac{p_d}{p_h}$ is approximately $13.3\%$, a small difference.

\section{The final curve}

Let $H$ be the set of all hospitalized individuals and $N$ the set of all non-hospitalized infected individuals, define
\begin{equation}
    E_H(t) = \sum_{i\in H} \mathbf{1}_{[ \tau_E^i, \tau_I^i )}(t)\text{ , } I_H(t) = \sum_{i\in H} \mathbf{1}_{[ \tau_I^i, \tau_{R}^i )}(t)\text{, }R_H(t) = \sum_{i\in H} \mathbf{1}_{[ \tau_{R}^i, \infty )}(t).
\end{equation}
Let $E_N(t)$, $I_N(t)$ and $R_N(t)$ be defined analogously. Also, let
\begin{equation}
    C_H = \sum_{i\in H} \mathbf{1}_{[ \tau_{A}^i, \infty )}(t)\text{ and }C_N = \sum_{i\in N} \mathbf{1}_{[ \tau_{A}^i, \infty )}(t),
\end{equation}
where $\tau_{A}^i$ is the first attendance date with a positive result.

Assuming no sub-notification among hospitalizations and deaths. Also, using $T$ cases on July 25th, we define 
\begin{equation}
     \alpha = \frac{T - C_H(t^*)}{ C_N(t^*) },
\end{equation}
where $t^*$ is July 25th, $C_H(t^*) = 119$ and $C_N(t^*)$ varies depending on the missing data reconstruction, the inference of test results and the individual timeline reconstruction. Then, the quantity $\alpha$ captures the ratio between overall mild cases and followed mild cases. On average, only $16.75\%$ of the patients with mild or no symptoms looked for medical help.

Finally, we reconstruct the curves:
\begin{equation}
    E(t) = E_H(t) + \alpha E_N(t)\text{ , }I(t) = I_H(t) + \alpha I_N(t)\text{ and }R(t) = R_H(t) + \alpha R_N(t).
\end{equation}

\begin{figure}[h]
    \centering
    \includegraphics[width=\linewidth]{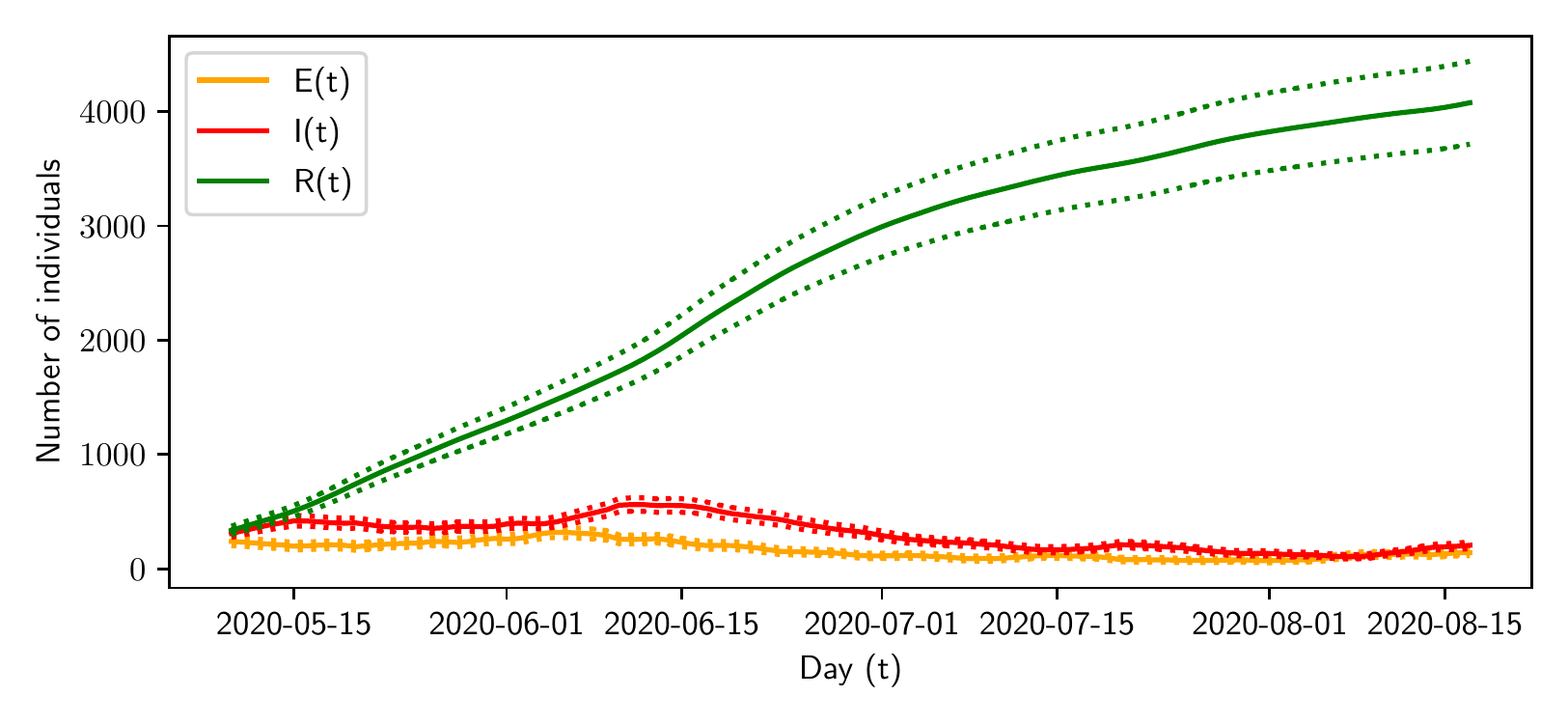}
    \caption{\textbf{Final estimated curve for exposed, infectious and recovered compartments.} The solid lines are the mean over all 400 trials and the dashed ones represent one standard deviation up and below.}
    \label{fig:finalcurve}
\end{figure}

The procedure has four stochastic steps: test type re-sample, test result correction, symptom onset date re-sample and individual timeline reconstruction. The final curve is given by the mean over 400 trials, see Figure \ref{fig:finalcurve}. Of course, the curve of susceptible individuals ($S(t)$) is given by the total population minus the sum $E(t)+I(t)+R(t)$.

\chapter{COMORBUSS - Stochastic Agent model}
\label{sec:COMORBUSS}

\section{COMORBUSS}

COMORBUSS is a bio-social agent model for the study of disease propagation in a community and the evaluation of mitigation measures. Let us clarify each part of this statement. 

\begin{figure}
    \centering
    \includegraphics[width = 0.85\textwidth]{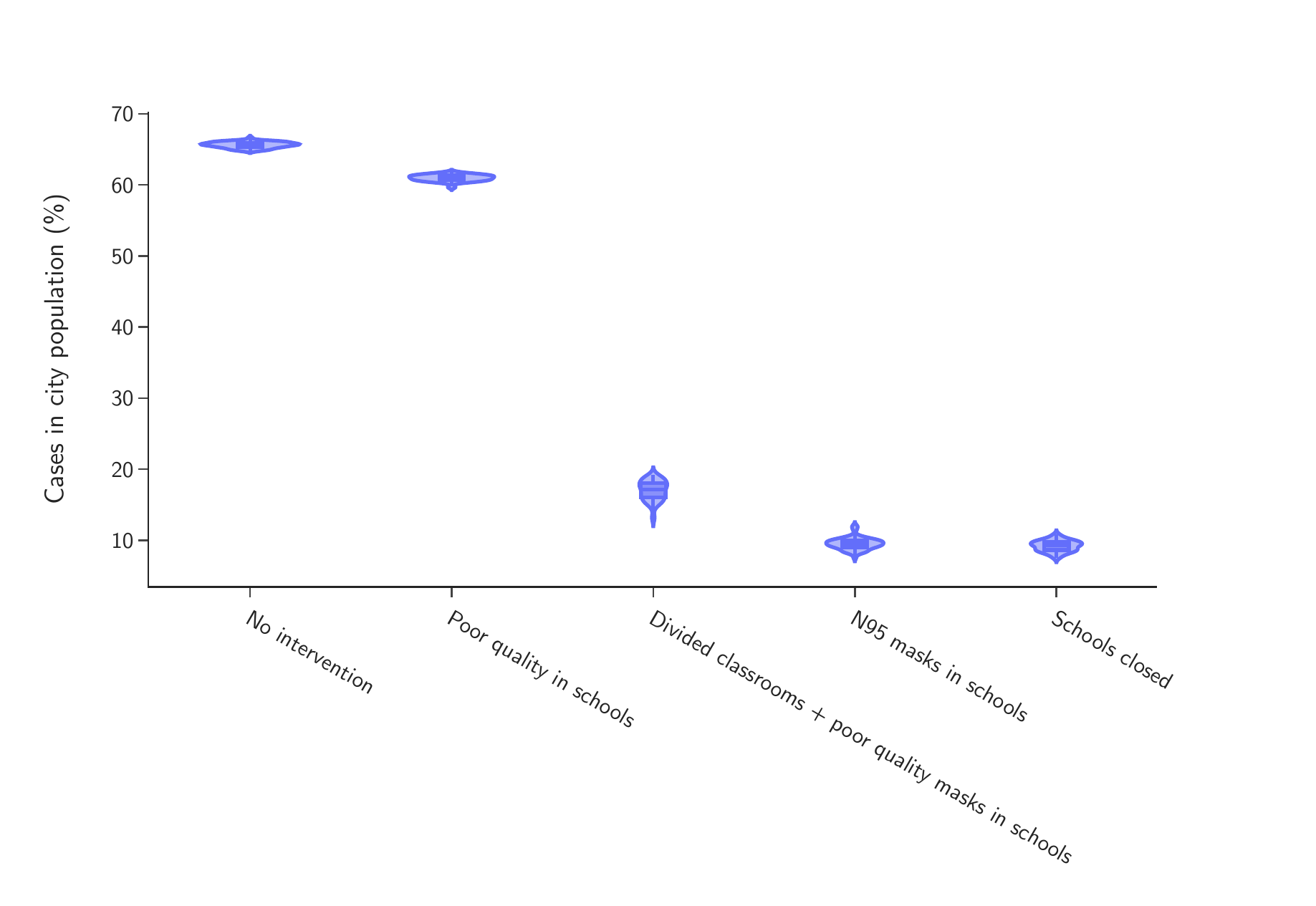}
    \caption{\textbf{Protocol efficacy.} Effectiveness of different mitigation policies measured by cases increase for Maragogi-AL.}
    \label{fig:effectiveness}
\end{figure}

An agent model is a one where we simulate individual agents which represent persons in the modeled community. These agents interact with each other and the environment according to a set of rules, and have their own characterization. This allows for the creation of models which captures the heterogeneity of the real community we are studying. Moreover, mitigation measures can be directly modeled by modifying the behaviour of the agents (e.g. quarantines, social isolation, reduction of students in classrooms) or the transmission of the pathogen (e.g. masks and vaccination). This way, the effectiveness of these mitigation policies can be directly measured and compared, see Figure \ref{fig:effectiveness}.

\begin{figure}
    \centering
    \includegraphics[width = 0.85\textwidth]{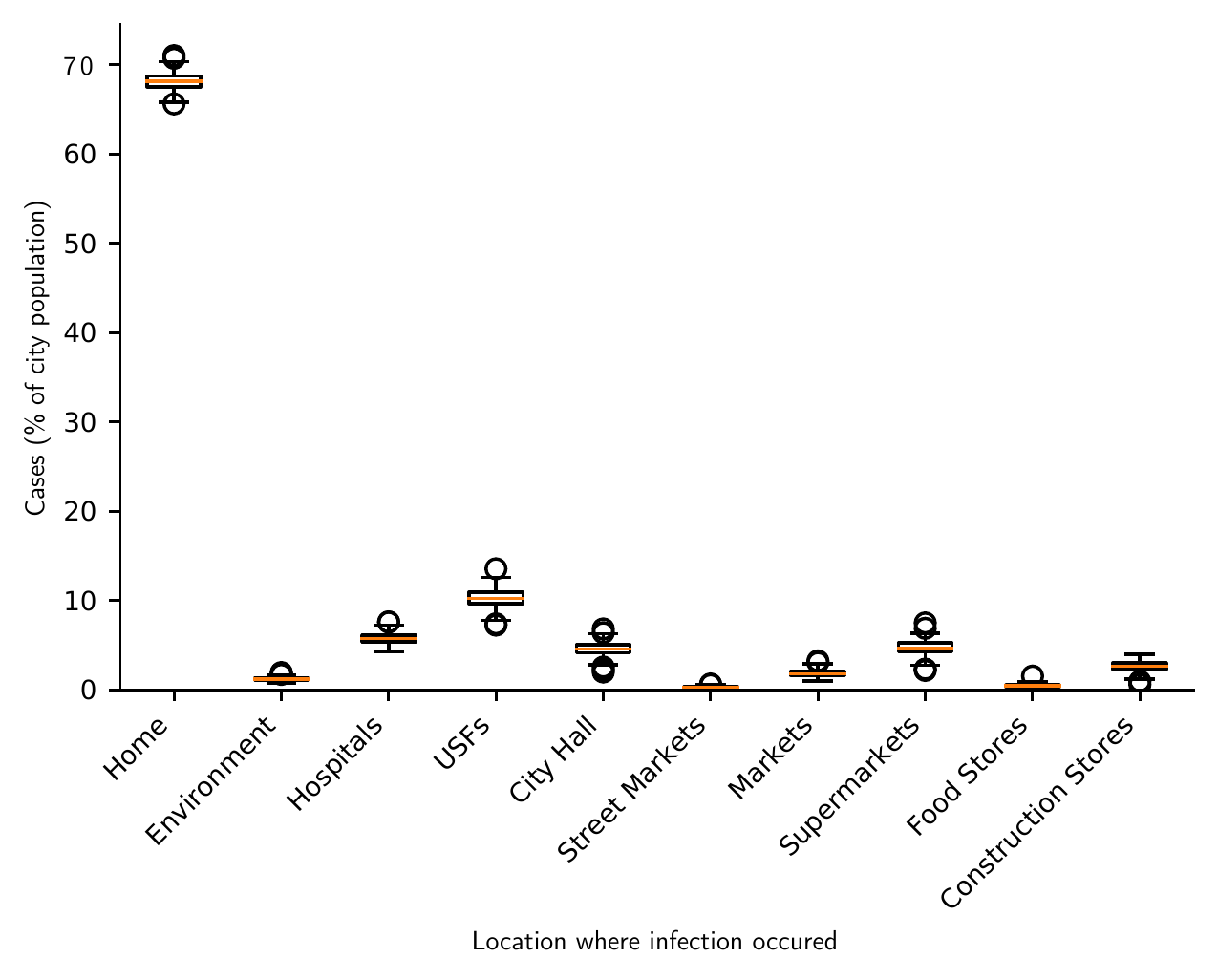}
    \caption{\textbf{Infection placement.} Percentile of infections that happen in each service category.}
    \label{fig:inf_location}
\end{figure}

By bio-social, we mean to emphasize that COMORBUSS at its core is driven by two stochastic models: one for disease progression and propagation based on the individual biology of the agents and the other for the social dynamics of the agents based on their identities and roles in the community. Connecting these two models is the core modeling assumption that disease transmission rides on social contacts produced by the community dynamics. As the social dynamics model drives the individual agents as workers or clients of the services which define the infrastructure of the community (such as hospitals, schools, markets, restaurants, stores etc.), the agents meet at these locations and possibly infect others. As transmission is contextualized by location and by the roles of the agents involved (e.g. client, worker), we can identify which are the services that contribute the most in driving the infection; see Figure \ref{fig:inf_location}.

COMORBUSS as an agent-based model possesses the following remarkable advantages which are derived from \textbf{our approach to directly model social dynamics} and the \textbf{omniscience the model guarantees to the analyst} :
\begin{itemize}
    \item individualized and heterogeneous description of the community;
    \item behaviour models for interventions and their quantitative assessment, even with partial compliance;
    \item realist decision making models with dynamic criteria for adoption of interventions;
    \item ability to produce counterfactual scenarios regardless of the complexity of the scenario, enabling direct comparison in experimentation;
\end{itemize}

All these advantages make COMORBUSS a valuable tool both in the evaluation of policies and in the development and testing of new ideas and methods in epidemiology.

\section{Community Model}
\subsection{Creation: initializing a mimetic community model stochastically}
We seek the average epidemiological behaviour and the associate variance for a city with a given demography. This is done by simulating multiple realizations of a stochastic model for the disease propagation in this community. In order to eliminate biases introduced by a single  societal network, we generate for each random seed a new community representation which captures the following real demographic information of a given city:
\begin{itemize}
    \item population size;
    \item age distribution (binned in groups of 5 years);
    \item household structure (size distribution and age distribution of members);
    \item service infrastructure;
    \item job allocation by age group.
\end{itemize}

\subsubsection{Creating households while preserving age distribution and average household size}\label{sec:household_init}

The agents are created in household groups which are defined sequentially and modified such that real age distribution and average household size are respected. In order to avoid unrealistic household structures (e.g. children living unsupervised) and to consider households with sizes far from the average, we have created and carefully curated an artificial dataset of households mimicking the households in the modelled city (see SI 3).

While the generated population is smaller than the desired population, a household is sampled from the reference population dataset. We then evaluate the average size of the households created so far: if it is smaller than the desired average household size, a new agent is added to this house; if it is  larger, an agent is randomly removed from this house. 


The probabilities used in the selection of agents to be added or removed is computed from the difference between the real age distribution and that of the current agent population. We then look at the resulting values for the age groups of the agent candidates.

\begin{itemize}
    \item If removing an agent: we consider as candidates for removal only the agents whose age groups had negative values in the difference between distributions. We then assign the absolute value of these differences to each agent and normalize them so that they sum to 1. Each value is then used as the probability of removing the corresponding agent.
    \item If adding an agent: we consider candidates for creation only agents whose age groups had positive values in the difference between distributions. We then filter these positive values and normalize them so that they sum to 1. These are used as the probabilities for selecting an agent of the corresponding age group for creation.
\end{itemize}

\begin{figure}
    \centering
    \includegraphics[width = 0.95\textwidth]{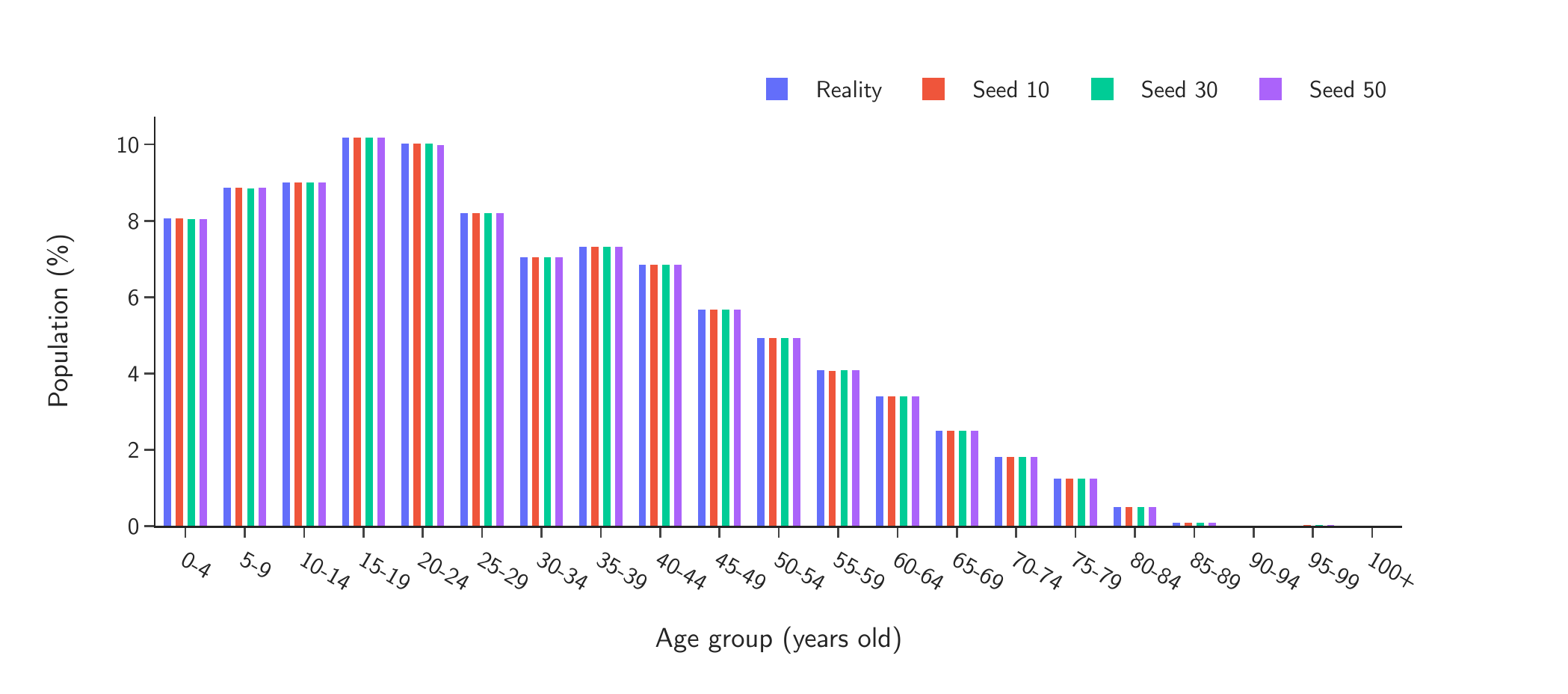} 
    \caption{\textbf{Age distribution.} Comparison of real age distribution of Maragogi-AL and for a few randomly generated populations using to our algorithm.}
    \label{fig:age_goups}
\end{figure}


\subsubsection{Household initialization of compartmental data}\label{sec:household_init_compartimental}

In contrast to ODE simulations using compartmental data, which only require the compartment values for initialization, a bio-social agent-based simulation also requires relating compartments with social characteristics in the community. For example, in a community with $250$ individuals initialized with $5$ infectious ones, having the $5$ agents living in the same house or having them living in $5$ different houses generates very different results. In the first case, the disease cannot spread more in the same house, while in the second, it can use the time infected individuals stay at home to spread to others. While a random initialization can still be used to generate a certain tendency in simulations, the high variation in the outcomes demands several realizations in order to reduce standard deviation. Since the preferred environment for spread is always individual's homes, see \cite{Curmei_2020_05_23}, the average results can also be misleading. The reason is that a random initialization of few agents will most likely position only one infected agent per home. Another drawback of this approach is that it ties simulation results, and therefore calibrated parameters, to the number of agents in the community, therefore making it difficult to export results to other cities of similar but still different attributes. 

Ideally, one should be able to relate compartment data to age, social role and household distribution in a time dependent manner. This level of information would allow for a complete disassociation of the probability of infection $p$, the main parameter calibrated in this work, to the community and its individuals. Unfortunately, it is clear that such data is not available in practice, and that collecting it in a meaningful representative way would be nearly impossible. We propose a synthetic half-way solution to this problem, which consists of using data gathered from social behavior and household structure to determine the time independent probability of a given compartment structure be present at a given home. 

The technique we use to synthetically detach the calibrated infection probability $p$ from most population characteristics is to answer the following question: given a home with $n$ individuals, such that it has at least one of its members with an active compartment state, what is the chance that a given compartment configuration is present during the disease life of that home? To answer this question, let us first clear out the meaning of some of these terms:
\begin{itemize}
    \item Active compartment state: exposed or infectious states.
    \item Disease life at a home: the period of time in which at least one individual from that home is at an active compartment state. 
    \item Compartment configuration: distributing codes for each compartment, such as \verb|S|, \verb|E|, \verb|I|, \verb|R|, a configuration is any member of the combinations of $n$ codes out of the $4$ possible ones. For example, with $n=3$, the configuration \verb|SEE| tells us that the home has one susceptible person and two exposed ones. After some time, the same home can have the following configuration: \verb|SEI|, meaning that one of the exposed persons became infectious.
\end{itemize}

\begin{figure}
    \centering
    \includegraphics[scale=0.8]{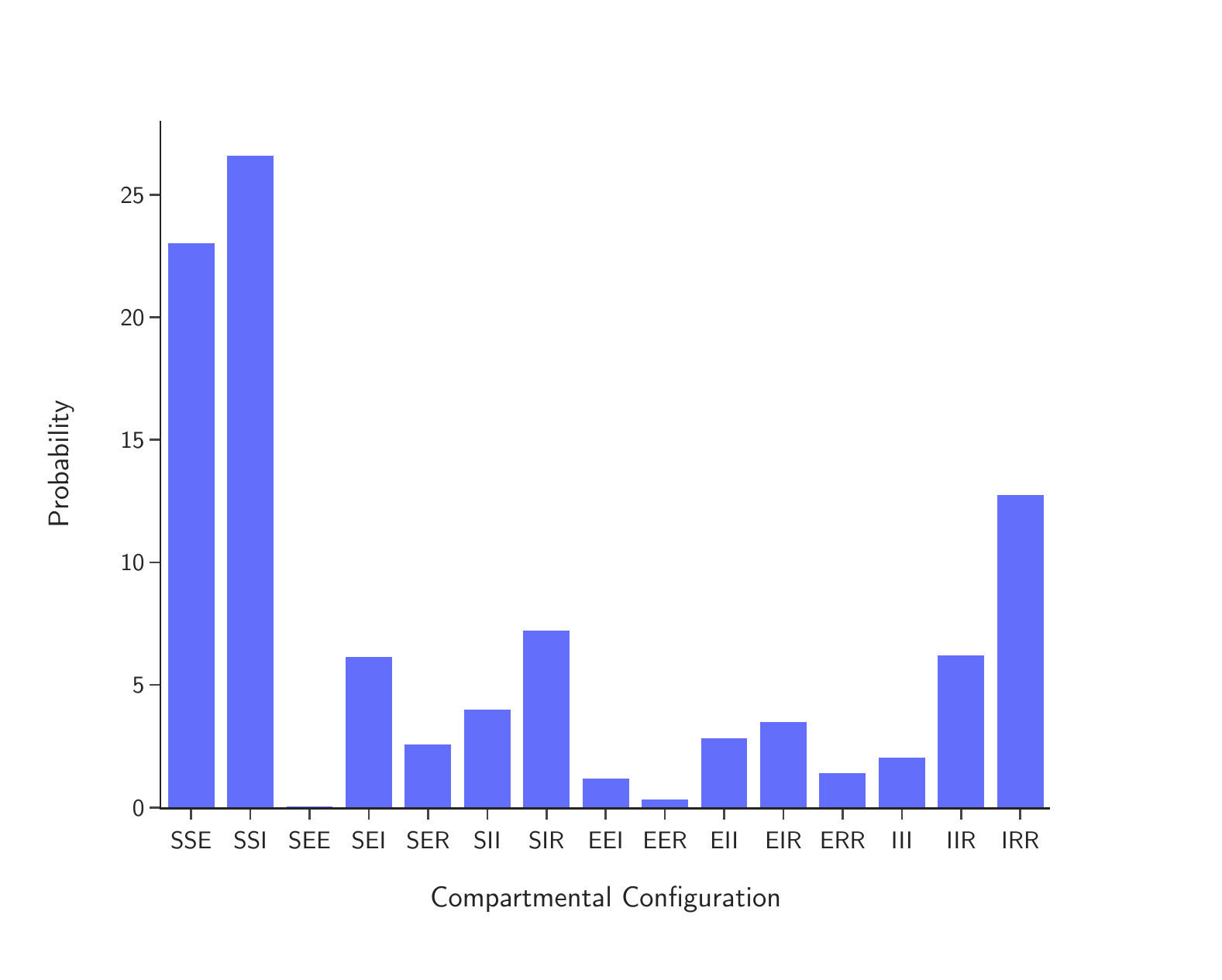}
    \caption{\textbf{Initial household state configurations.} Compartment configuration probabilities used for active homes occupied by tree people}
    \label{fig:exdic3ppl}
\end{figure}

To determine the probability of a compartment configuration occur at a home during its disease life, we divide the time a configuration is present at that home by the total time of its disease life. We do that for as many houses as possible, averaging out the probabilities for houses with same number of people. Figure \ref{fig:exdic3ppl} shows the values we used to initialize active houses with 3 people. From the figure we deduce that, if a house with 3 people is active, and all states are possible, then the most likely configuration to occur is \verb|SSI|, with about $26.6\%$ of chance. The second most likely configuration is \verb|SSE|, with about $23\%$ of chance. The least likely configuration is the one with two exposed individuals and one susceptible one, with about $0.05\%$ of chance, and so on. Notice that a random initialization would most likely have much higher probabilities towards \verb|SSE| and \verb|SSI|, being the remaining ones not present in most realizations of the community. It is also important to mention that, when the probability of infection changes, so does the compartment configuration probabilities. In general, whenever we speak about the probability of infection in this work, we are also including inherent compartment configuration probabilities that come along with it. 


The compartment configuration probabilities should be approximately the same for any point in time, and they also should not depend on the population number, only its household structure and overall social behaviour. As a result, we can safely use these probabilities to distribute compartment labels to individuals of a randomly selected active home, for as long as there are active compartment labels to distribute. After all active compartment labels are distributed, we randomly select houses to contain a given proportion of susceptible and recovered individuals. This proportion is also estimated from homes which have ended its disease life, therefore not containing active compartment individuals. 

Although the procedure just described allows for a consistent initialization of compartment states along homes in a community, collecting the data necessary to do so is still a hard task. However, if we suppose that the overall social behavior of the community is well captured, and also that the disease modelling inside homes is realistic, then we can conclude that the probability of infection itself determines the probability of each compartment configuration occur at an active home. This is the argument that allows us to use COMORBUSS to determine the compartment configuration probabilities. The idea is that, in order to simulate a community for a given probability of infection $p$, we first use another simulation to determine the compartment configurations for each possible size of a home. That is, we perform many realizations of a community disease spread using random initialization of states first. Next, we use houses whose disease life is fully captured by the simulations to determine the probabilities for each compartment configuration. Once these probabilities are derived, they are used to initialize compartment states of a second simulation, being that the one that approaches the real life's initial spread the most.

\subsubsection{Service infrastructure and job allocation}
Each service category (e.g. hospitals, supermarket, schools etc.) is created as a computer object sharing common defining and operating parameters. Inside each of these objects, we instantiate the same number of these service locations as are known to be had by the modelled city. One of the defining parameters is the average number of workers in the service category and the age groups that are known to function as workers. From this, when the service is created, we randomly select agents in the population of the appropriate age group and assign them as workers for that service. More detailed assignment procedures are in principle possible but are unavailable due to lack of required data. 

\subsection{Dynamics: stochastic model for community behaviour}
\subsubsection{Services as community drivers}
The core concept of COMORBUSS is the utilization of services to dynamically generate contacts in our model community. As such, each relevant social context is modelled as a service, even "the environment" as is dubbed random meetings on the streets and parks. The services which have been modelled in this work are
\begin{itemize}
    \item health facilities: hospitals, public health clinics;
    \item educational facilities: schools and day-cares;
    \item essential stores: street markets, markets, supermarkets, food stores, construction stores;
    \item city hall and environment.
\end{itemize}
\noindent A list of parameters which define a service is given and explained in the next section. 

An agent can have one of three roles when inside a service: it can be a worker, a visitor or a guest (someone whose standard address has been temporarily set to the service, due to hospitalization or quarantine). Every agent is assigned as a visitor to an instance of the service categories that it can visit. This agent will always visit the same instance, unless it is temporarily reassigned to a new instance. This can happen when a service is closed due to lack of workers (all workers being in quarantine or hospitalized). Visitors can in principle be assigned to multiple instances of that service class (e.g. visit more than one supermarket), but there was no reliable data sources upon which we could model more complex behaviour.

Each service has two defining restrictions: its working days and hours, as well as the age groups allowed to use it. Another key parameter is the average period of visitation for that service (e.g. one can say that any person visits the supermarket every week). From this we have the average frequency in days that the service is visited and, using the number of working hours of that service, we compute the hourly probability that an agent will visit that service.

During every hour a service is open, free visitors are randomly selected and sent to the instance of that service they are assigned to. If the agent is unable to make a visit (e.g. agent is working or visiting another service), the probability is accumulated to a later hour that the service is open and the agent is available to visit. In this way we organically produce "rush hours", such as when many workers visit the supermarket after their working hours. After concluding its visits, the agent is then returned to its address until it is selected again for some other activity.

Similarly, workers are sent to the instance of the service where they work during their working hours. One can also allocate the agents uniformly on different shifts. Guests are so far defined only via hospitalization or quarantines, so their mobility is restricted until the associated measure is completed. They are then returned to their household address, where normal social activities are resumed.

In this way, agents are driven between different locations and contexts according to their individual needs. At the same time, the collective behavior of the agents produces a complex and realistic model for the community dynamics. Any non pharmaceutical intervention can be modelled as temporary changes in the individual or collective behavior of the agents and its consequences can be measured directly.

\subsubsection{Visitation period}\label{sec:visitation_period}

Letting aside the interaction of agents inside the service for a moment, the visitation of agents is what contributes the most for the relevance of services in the disease spreading at the community. COMORBUSS models the visitation of agents to a service by randomly picking them according to a probability $p_{v}$. This happens at every time step that the service is opened and that agents are free for visitation, meaning that they are not resting at home nor visiting any other service. The probability $p_{v}$ is then given by the inverse of the visitation period $v_p$: $p_{v}=1/v_p$, where $v_p$ is a measure of how many time steps an agent takes to return to a service, given that it is opened\footnote{A direct implication of this definition is that the visitation period cannot have time length lesser than that of a time step}. To make the visitation period independent on the opening of the service and also on the magnitude of time steps, we assume that it is provided in consecutive days, and then we convert it to time steps. The conversion formula is given by 
\begin{equation}
    v_p = \frac{d_o}{7}\frac{h_o}{\Delta t}v_{pc},
\end{equation}
where $d_o$ is the total number of days a service is opened on a week, $h_o$ is the total number of hours a service opens for a day, $\Delta t$ is the time step in hours, and $v_{pc}$ is the visitation period in consecutive days. To calculate $v_{pc}$, it suffices to know the total number of visitors $v_w$ a service receives during a week, and the total number of agents $v_t$ that can in fact visit the service. With these two values the visitation period in consecutive days is given by 
\begin{equation}
    v_{pc} = 7\frac{v_t}{v_w}. 
\end{equation}


\subsection{Location contextualized contact networks}\label{sec:contact_networks}
By collecting the list of visitors, workers and guests at an instance of a service at any given time, we naturally know the collective history of the community and the sets of agents that can interact. However, how these agents interact is closely associated with the social context at that time. As examples: one does not interact closely between tables in a restaurant while the waiter interact with the set of tables it is responsible for as well as coworkers; in a classroom or factory, people are rigidly placed in space for most of the time. 

We therefore need to consider the social context of the agents in the process of taking the list of agents in a location and producing a contact network. COMORBUSS identifies each particular service having its own network structure, so distinct network models are built when representing restaurants, markets, hospitals or schools.

All network models share as a common feature to contain in each service two types of individuals: \emph{workers} and \emph{guests}. \emph{Workers} of a particular service are individuals which stay in this service over a daily time period during a realization of the stochastic community. On the opposite, \emph{guests} are individuals which visit a single time step that service respecting a frequency of visitation over the simulation. 

Both types of individuals are specialized for each service to mimic realistic features one may find in real-world services. For instance, waiters in restaurants are modelled as workers having contacts with visitors. The same idea is applicable to cashiers in markets. Hence, these observations also must be taken into account in the modelling of contact networks. Below we detail the network model for each service.

\subsubsection{Standard networks: houses and generic services}

In houses or generic services, no network configuration can be assumed. As a result, we utilize a standard network model in order to generate contacts. The contacts in this model are randomly distributed for each agent according to a given average number. This average value may change as the number of agents increase or decrease inside services, as discussed in Section \ref{sec:aggl}. Nevertheless, the contact networks generated are still dynamic, varying as agents are added or removed.

The contact networks are, with few exceptions, generated using Erdos-Renyi model, where the probability
$p_{ER}$ of an edge being added is given by $p_{ER} = c_{avrg} /(N − 1)$. Here $c_{avrg}$ is the average number of contacts and $N$ is the total number of nodes in the graph. The parameter $c_{avrg}$ depends on the definition of contacts, and in this work we assume it to be the following: “two people at two meters or less away from each other for the duration of one hour”. 

\subsubsection{Contact varying with agglomeration}\label{sec:aggl}

Any contact networks needs a fundamental parameter, the average number of contacts (vertices) across the nodes. By default, this input parameter is fixed for each type of network. However, its variation over time number may need to be considered in some social contexts due to high variation of occupational density of people in that place. For example, in the case of markets, there are rush hours in which agglomeration is higher. It is also common in this type of service that there are considerable less clients in the beginning or end of the work day. To deal with the non-uniformity of the number of agents inside each service, we propose a formula to adjust the average number of contacts. The idea comes from supposing that the opportunity for a contact is directly linked to the available space to the agents.

Suppose that two of $N$ agents get in contact with each other whenever they share some specified area $A$ around their position in space. The expression relating $c_{avrg}(N)$, $A$ and $N$ is given by
\begin{equation}\label{eq:rand_walk_avrg_cont}
    E(c_{avrg}(N)) = \frac{\frac{N(N-1)}{2}A}{N} = \frac{N-1}{2}A,
\end{equation}
where $E$ is the expectation operator. This formula comes from assuming random walking of $N$ agents inside a given service with transit area $A$. To avoid knowing the transit area, we suppose that a sample $N_0$, $c_{avrg,0}:=c_{avrg}(N_0)$ is collectable, and then we approximate $A$ through the formula
\begin{equation}
    A\approx \frac{2c_{avrg,0}}{N_0-1}.
\end{equation}
As a result, the expression for the mean number of contacts $c_{avrg}(N)$ varying with the number of agents $N$ is  
\begin{equation}\label{eq:aggl_cont}
    c_{avrg}(N) \approx c_{avrg,0}\frac{N-1}{N_0-1}.
\end{equation}

Equation \eqref{eq:aggl_cont} is used in some of the contact networks introduced below. For example, in the case of markets, supermarkets or street markets, the formula can be used to adjust the average number of contacts among visitors shopping in the services. In the case of hospitals, the formula can be used in the average number of contacts among visitors. Another place where such expression is useful is in the contact network for homes. Assuming equally sized homes, one can infer that the more people at home, the more contacts. Since the number of people at home varies along the day, such formula is well fitted to capture the dynamics of movement inside a home.

\subsubsection{Networks for environment layer}\label{sec:env_net}

The dynamic of contacts in environment layers is very individual-specific, and therefore we approximate it by random walking. The formula for the average number $c_{avrg}$ of contacts among $N$ agents in the environmental layer with transit area $A$ is given by equation \eqref{eq:rand_walk_avrg_cont}. The transit area $A$ is in this case given by:
\begin{equation}
    A = \frac{\pi r^2}{A_u},
\end{equation}
where $r$ is an \textit{infection radius}, and $A_u$ is the urban area available in the environment layer. The infection radius is given by half of the largest distance between two agents such that they can be considered to be in contact. We assume $2$ meters as a default value.

\begin{figure}
    \centering
    \includegraphics[scale=0.8]{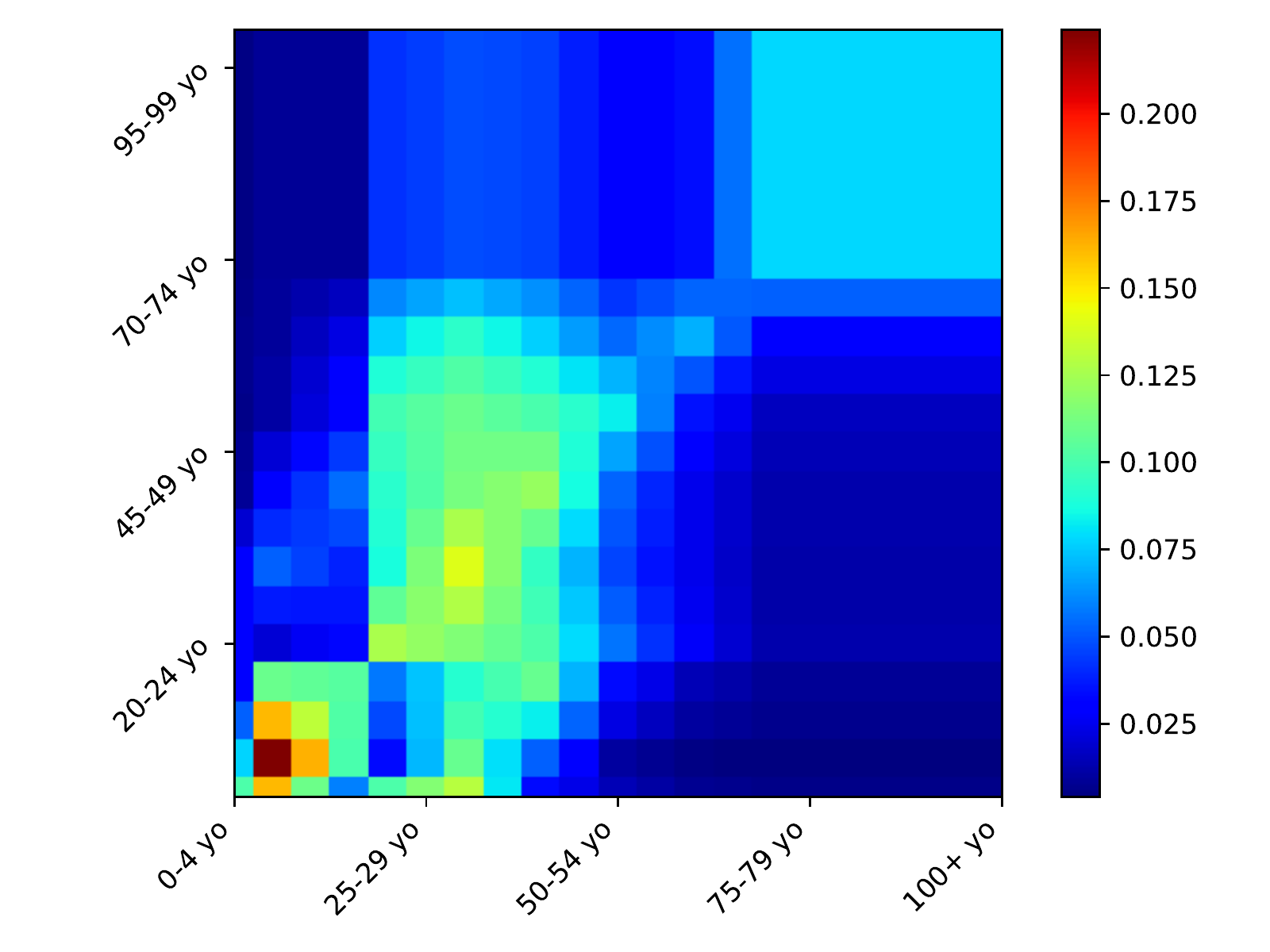}
    \caption{\textbf{Contact probability matrix.} Color map representing probabilities that if a person from an age group in the y-axis met someone, that person belongs to the age groups in the x-axis.}
    \label{fig:waifw}
\end{figure}

Although at random, contacts may follow some tendency according to the age of agents. We have used the probabilities exposed in Figure \ref{fig:waifw}, which have been derived from Table 2 of \cite{DelValle2007}.

\subsubsection{Network for restaurants}

Waiters are restaurant workers who have the greatest potential of becoming disease super spreaders inside their work place. This happens because they get in contact, as a group, with every visitor who enters the restaurant. As a result, waiters define a special group of workers which must have special treatment regarding their contact network.

Taking into account the social roles of waiters in restaurants, we model contacts in three categories:
\begin{itemize}
    \item \emph{visitor-visitor} contacts. 
    \item \emph{waiter-visitor} contacts. 
    \item \emph{worker-worker} contacts.  
\end{itemize}

Having these categories in mind, the contact networks for restaurants are configured by setting the following parameters: the portion of workers who are waiters, the average number of contacts among workers, and the mean number of persons seating around the same table. Because this last parameter is usually difficult to estimate, it can be discarded, in which case tables are evenly distributed among waiters in the restaurant. 

The contact network for workers is randomly created, always respecting the mean number of contacts provided as input. Among these workers are those composed of waiters, who get in contact with every visitor on the tables they serve. These visitors in turn get in contact with everyone else in the same table. 

\begin{figure}
    \centering
    \includegraphics{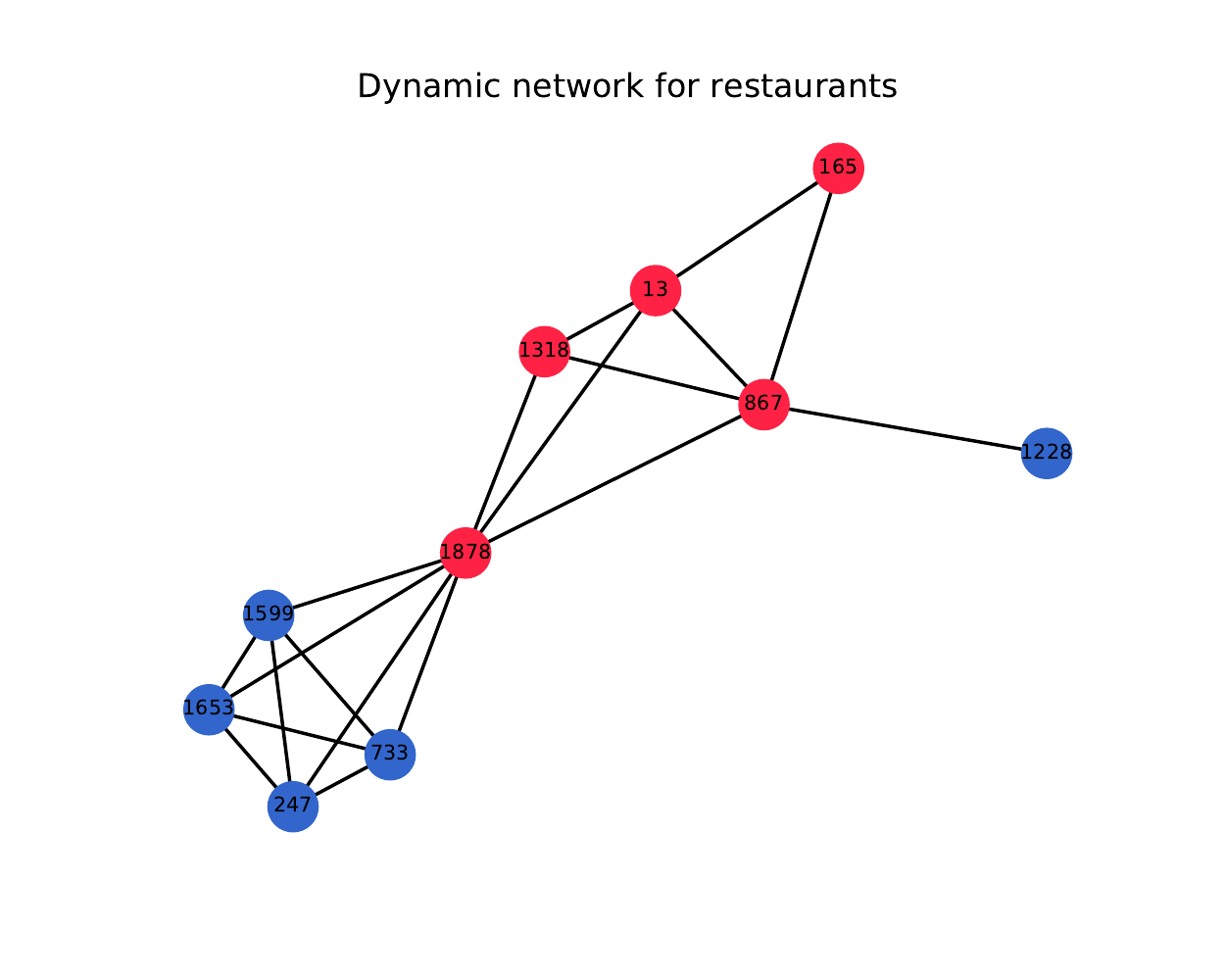}
    \caption{\textbf{Example of a dynamic network for restaurants.} Agents (with their ids in circles) in red are workers, agents in blue are visitors. Numbers inside each circle represent the identification number of that agent.}
    \label{fig:rest_net}
\end{figure}

Figure \ref{fig:rest_net} shows an example of a network for a given restaurant with $5$ visitors, $2$ waiters and $3$ other workers. Notice that only the waiters, identified by ids $1878$ and $867$, are those who get in contact with the visitors. It is clear, however, that other workers get in contact among themselves. The same thing happens to visitors in the same table.

\subsubsection{Network for markets}\label{sec:markets_net}

The contact network for markets is similar to that of restaurants in the sense that there exists a class of workers which needs different treatment: the cashiers. While other workers usually do not get in contact very frequently with visitors, every visitor mandatorily gets in contact with a cashier, either directly or indirectly through shared surfaces, such as shopping belts or credit card machines. Second order contacts include those among visitors and among workers. 

The contacts between workers and visitors is randomly created, respecting a given average of contacts provided as input. The contacts between visitors and cashiers is also random, but in this case each visitor is assigned to a cashier. Cashiers are fixed agents, which comprise a fixed proportion of all workers in markets. 

\begin{figure}
    \centering
    \includegraphics{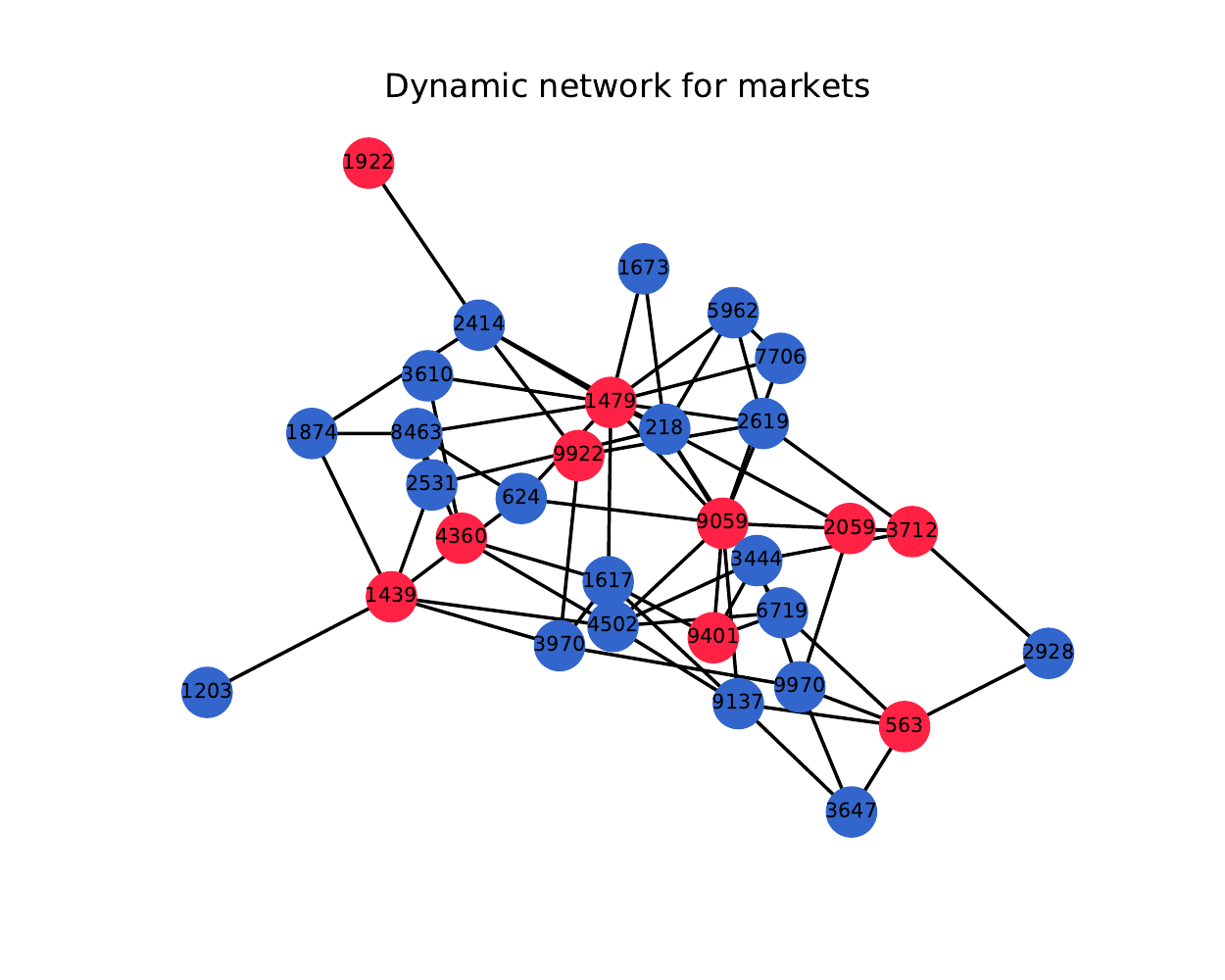}
    \caption{\textbf{Example of a dynamic network for markets.} Agents (circles) in red are workers, agents in blue are visitors. Numbers inside each circle represent the identification number of that agent. A fraction of the workers is designated as cashiers and each client passes through one of them.}
    \label{fig:market_net}
\end{figure}

Figure \ref{fig:market_net} shows an example of a network for a market. Notice that every visitor (blue agent) gets in contact with at least one worker. Cashiers are workers (red agents) who get in contact with many visitors. Example of cashiers in the figure are those with ids $1479$ and $9059$. Example of non-cashiers are those with ids $1922$ and $2059$

\subsubsection{Network for schools}

Schools have two different network models: one for class time, and one for break time. During classes, the nature of contacts among students can be very geographic, since students tend to stay seated during long periods of time. During breaks, students are free to walk around public spaces inside the schools. As a result, the distinction between two types of networks is needed. 

During class time, we propose a network that connects agents according to the neighbors nearby, where students are assumed to follow a geographical disposition of a rectangle. Teacher are treated separately, since they usually move more frequently. The contact frequency between a student and a teacher may vary according to the age of the student. For example, for students who are toddlers, the contact is frequent, but for university students, direct contacts are unusual. The parameters for this type of network are the number of students in a class, and the average number of contacts between teachers and students.

During break time, we propose a simple network in which students get in contact at random. The factor that influences the most this type of network is the number of classes allowed to have a break together, as well as the different ages of the students. The parameters for this network are the number of classes to have break together and the average number of contacts among students. 

\begin{figure}
    \centering
    \includegraphics{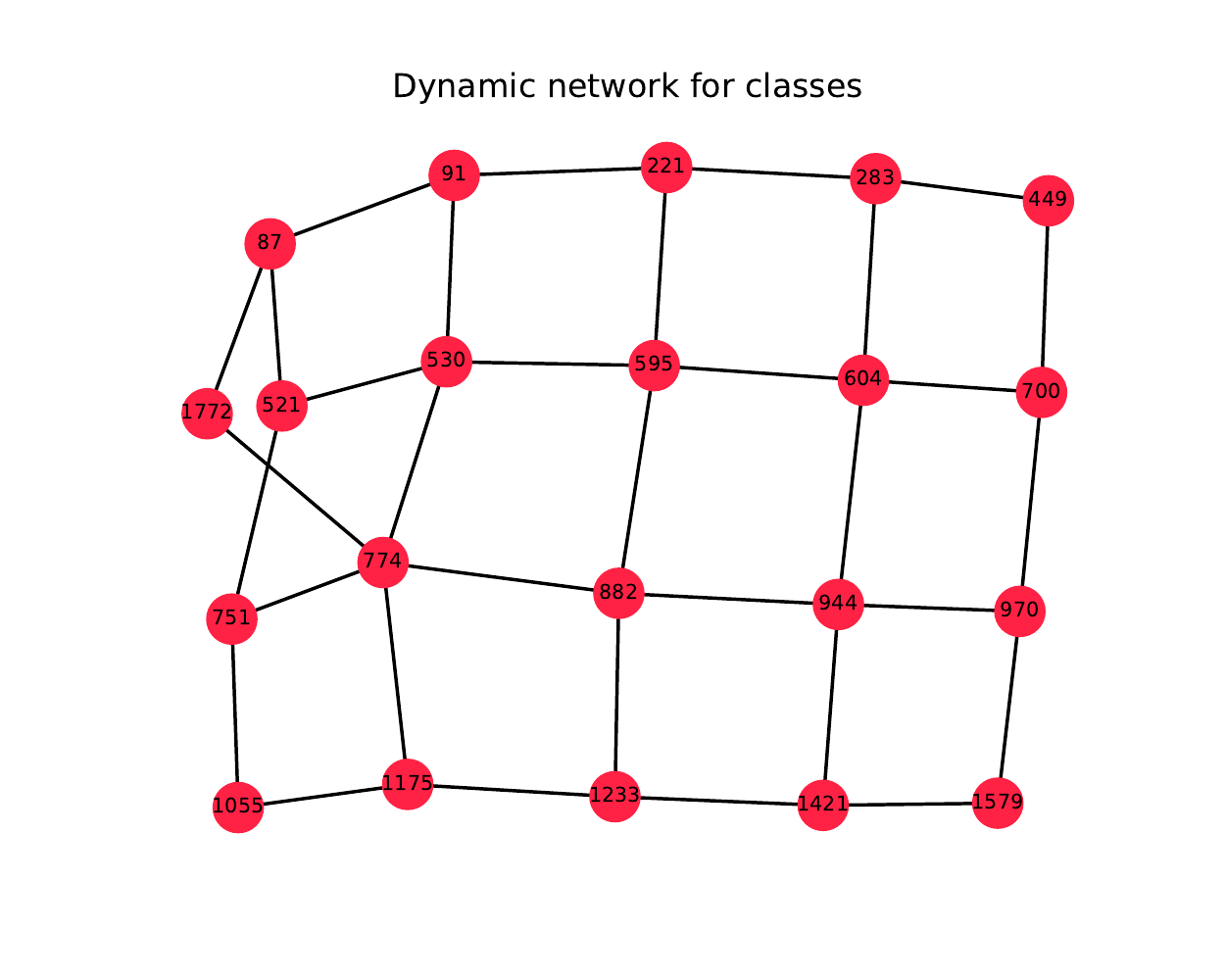}
    \caption{\textbf{Example of a dynamic network for classes in schools.} Students are geographically positioned in lines, with one teacher in charge.}
    \label{fig:schools_net}
\end{figure}

Figure \ref{fig:schools_net} shows a network for classes inside a school. The teacher is identified by the id $1772$.

\subsubsection{Network for hospitals}\label{sec:hospitals_net}

Networks for hospitals have, in addition to workers and visitors, admitted persons (hereby labelled guests), which stay in the facility for long periods of time. While these people are admitted, they get in contact with only a few hospital workers. The workers, on the other hand, get in contact with other workers, with some getting in contact with visitors as well. Visitors are yet another type of individuals which comprise those who seek help in the occasional sickness, as well as those only visiting admitted persons. The need to distinguish between three types of agent makes this type of network more complex than those introduced before. Another source of complexity is the fact that some workers are assigned to deal with a specific disease in a pandemic scenario, an attempt to contain the spread of the disease among workers. 

Contacts for visitors are adjusted by providing the average number of contacts among themselves, to the workers of the hospital, and to the admitted persons. Contacts among workers take into account the two classes of workers: typical workers and disease workers. The average number of contacts among typical workers, among disease workers, and between typical workers and disease workers must be provided. This last number is typically very small. Finally, the average number of contacts between admitted persons and disease workers is a key parameter, which can determine the spread of the disease in the hospital. 

\begin{figure}
    \centering
    \includegraphics{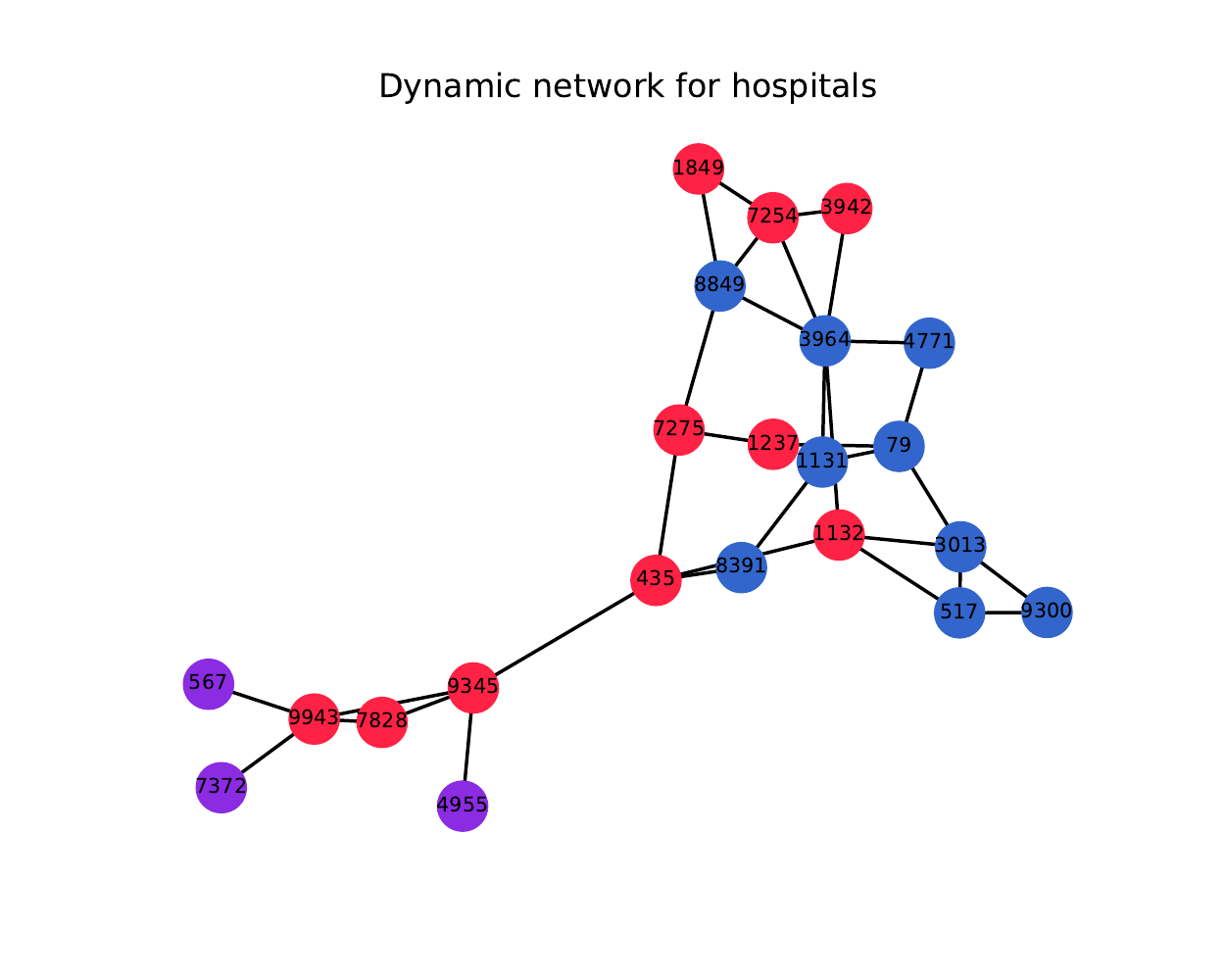}
    \caption{\textbf{Example of a dynamic network for hospitals.} Agents (circles) in red are workers, agents in blue are visitors, agents in purple have been admitted to the hospital and are placed in a COVID-19 dedicated ward. Numbers inside each circle represent the id of that agent.}
    \label{fig:hosp_net}
\end{figure}

Figure \ref{fig:hosp_net} exemplifies the network for hospitals. Agents with ids $567$, $7372$, and $4955$, in purple, have been admitted to the hospital. Agents $9943$, $7828$ $9345$ are disease workers, the only workers to get in contact to the admitted people. However, they may as well get in contact with other workers, in the figure, exemplified by the contact between agent $9345$ and agent $435$. This last worker may get in contact with another worker, exemplified by its connection to agent $1132$. Workers also get in contact with visitors, which can be seen by the connection between agent $435$ and agent $8391$. Finally, several visitors (in blue), also get in contact among themselves, as exemplified by the connection between agent $517$ and agent $9300$.

\subsection{Service parameters}
Every service is defined by the following parameters:
\begin{itemize}
    \item \verb|name|: Name of the service;
    \item \verb|number|: Number of instances of this service;
    \item \verb|days|: Days of the week the service is open to visitation;
    \item \verb|hours|: Hours of the day the service is open to visitation;
    \item \verb|visitation_period|: Mean period in days each agent visits this service;
    \item \verb|age_groups|: List of age groups that visits this service;
    \item \verb|workers|: Parameters to select workers of different types in this service. Each type can be configured with:
    \begin{itemize}
        \item \verb|name|: Name of the worker type;
        \item \verb|number|: Number of this worker type by instance;
        \item \verb|shifts|: Shifts available for this type of worker, workers are uniformly distributed between shifts;
        \item \verb|age_goups|: List of age groups that can be this worker type.
    \end{itemize}
    \item \verb|rooms|: Rooms to distribute workers, each room type is defined with a fixed number for each worker type, rooms of each type are created until there is no more workers of the required type available. At the end all remaining workers are placed in the "public" room that is the same room used by visitors;
    \item \verb|net_type| and \verb|net_par|: Type of network and it's parameters to be used to generate contacts in this service;
    \item \verb|inf_prob_weight|: Weight applied to the infection probability in this service (used to reduce the infection probability in outdoor services);
\end{itemize}

\subsection{Transportation layer}
\label{sec:comorbuss_transport}

A layer for transportation can be optionally activated in COMORBUSS. This layer intercepts all changes in placement during a simulation and places particles in a transport network for a set window of time. In this network the population is divided in two groups which are randomly assigned on initialization according to the percentage of the population which used the public transportation services:

\begin{itemize}
    \item \textbf{private transport:} this group is isolated during the time the particle is in the transport layer;
    \item \textbf{public transport:} this group is described by smaller non connected graphs, the size of which are defined by an input distribution with mean corresponding to the average number of users in each vehicle of the public transportation system of the modeled city. The contact networks in each vehicle employ an Erdös-Renyi generator with mean number of contacts taken as input from the user.
\end{itemize}

\noindent After the desired time in the transportation layer, the particles are placed at their destination. Without the transportation layer enabled, all particle movement is instantaneous.

\section{Epidemiological Model}
\subsection{Progression: stochastic compartmental model for the disease}
At any time, the state of an agent with respect to the modeled disease falls into one of the following compartments:
\begin{itemize}
    \item ($S$) Susceptible: the susceptible portion of individuals in the population. This portion of the population comprehends persons that had never had contact with the disease, and therefore they are susceptible to an infection.
    \item ($E$) Exposed: the exposed (or incubating) portion of individuals in the population. Individuals in this scenario have already had contact with the disease, but are still in the incubation stage of the disease. This means that they had been infected but are not infectious.
    \item ($I$) Infectious: the agent carries the virus and is infectious. The disease itself can manifest in different ways, which are sub-categorized as: 
    \begin{itemize}
        \item ($P_s$) Pre-symptomatic: particles have already become infectious, but they have not yet developed a viral load large enough to show symptoms.
        \item ($A_s$) Asymptomatic: this type of particle has passed activation of the disease, but will never show symptoms. However, they are still infectious.
        \item ($S_y$) Symptomatic with mild symptoms: the population in this portion are those infectious that show mild symptoms.
        \item ($S_s$) Symptomatic with severe symptoms: the population in this portion are those infectious that show severe symptoms.
    \end{itemize}
    \item ($R$) Recovered: the recovered particles have gone through all the stages of the disease, and that had overcome the disease.
    \item ($D$) Deceased: the deceased particles have gone through all the stages of the disease, developed severe symptoms and that have died due to it.
\end{itemize}

\begin{figure}[h]
    \centering
    \includegraphics[width=0.7\textwidth]{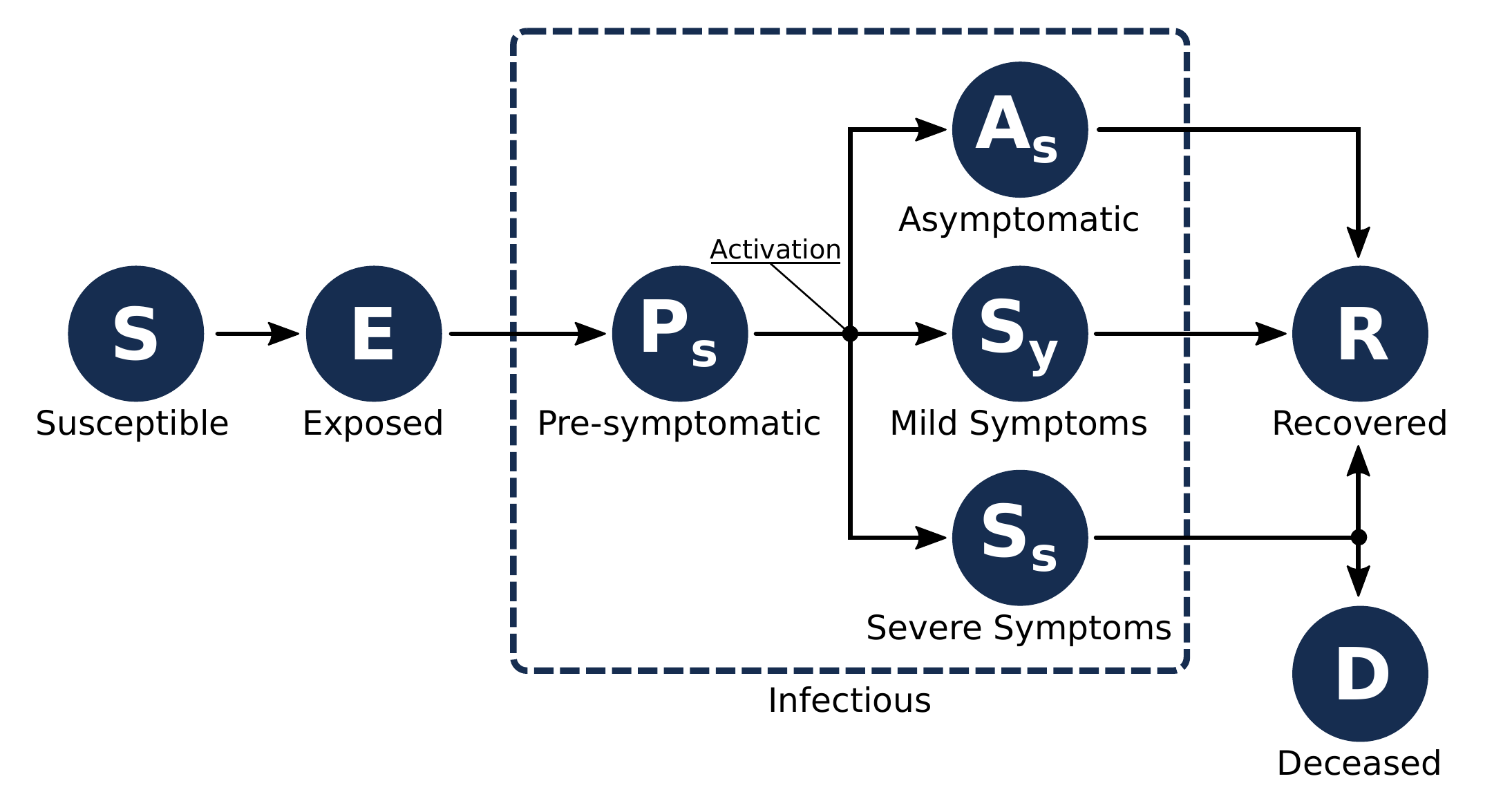}
    \caption{\textbf{Disease progression.} Diagram illustrating how agents can transition between states of the disease.}
    \label{fig:state_transition}
\end{figure}

Upon contracting the virus (becoming exposed), agents follow the transitions diagram depicted in Figure \ref{fig:state_transition}. The transition between states is stochastic, with transition probabilities as the inverse of the average period on which people remain in that compartment, according to \cite{kerr2020covasim}. The values and references for these periods can be found in the \href{https://gitlab.com/ggoedert/comorbuss/-/blob/paper_school_protocols/schools-paper-scripts/maragogi_base_conf.py}{maragogi\_base\_conf.py} file\footnote{\url{https://gitlab.com/ggoedert/comorbuss/-/blob/paper_school_protocols/schools-paper-scripts/maragogi\_base\_conf.py}}. After becoming infectious, an agent remains pre-symptomatic for two days, after which there is the activation event when it is decided if the disease will manifest as asymptomatic or symptomatic with mild or severe symptoms. 
Infectious agents recover with a probability estimated from the average duration of the infection; note that the duration of the disease in the case of severe symptoms is longer and such agents can instead convert to the Deceased compartment with a probability dependent of the age group of the agent, see details in Section \ref{sec:data_processing}.

\subsection{Transmission}
\subsubsection{Standard: contact through location-contextualized network}
The standard transmission model for COMORBUSS is based on the contact networks in a location. The first condition for transmission is that a susceptible agent be in contact with an infectious one. Provided such a meeting happens, the susceptible agent converts to the exposed compartment if a random number drawn from a uniform distribution (in unit interval) is less of equal than then probability of an infection occurring. This probability is the product between the infection probability which is produced by the calibration of the model, the susceptibility of the susceptible agent (which depends on its age group and vaccination status) and a correction parameter which accounts for contacts that do not last the entire time step of an hour.

If this random decision process results in a new infection, the compartment of the previously susceptible agent is rewritten to exposed, and the location, time and source of the infection are recorded in an infection tree.

\subsubsection{Specialized: aerosol transmission model in indoor locations}
In many closed locations where people are present for long periods of time, the main form of infection is not via direct contact with an infections person, but by inhaling infectious particles which are suspended in the air and accumulate over time. Naturally, the modeling of this process requires more detailed information on that location, as it depends on its volume and its rate of air exchange with the outside. These details are not readily available for most services, but for the purposes of this study we acquired the data of the two major schools in the modeled city. 

We developed a modified Wells-Riley model which takes into account different parameters for teachers and students. Not only do we consider that these two groups may have different masks, but teachers also release more infectious particles since they speak loudly and continuously. 

COMORBUSS naturally tracks all the agents in each classroom and identifies which are infectious. By solving a differential equation for the concentration of infectious particles over time, we compute the balance of absorbed and released particles by students and teachers. We then compute the dose absorbed by each agent in the last time step, and from this dose we evaluate the probability of that agent being infected. The modeling details are provided in Section \ref{sec:airborne}. Once an infection is produced, we randomly select a source among the infectious individuals in that room and store all the details of this new infection and the usual infection tree.

\subsection{Disease-defining Parameters}

\begin{itemize}
    \item \verb|inf_probability|: Probability of an infectious particle to pass the infection in an encounter.
    \item \verb|susceptibility|: Susceptibility of an particle (defined by age group), the final probability of an infection to occur in an encounter will be given by the \verb|inf_probability| of the source particle multiplied by the \verb|susceptibility| of the susceptible particle.
    \item \verb|inf_duration|: Mean duration of a asymptomatic or mild symptomatic infection (infectious state).
    \item \verb|inf_severe_duration|: Mean duration of a severe symptomatic infection (infected state).
    \item \verb|inf_incubation|: Mean duration of the incubation period (exposed state).
    \item \verb|inf_sympt_timeto|: Time between the the transition to the infectious state and the activation of symptoms.
    \item \verb|inf_prob_sympt|: Probability of a infected particles to develop symptoms (defined by age group).
    \item \verb|inf_severe_sympt_prob|: Probability of a infected particles to develop severe symptoms (defined by age group).
    \item \verb|inf_severe_death_prob|: Probability of a infected particles to die (defined by age group).
    \item \verb|inf0_perc|: Percentage of particles in each infection compartment at the start of the simulation. This is obtained from sampling of the distribution of cases in the initial step inferred in the calibration process.
    \item \verb|inf0_perc_symp|: Percentage of infected particles in each symptoms compartment at the start of the simulation.
\end{itemize}

\noindent To see the used values for these parameters see the \href{https://gitlab.com/ggoedert/comorbuss/-/blob/paper_school_protocols/schools-paper-scripts/maragogi_base_conf.py}{maragogi\_base\_conf.py} file\footnote{\url{https://gitlab.com/ggoedert/comorbuss/-/blob/paper_school_protocols/schools-paper-scripts/maragogi\_base\_conf.py}}.

\section{Interventions}
\subsection{Non Pharmacological Interventions}
COMORBUSS encompasses various Non Pharmaceutical Intervention (NPI) models, such as
\begin{itemize}
    \item individualized quarantine;
    \item generalized lockdown;
    \item social isolation (reduction in social activity);
    \item service based interventions;
    \item contact tracing;
    \item testing campaigns.
\end{itemize}
The scenarios simulated in this study are based on the first wave of infection in Maragogi, hence services related to tourism are closed. The other standard NPI adopted in the base scenario is social isolation based on telephonic triangulation data processed by \cite{Inloco_data} to provide the daily percentage of people which stayed home. This is modelled by randomly selecting at the beginning of each simulation day the desired number of agents and confining them to their homes for that day. 

Standard testing policy is the sorological testing of symptomatic agents. Diagnosed agents are quarantined at home if presenting mild symptoms or are hospitalized if their symptoms are severe. Quarantines and hospitalizations are lifted when agents leave the infectious compartment by recovering or dying.

\subsubsection{Intervention in School Dynamics}
We implemented and combined the following NPIs in the context of schools:
\begin{itemize}
    \item reduced workload: daily teaching hours are reduced from 4 to 2 hours;
    \item alternating groups: students are separated into two groups which attend the classroom in alternating days;
    \item masks: students and professors are supplied masks with given penetration factors;
    \item active monitoring: suspicious cases are monitored and intermittent closing is declared upon discovery of cases
    \begin{itemize}
        \item suspicious cases are students, professionals or their relatives which present symptoms;
        \item suspicious cases are tested and if the diagnose is positive the student is quarantined;
        \item the classroom associated to the quarantined person is closed for 14 days;
        \item if using alternating groups, only the group associated to the quarantined person is suspended;
        \item if more than one classroom is closed in the span of a week, the whole school is closed for a week.
    \end{itemize}
\end{itemize}

The effects of these NPIs and their combinations are the main results of this work.

\subsubsection{Aerosol transmission model: masks and air exchange}
Interventions in the aerosol model are made via parametrization of the Equation \eqref{eq:mass_balance2} in Section \ref{sec:airborne}. We introduce values for the penetration factor of masks $p_m^i$ used by students and professionals, and test the efficacy of different scenarios under various values of volume flow rate $\Lambda$ of air with the exterior. For reference, we highlight documented or recommended values of these parameters.

\subsection{Vaccination Model}
The vaccination model used in this study is a simple binary model for infection. Vaccinated agents can become immune (susceptibility 0) with a probability given by the vaccine effectiveness after a given period. If the vaccination of an agent does not lead to immunity, its susceptibility remains unchanged. In the present work we assumed a worst case scenario where vaccines were not widely available and were prioritized to the teachers and staff. We assume that these professionals are already immune upon the reopening of the schools.

\section{Data and Code availability}
\subsection{Distribution and Documentation}
COMORBUSS has a project webpage under the link \href{https://comorbuss.org/}{https://comorbuss.org}, where all developments, results and links are assembled.

The source code for COMORBUSS is available in the repository \href{https://gitlab.com/ggoedert/comorbuss}{https://gitlab.com/ggoedert/comorbuss} under the licende \href{https://www.gnu.org/licenses/agpl-3.0.en.html}{AGPLv3}. The version of the code together with all required input files and simulation scripts is available under the Tag Paper\_Maragogi\_Schools.

The full documentation of the COMORBUSS library is available via \href{https://docs.comorbuss.org/}{https://docs.comorbuss.org/} under the license \href{https://creativecommons.org/licenses/by-sa/4.0/}{CC BY-SA 4.0}.

\subsection{Computing language and Dependencies}
Here we specify all the versions of the computer libraries used for the present work. COMORBUSS is written in Python (version 3.7.7) and requires the following modules:
\begin{itemize}
    \item numpy v.1.18
    \item matplotlib v.3.1.3
    \item pandas v.1.0.5
    \item seaborn v.0.10.1
    \item h5py v.3.10
    \item h5dict v.0.2.2
    \item scipy v.1.5.0
    \item portion v.2.0.2
    \item networkx v.2.5.1
    \item tqdm v.4.46.0
    \item numba v.0.53.1
\end{itemize}

\chapter{Parameter Estimation and Calibration}
\label{sec:calibration_initialization}

In this section we describe how to use data gathered from Maragogi-AL city to estimate some main parameters of the model, such as those in the definition of services. We also specify the calibration procedure used to approximate the poorly estimated or unknown parameters whose variance influences SEIR curves the most.

In the sections to come, we first focus on the estimation of some key parameters in the definition of services. Section \ref{sec:househols_indoor_outdoor} gives reasoning to the choice of average number of contacts inside homes, and also to the relation between indoor and outdoor infection probability values. Sections \ref{sec:visit_perio} and \ref{sec:network_params} are intended to explain the estimation of the most relevant service parameters with respect to disease transmission: the visitation period and the network parameters. Finally, the subsequent two sections detail the calibration process and the sensitivity of results with respect to the population size, respectively.

\section{Estimation of parameters for household and indoor/outdoor environments}\label{sec:househols_indoor_outdoor}



Regarding the specific epidemic history of Maragogi-AL city, one notices that a large portion of the population stayed at home during the period we consider in this study. This fact is confirmed from the \textit{Inloco} geolocation data (currently under the name of \textit{Incognia} \cite{Inloco_data})\footnote{The company uses high resolution smartphone geolocation data to generate the social isolation index time series, see \cite{Inloco_data}. We must point, however, that due to geographic limitations, the regional cellphone signal is not captured with high quality, causing underestimation of the social isolation index.}. From the high level of social isolation in this period, we assume that the transmission rate at homes was higher in comparison to the one at other environments such as essential services. The household transmission rate, denoted $R_h$, is introduced in Cumei and co-authors \cite{Curmei_2020_05_23}, and is defined by the average number of new infections caused by an infected individual inside the its household. Given the intense social isolation in Maragogi-AL, we have used the largest value of $R_h$ estimated by \cite{Curmei_2020_05_23} as our reference. This leads to choosing the average number of $1$-hour contacts $c_{homes}$ inside a home so that the total number of new infections in houses during the period considered is about $70\%$ of the total. In our simulations, we have used $c_{homes}=0.7$. 

Transmission rates also vary considerably for indoor and outdoor environments\footnote{Outdoor environments in Maragogi-AL, for the sake of our study, include the environmental layer (see Section \ref{sec:env_net}) and street markets}. From the meta-analysis of Nishiura and co-authors \cite{Nishiura_2020_indoor_inf}, it is inferred that indoor environments increase $18.7$ times the probability of disease contagion with respect to outdoor ones. We consider this aspect into our simulations, multiplying the infection probability inside outdoor environments by a weight equals to $1/20$.



\section{Estimation of service's visitation period}\label{sec:visit_perio}

The visitation period of services is one of the parameters that influences the most the disease spreading at a given city, since it controls the influx of agents inside services. In Section \ref{sec:visitation_period} we define visitation period, and also how to estimate it. In this section we describe how we collected the data for the actual values of this parameter for each service in the city of Maragogi-AL.

\textbf{Street market}: street market opens at Saturdays, from 6:00 to 12:00. During the pandemics, an average of $3000$ people visited the street fair every day it opens, see Section~\ref{sec:maragogi_is_typical}. Assuming that all people from age groups 5 and above can visit the fair, the calculation for $v_{pc}$ is 
$$
v_{pc} = 7\frac{v_t}{v_w} = 7\frac{20884}{3000} \approx 48.73 ~\mathrm{days}. 
$$

\textbf{Hospitals}: the total number of hospital attendances in each unit of hospital (UPA and SAMU) from April, 29th to June, 28th 2020 was $3579$ and $304$, respectively. To estimate the actual number of people who visited the hospitals during this period of time, we must take into account other people accompanying these attendees. In order to do that, let us suppose that at least children from the first 3 age groups are accompanied by an adult, and that the same happens to elderly from the 13th age group and above. 

Assuming that everyone in Maragogi had the same number of contacts with the disease, we estimate from the susceptibility $p_{sus}$ and the probability of developing severe symptoms $p_{sev}$, what is the portion $po$ of attendees that brought another person with them to the hospital:
$$
po = \frac{\langle p_{sus} * p_{sev} ,pop_{ce}\rangle }{\langle p_{sus}*p_{sev},pop_{t}\rangle } \approx 0.2496,
$$
where $*$ is the point-wise multiplication of vectors, $\langle \cdot , \cdot \rangle$ is the inner product, $pop_{t}$ is the vector of all people from all age groups, and $pop_{ce}$ agrees with $pop_t$ for children or elderly, but has null entries otherwise. 

The attendees of UPA do not all come from Maragogi, but the hospital estimates that at least half of them do. Taking all this information into account, we estimate the visitation to the hospitals to be
$$
v_{pc} = (180-120)\frac{32702}{(0.5*3579+304)*1.2496} \approx 750 ~\mathrm{days}.
$$

\textbf{USFs}: the USFs open during the week only, but they receive much more people than hospitals. From the day 130 to 210 of 2020, they have attended a total of $7334$ people. Taking into account people that come accompanied, the visitation period for the USFs is
$$
v_{pc} = (210-130)\frac{32702}{7334*1.2496}\approx 285.5
$$

\textbf{Supermarkets}: to account for visitation routines in supermarkets, $5$ of the largest of its kind have been interviewed. The supermarkets ``Preço bom'' have reported $3000$ attendances every week, while supermarkets ``Supermar'', ``Mercado Nacional'' and ``Mercadinho Durare'' have reported an average of $550$ attendances weekly. As a result, the visitation period for this category of services in days is
$$
v_{pc} = 7\frac{20884}{3550} \approx 41.18,
$$
as long as all age groups from the fifth and above are considered consumers. 

\textbf{Markets}: given the big difference in the contact network for supermarkets and other kind of markets, we decided to separate them into two distinct types of service. For markets, which are more local and smaller in size, we gathered information from two representatives, namely markets ``Mini Carrefour'' and ``Mercadinho do Beto''. These two markets reported an average of $50$ visitors per week. We then assumed similar visitation for all other $37$ instances, which then allowed us to estimate the following visitation period, in days:
$$
v_{pc} = 7\frac{20884}{50*39} \approx 74.97.
$$

\textbf{Food Stores and Construction stores}: the other types of services that received people and that remained opened during the period considered were grouped into two categories: 
\begin{itemize}
    \item Food stores: all other types of services that sell specialized food, such as fruits and vegetables and beverages. This category also includes pharmacies. 
    \item Construction stores: all types of stores that sell maintenance equipment, such as those for civil engineering, household equipment, vehicle parts, etc.
\end{itemize}
These two types of stores are small and we assumed their visitation periods were twice and four times longer than the visitation period of markets, respectively. 

\section{Estimation of service's contact network parameters}\label{sec:network_params}

In this section we describe the data used in the contact network parameters for services of Maragoggi-AL, according to their definition given in Section \ref{sec:contact_networks}.

Contacts are a way to quantify the opportunity for disease spreading, if agents getting in contact have the proper compartmental state. In this work we have assumed a contact to have the following definition: ``two people at two meters or less away from each other for the duration of one hour''. 

A general procedure to quantify the network parameters, as described in the following sections, is
\begin{enumerate}
    \item estimate the average amount of time $t _{cont}$ people at two meters or less away from each other;
    \item quantify the number of contacts by taking into account $t _{cont}$ instead of one hour;
    \item derive a weight $w_{cont}$ to be multiplied by the parameter values, scaling contacts to the duration of one hour.
\end{enumerate}
As an example, $w_{cont}$ would be given by $w_{cont}=1/12$ if $t _{cont}$ is $5$ minutes. That would mean that a person having $12$ contacts of five minutes would equal having one of one hour. 


\textbf{Street Markets}: for street markets, we have used the model described in section \ref{sec:markets_net} with some few modifications:
\begin{itemize}
    \item All workers are cashiers, hereby denominated sellers;
    \item Visitors can get in contact with more than one cashier/seller.
\end{itemize}
Since there only two categories of agents inside street markets, the cashiers and the visitors, we need to approximate three parameters: the average number of contacts among sellers, the average number of contacts among visitors, and the average number of contacts between the two of them. 

Before obtaining values for the average number of contacts, we need to estimate its average length of time $t _{cont}$. We have done so using recordings of individuals collected by drones at one of the days that the street market opened. By following the routine of anonymous people inside the street market, we have calculated $t _{cont}$ to be $5$ minutes. We are now in position to estimate the average number of contacts:

\noindent\textbf{Contacts among sellers}: there were $185$ sellers in street markets during the time considered, being distributed along $120$ stands. $55$ of these stands were owned by one seller, and $65$ of them had two sellers as owners. We assume that the stands with two sellers were constantly in contact, and that an average of $3$ contacts of $5$ minutes happened among sellers of different stands each hour. As a result, the average number of contacts $\overline{c}_{sellers}$ between sellers per hour is
\begin{equation}
    \overline{c}_{sellers} \approx \frac{65*12+3*120}{120}\approx 6.
\end{equation}

\noindent\textbf{Contacts between sellers and visitors}: from frames collected by drones during the opening hours of the street market, we have estimated that about $300$ people stayed around stands every hour. We have also estimated that visits took $50$ to $60$ minutes in average. As a result, an average of $2$ visitors were found around $60$ stands, while $3$ visitors stayed constantly close to $60$ stands. In the worst case scenario, we have all groups of three people getting in contact with $2$ sellers, $5$ groups of two people getting in contact with $2$ sellers, and the remaining $55$ groups of two people getting in contact with $1$ seller. As a result, the maximum number of contacts of $5$ minutes that visitors have with sellers $c_{vis\rightarrow sell}^{max}$ is
\begin{equation}
    c_{vis\rightarrow sell}^{max} \approx 12\frac{60 * 3 * 2 + 5 * 2 * 2 + 55 * 2 * 1}{458}\approx 12.8,
\end{equation}
where $458$ and  is the average number of visitors present in the street market per hour, considering visits of $55$ minutes. Analogously, in the best case scenario, $55$ groups of three visitors are found around stands of one seller, $5$ groups of three visitors stay close to stands with two sellers, and the remaining $60$ groups of two visitors get in contact with two sellers per hour. In this case we have that the minimum number of contacts of $5$ minutes that visitors have with sellers $c_{vis\rightarrow sell}^{min}$ is
\begin{equation}
    c_{vis\rightarrow sell}^{min} \approx 12\frac{55 * 3 * 1 + 5 * 3 * 2 + 60 * 2 * 1}{458}\approx 8.3.
\end{equation}
Taking the average between the worst and best case scenarios, we have that the average number of contacts $\overline{c}_{vis\rightarrow sell}$ that visitors have with sellers is
\begin{equation}
    \overline{c}_{vis\rightarrow sell}\approx \frac{c_{vis\rightarrow sell}^{max}+c_{vis\rightarrow sell}^{min}}{2}\approx 10.6.
\end{equation}

\noindent\textbf{Contacts among visitors}: from the data acquired through drone observations, we know that about $3000$ people attend the street market when it opens. In addition, since visitors take about $55$ minutes doing shopping, we also know that about $458$ people visit the street market per hour. Out of those people, some are doing shopping and some are assumed to be randomly walking in the transit area of the fair. The remaining few formed clusters of people socializing. From our observations, the average number and amount of people in each cluster is
\begin{itemize}
    \item $1$ cluster of $5$ people: $5\times 4/2=10$ $1$-hour contacts;
    \item $3$ clusters of $4$ people: $3(4\times 3)/2=18$ $1$-hour contacts;
    \item $11$ clusters of $3$ people: $11(3\times 2)/2=33$ $1$-hour contacts;
    \item $23$ clusters of $2$ people: $23(2\times 1)/2=23$ $1$-hour contacts.
\end{itemize}
The number of $1$-hour contacts happening in stands where $2$ visitors could be found is $60(2\times 1/2)=60$, and the total number of contacts happening in stands where $3$ visitors could be found is $60(3\times 2/2)=180$. Finally, for the random walking of the remaining $62$ people, we assume an infectious radius of $2$ meters. Whenever agents are found at less than this distance away from each other, we count a contact. However, since the transit area of the fair is approximately $1607\,m^3$, it makes sense to actually consider this type of contact only for a number of people larger than $1607/4 \pi \approx 128$. As a result, only the above two types of contacts are considered, and therefore
\begin{equation}
    \overline{c}_{visitors}\approx 12\frac{324}{458}\approx 8.5,
\end{equation}
where $\overline{c}_{visitors}$ is the average number of $5$-minute contacts happening among visitors per hour. 

\textbf{Hospitals and other health facilities}: for the networks of hospitals and other health facilities (that do not treat diagnosed individuals), data has been acquired from the city hall. For this type of services, the contact network employed is that introduced in Section \ref{sec:hospitals_net}, for which we have the following parameters:
\begin{itemize}
    \item $p_{dis. w.}$: percentage of hospital workers that deal specifically with the pandemics disease in question;
    \item $\overline{c}_{workers}$: average number of $1$-hour contacts among non-disease workers;
    \item $\overline{c}_{dis. w.}$: average number of $1$-hour contacts among disease workers;
    \item $\overline{c}_{dis. w. \rightarrow w.}$: average number of $1$-hour contacts from disease workers to non-disease workers;
    \item $\overline{c}_{visitors}$: average number of $1$-hour contacts among visitors;
    \item $\overline{c}_{vis.\rightarrow w.}$: average number of $1$-hour contacts from visitors to non-disease workers.
    \item $\overline{c}_{guests\rightarrow dis. w.}$: average number of $1$-hour contacts from guests (admitted persons) to visitors.
\end{itemize}
According to data from city Hall, the values of the parameters above for campaign hospitals are: $p_{dis. w.}=0.19$, $\overline{c}_{workers}=2$, $\overline{c}_{dis. w.}=2.9$, $\overline{c}_{dis. w. \rightarrow w.}=0.2$, $\overline{c}_{visitors}=2$, $\overline{c}_{vis.\rightarrow w.}=1$, and $\overline{c}_{guests\rightarrow dis. w.}=0.15$. For other types of health facilities, the difference is that there are no disease workers dealing specifically with admitted persons. Therefore, the non-zero values for the above parameters are: $\overline{c}_{workers}=2$, $\overline{c}_{visitors}=2$, and  $\overline{c}_{vis.\rightarrow w.}=1$.

\textbf{Markets, supermarkets, food stores and construction stores}: the data used in the network parameters of these services has also been collected from city hall estimates. The type of network used here is that presented in Section \ref{sec:markets_net}, whose main parameters are
\begin{itemize}
    \item $\overline{c}_{workers}$: average number of $5$-minute contacts among workers;
    \item $\overline{c}_{visitors}$: average number of $5$-minute contacts among visitors;
    \item $\overline{c}_{vis. \rightarrow w.}$: average number of $5$-minute contacts from visitors to workers;
    \item $p_{cashier}$: percentage of workers that are cashiers.
\end{itemize}
For supermarkets, the parameters above have been estimated to be equal to: $\overline{c}_{workers}=3$, $\overline{c}_{visitors}=3$, $\overline{c}_{vis. \rightarrow w.}=0.25$, and $\overline{c}_{vis. \rightarrow w.}=0.22$. For the remaining services, those parameters are: $\overline{c}_{workers}=3$, $\overline{c}_{visitors}=3$, $\overline{c}_{vis. \rightarrow w.}=0.25$, and $\overline{c}_{vis. \rightarrow w.}=0.29$. 

\textbf{City hall:} for the city hall we have used the standard Erdos-Renyi model, where the probability
$p_{ER}$ of an edge being added is given by $p_{ER} = c_{avrg} /(N − 1)$. Here $c_{avrg}$ is the average number of contacts and $N$ is the total number of nodes in the graph. However, the value of $c_{avrg}$ has been calibrated along with the probability of infection due to lack of information, and also due to the number of workers being high in comparison with the remaining services ($355$ against $916$). The calibrated value ended up being $c_{avrg}=0.61$ $1$-hour contacts per hour. See section \ref{sec:optimization}.

\textbf{Environmental layer}: for the environmental layer, which comprises agents out of home who are not in either of the other services, we have used the network model explained in Section \ref{sec:env_net}. The only customizable parameter in this network is the urban area, which in the case of the Maragogi-AL city is: $7.654\,km^2$. 

\textbf{Schools}: for schools we have used an entirely different transmission model which is not based on physical contacts, but rather aerosol transmission of infectious particles. See Section \ref{sec:airborne} for details.

\section{The optimization program}\label{sec:optimization}

After calculating or estimating directly all parameters we consider relevant for simulating a community, we are left with the task of approximating the infection probability $p \in [0,1]$ and the mean number of 1-hour contacts $c \in \mathbb{R}_+$ between workers in the City Hall. Since the infection probability is a very behavior-dependent parameter, it is difficult to approximate it directly. Similarly, the contact network inside the City Hall could not be assumed from a-priori information. To find parameter values that best fit disease data, we use an optimization program to estimate these parameters for the period considered. In this section we describe the methodology used in this optimization step, and we also provide numerical evidence that it is in fact well suited for the task. 

Let $\hat{x}$ be a candidate for approximating $x=(p,c)$. We evaluate how close $\hat{x}$ is of $x$ by using the Wasserstein distance as a goodness-of-fitness as following:
\begin{itemize}
    \item Let $\mathcal{D}$ be a set of time markers (in our case, days), and let $\mathcal{E}$ be defined by
    \begin{equation}
        \mathcal{E} = \{ (s,e,i,r) \in [0,1]^4 : s+e+i+r = 1 \}.
    \end{equation}
    Then $\mathcal{X} = \mathcal{D}\times\mathcal{E}$ contains any SEIR curve evaluated at times in $\mathcal{D}$. We define $\{X_i\}_{i=1}^{n}$ as the possible SEIR trajectories generated by our SEIR curve reconstruction (see SI \dataprocessing) and $\{X^y_j\}_{j=1}^m$ be $m$ i.i.d. trajectories generated by our model when we use the parameters in $(y_1,y_2)=y \in [0,1]\times \mathbb{R}_+$ as the infection probability $p=y_1$ and the mean number of 1-hour contacts $c=y_2$. We also set $\hat{\nu}$ as the empirical measure given by $\{X_i\}_{i=1}^n$ and $\hat{\mu}^y$ to be the empirical measure obtained from $\{X^y_j\}_{j=1}^m$. 
    
    \item The $L_1$-Wasserstein distance between $\hat{\nu}$ and $\hat{\mu}^y$ is given by
    \begin{equation}
        W_1(\hat{\nu}, \hat{\mu}^y) = \inf_{\gamma \in \Gamma( \hat{\nu}, \hat{\mu}^y )} \int_{\mathcal{X} \times \mathcal{X}} || X - Y ||_1 d\gamma(X, Y),
        \label{eq:wassersteindist}
    \end{equation}
    where $\Gamma( \hat{\nu}, \hat{\mu}^y )$ is the set of all couplings of $\hat{\nu}$ and $\hat{\mu}^y$. In our case, with empirical measures having finite support, one can evaluate Equation \ref{eq:wassersteindist} using linear programming, so we employ the solution implemented on the \textit{Python Optimal Transport package} \cite{flamary2021pot}.
    \item We evaluate
    \begin{equation}
        \hat{x} = \textit{argmin}_{y} W_1(\hat{\nu}, \hat{\mu}^y)
    \end{equation}
    in three steps: 1) using a population size of $N=10000$, we perform a grid search to narrow down the search space; 2) still using $N=10000$, the search for an optima in the narrowed space is performed by applying Nelder-Mead algorithm; 3) we apply Nelder-Mead with $N=32702$ (the real population size). The first two steps using a small population size reduces the computational cost of the process, and the last one corrects any artefact produced by the rescaling to $N=10000$.
\end{itemize}

\begin{figure}
\centering
\subfloat[Susceptible]{{\includegraphics[scale=0.58]{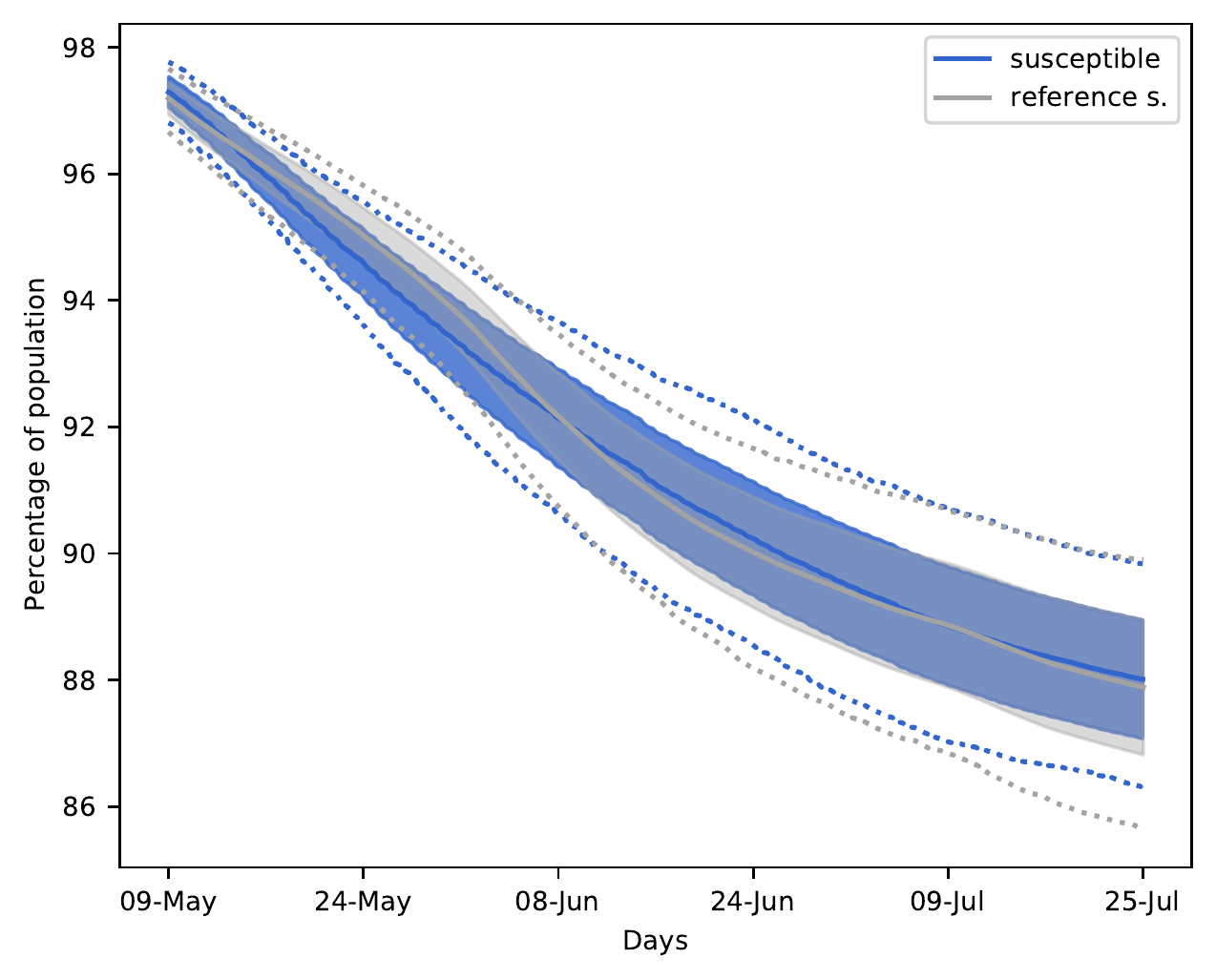}}}%
\subfloat[Recovered]{{\includegraphics[scale=0.58]{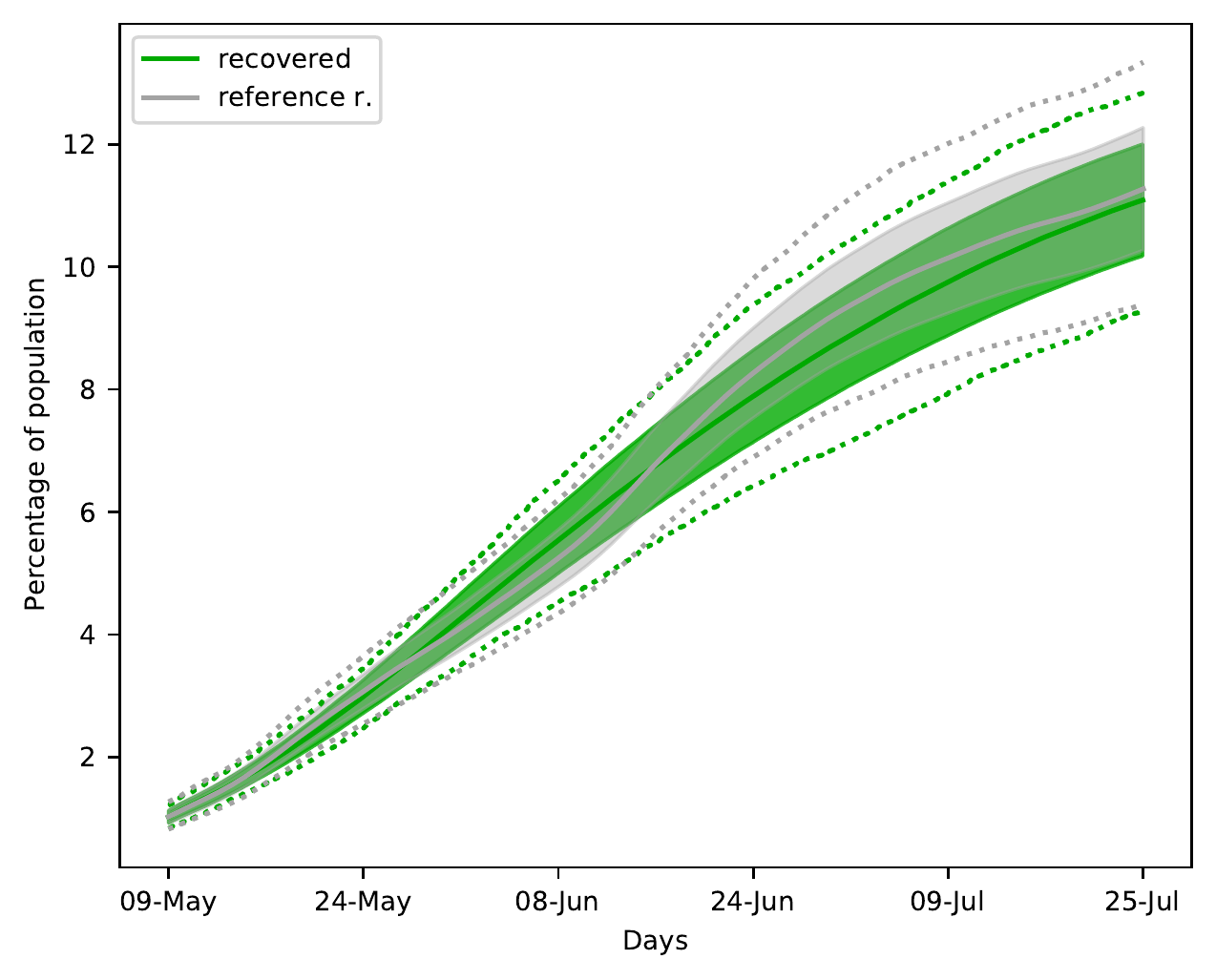}}}
\caption{\textbf{Reference susceptible and infectious compartmental distribution curves for the city of Maragogi-AL in comparison to their calibrated versions calculated from COMORBUSS.} The recovered curves also include the deceased compartment. The solid curves represent the mean over $\nseeds$ samples, the dotted curves limit a $95\%$ percentile of the distribution, and the colored clear region is bounded by two shifted mean curves. These shifted curves are obtained by summing and subtracting the point-wise standard deviation over the $\nseeds$ samples.}\label{fig:caliSRdist}
\end{figure}

\begin{figure}
\centering
\subfloat[Exposed]{{\includegraphics[scale=0.58]{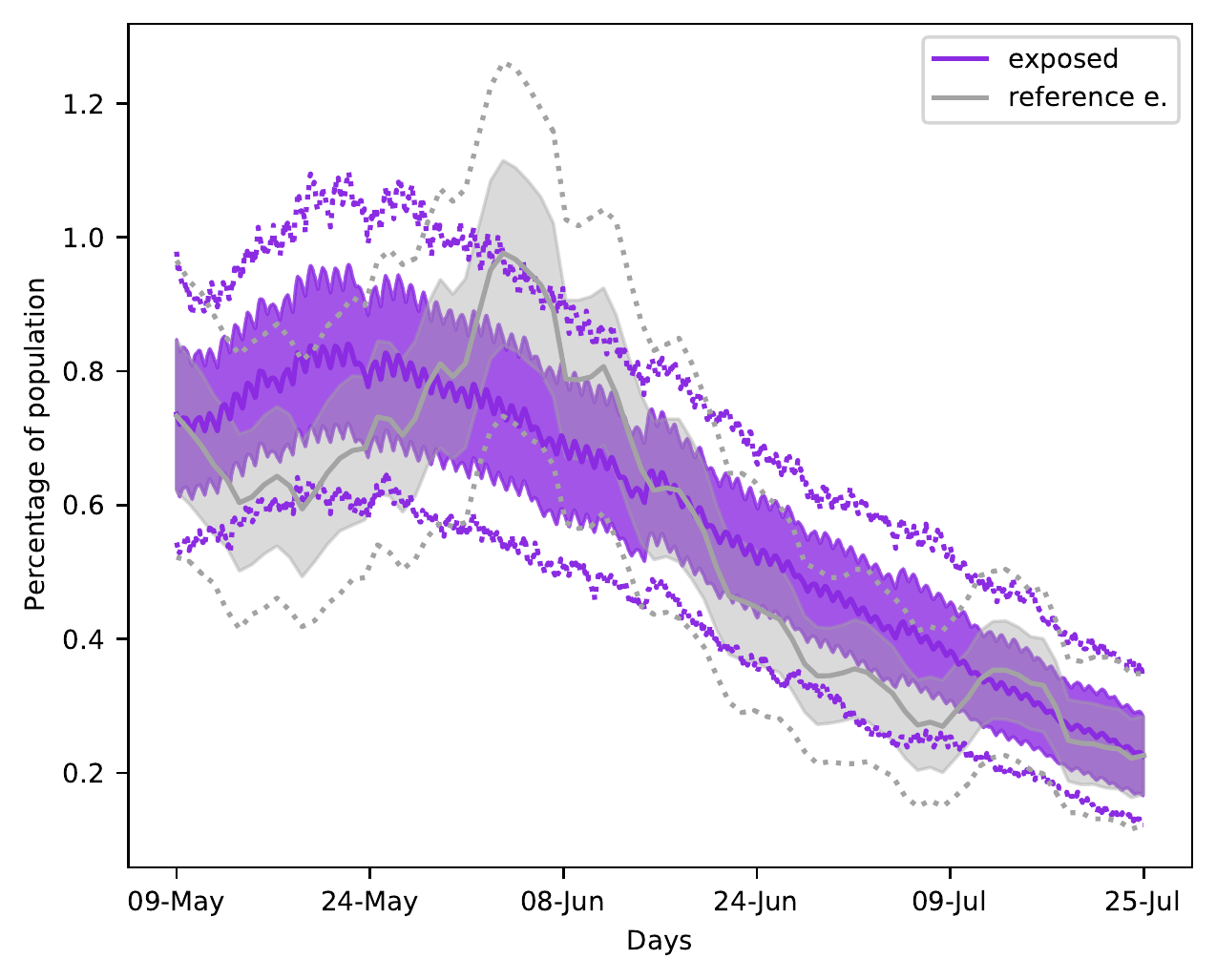}}}%
\subfloat[Infectious]{{\includegraphics[scale=0.58]{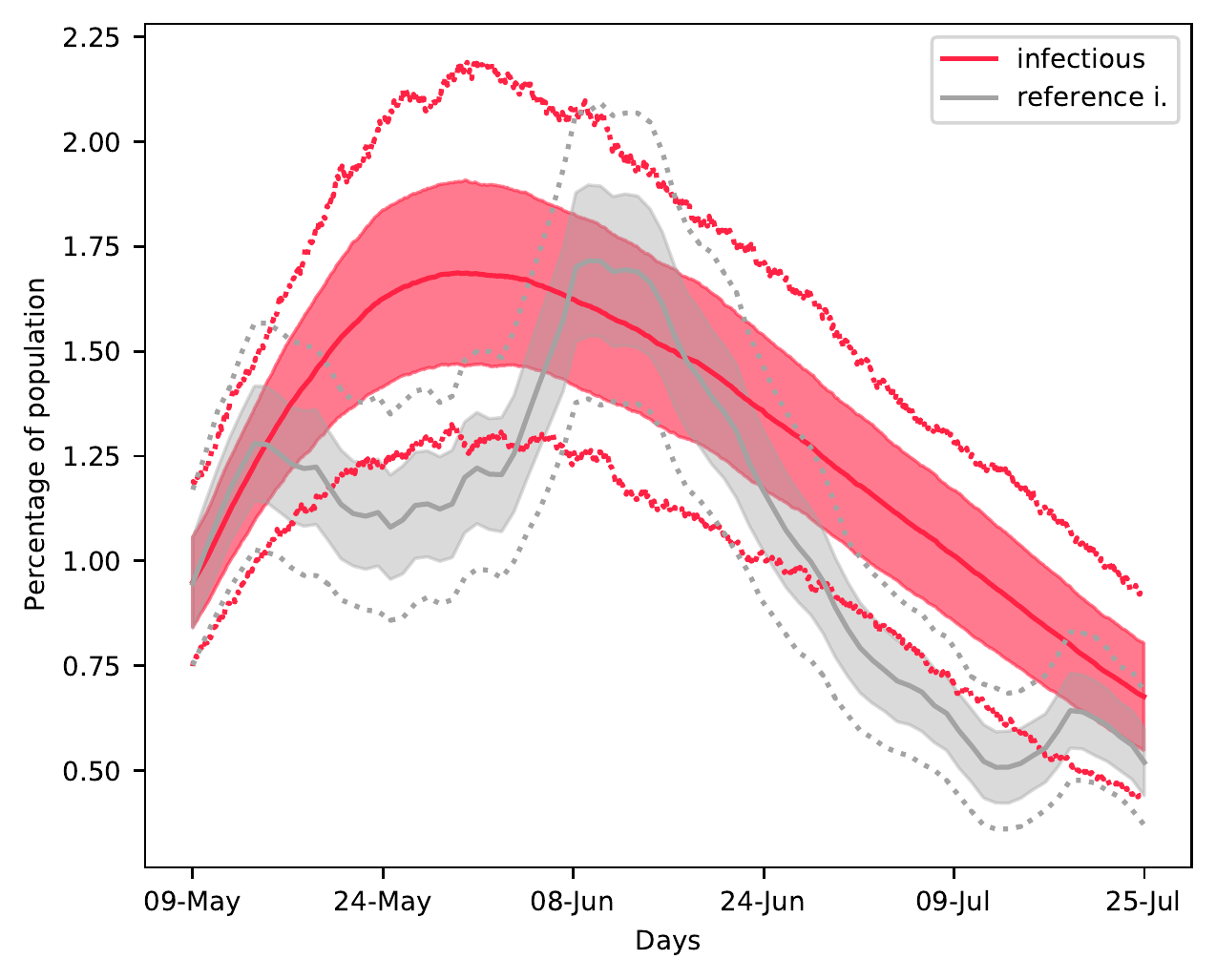}}}
\caption{\textbf{Reference exposed and infectious compartmental distribution curves for the city of Maragogi-AL in comparison to their calibrated versions calculated from COMORBUSS.} The solid curves represent the mean over $\nseeds$ samples, the dotted curves limit a $95\%$ percentile of the distribution, and the colored clear region is bounded by two shifted mean curves. These shifted curves are obtained by summing and subtracting the point-wise standard deviation over the $\nseeds$ samples.}\label{fig:caliEIdist}
\end{figure}

In practice, we use $n=m=\nseeds$ and $\mathcal{D}$ as the days between May 9th and July 25th 2020. 
The calibration procedure just described generates the following approximations for $(p,c)$: $p\approx 0.1356$ and $c\approx 0.6116$. The $L_1$-Wasserstein distance between the approximated optima and the reference curves is then given by $W_1(\hat{\nu}, \hat{\mu}^{(0.1356,0.6116)})=9.3\times 10^{-3}$. The resulting SEIR curves are compared to the reference curves in Figures \ref{fig:caliSRdist},\ref{fig:caliEIdist}. We notice a very good fit, specially for the susceptible and recovered compartments. These compartments are, in fact, usually the ones obtained with the highest accuracy for the reference curves. 

\textbf{The quality of $\hat{x}$.} The Wasserstein distance is a widely used goodness-of-fit measure \cite{sommerfeld2016inference, arjovsky2017wasserstein} for determining how close are two distributions. It has well-known concentration bounds when the measures are empirically approximated, which is exactly our case (see \cite{dedecker2019behavior, arjovsky2017wasserstein}). A good indicator of quality for the estimate $\hat{x}$ is how closely can one recover a calibration parameter when the input SEIR data is generated by COMORBUSS  by using a given value for this parameter. We check this property experimentally by using the infection probability as the aforementioned calibration parameter. The experimental protocol is as follows:
\begin{itemize}
    \item Let $p \in [0,1]$ be fixed and $S_1, \cdots, S_{50}$ be $50$ disjoint sets of $\nseeds$ seeds each (we took $S_1 = \{1001, \cdots, 1\nseeds\}$, $S_2 = \{1\nseeds +1, \cdots, 1000+2\cdot \nseeds\}$, etc.);
    \item We set in COMORBUSS the infection probability as $p=0.15$, the mean number of 1-hour contacts in the City Hall as $c=0.3$, and the population size as the full value $N=32702$, and we run simulations using the seeds from $S_i$, $i=1,\cdots , 50$. This procedure generates $50$ empirical measures $\hat{\nu}^{(0.15,0.3)}_i$, $i=1,\cdots , 50$;
    \item For each $\hat{\nu}^{(0.15,0.3)}_i$, $i=1,\cdots , 50$, we solve
    \begin{equation}
        \hat{x}_i = \textit{argmin}_{y\in [0,1]} W_1(\hat{\nu}^{(0.15,0.3)}_i, \hat{\mu}^y).
    \end{equation}
    To simplify the procedure and to reduce computational cost, we fix $y_2$, the second coordinate of $y$ as $y_2=0.3$. That is, we effectively only calibrate for the infection probability in this test. Nevertheless, this showcase the effectiveness of the calibration procedure proposed. 
\end{itemize}

After trying to recover $p=0.15$ as the infection probability using the procedure just described, we obtain the following approximation $\hat{p}$ for $p$: $\hat{p}=0.147\pm 0.0008$. We notice that the approximation for $p$ is very close to the original value we attempted to recover. This simulation asserts not only that the optimization program is good for approaching the real observed value for $(p,c)$, according to the input data, but also that the scaling made in COMORBUSS for the population size is effective (see section below). 

\section{Remarks about the population size}

As mentioned before, the most critical parameter to controlling computational time is the population size $N$ (see \COMORBUSS). As a result, understanding the impact of this parameter with respect to changes in the results is essential.

The sensitivity analysis on the population size $N$ usually focus on how the distribution of the final epidemic size (i.e., the distribution of the total number of cases after the epidemic ends) evolves with $N$. The dependence of a classical stochastic compartmental SEIR model with respect to $N$ has been analysed in \cite{greenwood2009stochastic, bibbona2017strong}.  In \cite{greenwood2009stochastic}, the authors provide experimental evidence that although the aforementioned distributions converge as $N$ grows, their convergence is slow. This fact is verified by noticing that even with $N$ in the order of $10^4$, one can still spot significant differences as $N$ grows.

We designed a similar experiment for our model. Unlike the compartmental model, our stochastic agent-based model constructs an entire city and assigns individuals to networks (e.g. family structures, schools networks, services networks). So, as discussed in \COMORBUSS, approximating real populations using values of $N$ distinct from the real population size may incur in rescaling errors. By looking at the total individuals assigned to each relevant social activity modeled in the city of Maragogi-AL, we determined that the minimum population size necessary for keeping at least one individual at each social role is $N=1000$. 

\begin{figure}[h!]
    \centering
    \includegraphics[width=.9\textwidth]{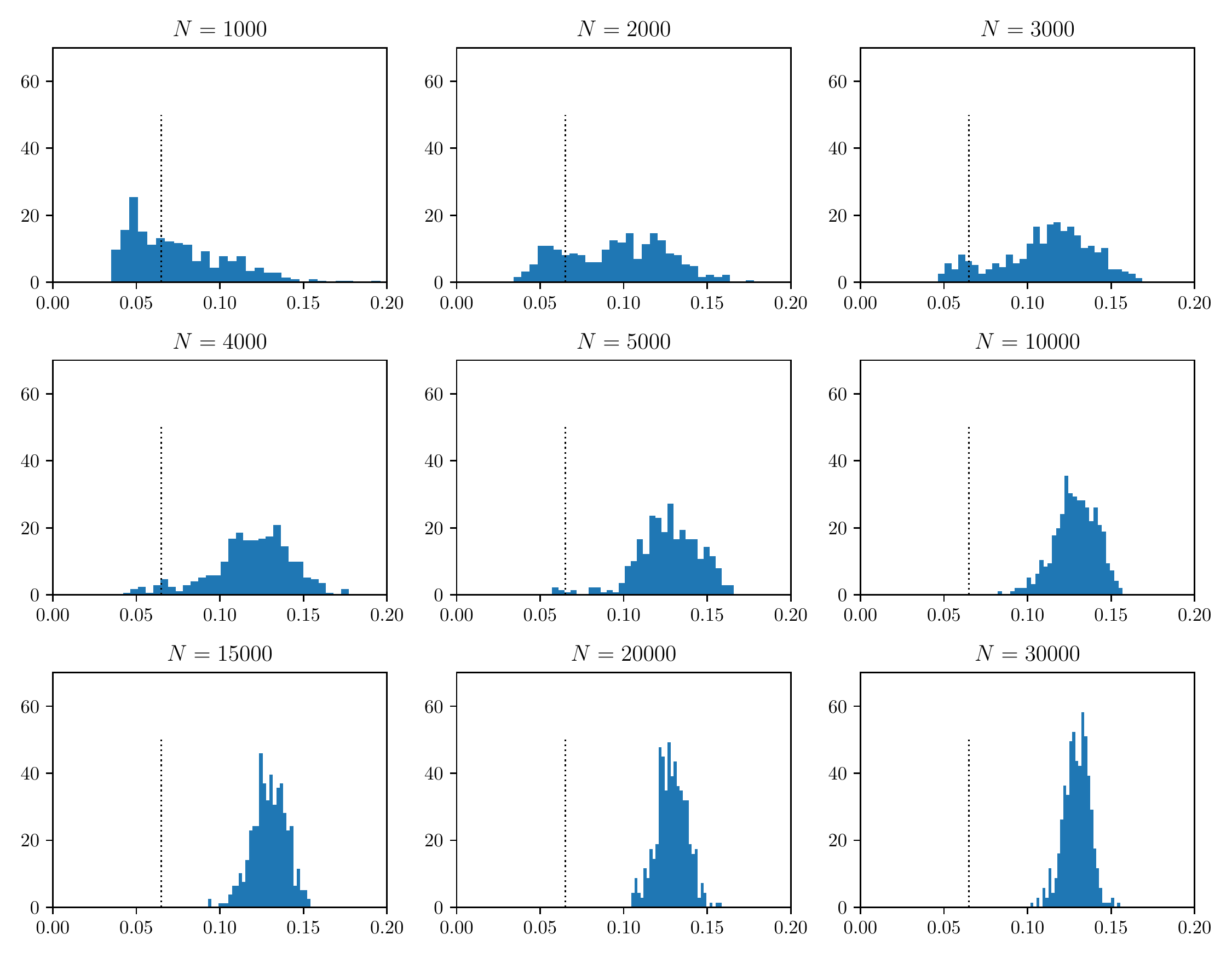}
    \caption{\textbf{Histograms of the final epidemic size for different values of $N$.} The $y$-values are normalized so that the histograms represent a distribution. For low values of $N$ the histograms are shifted towards the left side of the vertical dotted line, while for high values of $N$ the tendency flips to the right hand side of the line. The variance decays as $N$ grows, but the shape of the distribution still changes even for high values of $N$. Low values of $N$ also show evidence of bi-modal behavior.}
    \label{fig:finalepsize}
\end{figure}

To test how the final epidemic size changes with respect to $N$, we evaluate the results obtained from COMORBUSS by setting $N\in \{1000,2000,3000,4000,5000,10000,15000,20000,30000\}$. We make $\nseeds$ simulations for each value of $N$, and each simulation is kept running until the sum of exposed and infectious individuals become zero. After that, we evaluate the percentage of the population that was infected, calling it the final epidemic size. Results are shown in Figure \ref{fig:finalepsize}. 

The outcome of our tests, displayed in Figure \ref{fig:finalepsize}, agrees with the results exposed in \cite{greenwood2009stochastic}. We understand these results from a probabilistic perception. For small population sizes, statistical fluctuations are more significant, since probabilistic events such as spreading the disease or recovering from it occur less frequently. This can lead to rapid decay on epidemic measures in more realizations of the community, causing even bi-modal distributions for the final epidemic size (see Figure \ref{fig:finalepsize}). On the other hand, for large population sizes, the number of agents is prone to sustain the epidemic for a longer period of time. This is due to having a larger number of probabilistic events, which smooth out probabilistic fluctuations. This behavior helps shifting the distribution of the final epidemic size towards its right-sided mode (the process is clearly seen in Figure \ref{fig:finalepsize}, where the histograms tend to the right hand side of the vertical dotted line as $N$ increases). 

In \cite{greenwood2009stochastic} the authors point out that the final epidemic size distributions display a bi-modal behavior with two peaks. Our simulations also give evidence of the bi-modal structure, especially for small populations (see Figure \ref{fig:finalepsize}). This shows that COMORBUSS is capable of incorporating classical properties of stochastic compartmental models.

\begin{figure}[h!]
    \centering
    \includegraphics[width=.9\textwidth]{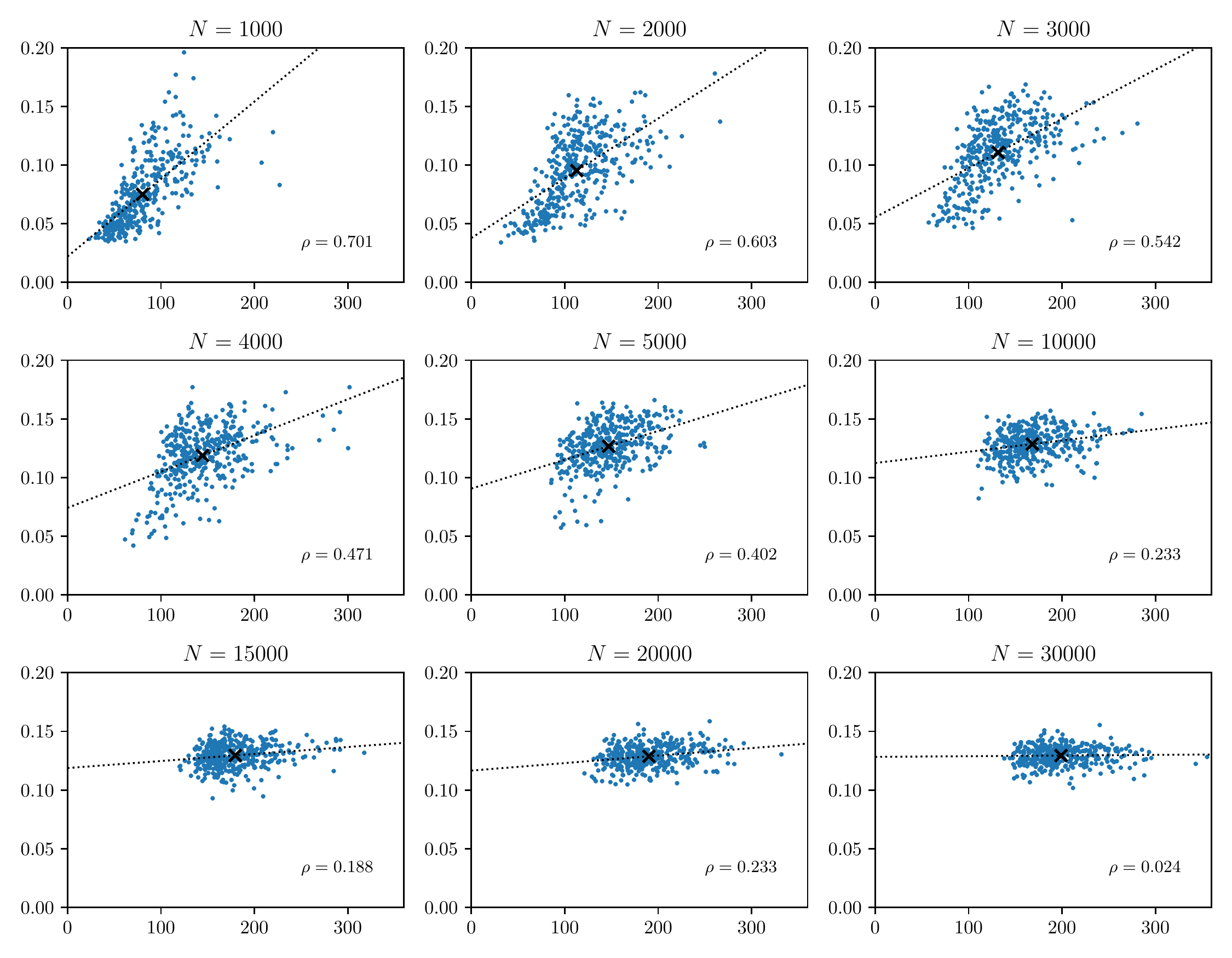}
    \caption{\textbf{Final epidemic size ($y$-axis in $\%$) vs number of days until the epidemic ends ($x$-axis in days) for different values of $N$.}The initial condition is $(S,E,I,R) = (.971,.007,.01, .012)$ for all realizations of the community. The \textbf{X} marker inside the clouds is the average over all points. The dotted line is a linear regression on the data, and $\rho$ is the correlation between both variables (epidemic size and its total duration).}
    \label{fig:dayvsprev}
\end{figure}

Figure \ref{fig:dayvsprev} helps to summarize how $N$ affects the model's behavior, which we can outline as two regimes:
\begin{itemize}
    \item \textit{Low values} ($N$ \textit{of order} $10^3$). Here, the epidemic has a more unpredictable behavior (the clouds are less concentrated) and it finishes sooner without infecting a large number of people (the clouds in the figure are shifted southwest). In fact, the average final epidemic size and average final day for $N=1000$ were $7.5\%$ and $80.3$ days, respectively. For $N=30000$ they were $12.9\%$ and $199.0$ days, respectively;
    
    \item \textit{High values} ($N$ \textit{of order} $10^4$). Here, the epidemic has a more predictable behavior (the clouds in the figure are concentrated) and also a longer duration. While for low values of $N$ a longer duration is associated with higher epidemic sizes, this behavior is softened by high values of $N$: the correlation $\rho$ between both variables decreases as $N$ grows.
\end{itemize}

These results point out that one must avoid to approximate population size of order $10^4$ using population sizes of order $10^3$ whenever possible. Approximations between same magnitudes are possible, as population sizes of $10000$, $20000$ and $30000$ display average final epidemic sizes of $12.85\%$, $12.87\%$ and $12.94\%$, respectively. Other variables are not so robust with respect to changes in the population size. For instance, the average total duration of the pandemic for a population size equal to $10000$, $20000$ and $30000$ were $168.2$, $190.1$ and $199.0$ days, respectively.

As a rule of thumb we choose to approximate the population of Maragogi-AL (32702 individuals) using $N=10000$ on the most computationally expensive and repetitive routines, such as the calibration process described above. For less expensive routines, such as those comparing different opening scenarios for schools, we make no approximation ($N=32702$). 

\chapter{Airborne Transmission model}
\label{sec:airborne}

\subsection{Aerosol-based model for infections in a closed environment}
\label{sec:aerosol_model}

\subsubsection{Relevant length and time scales for aerosol particles}
\label{sec:scale}

Aerosol particles carrying pathogens are expelled by infected individuals in a range of radii varying from $0.1~\mu m$ to $1 ~mm$. The majority of these particles lie in the submicron scale and the droplets size distribution depends on the breathing activity, varying from $0.1 ~\mu m$ to $5.0 ~\mu m$ with a peak around $0.5 ~\mu m$ \cite{MORAWSKA_2009_droplets_size}.

The pathogens being carried by airborne droplets have a typical lifetime inside the enclosed space, so we consider the pathogen concentration damping rate $\lambda_c$. This rate depends on the radius $r$ of airborne droplets \cite{Bazant_PNAS,Bazant_2021_co2}, and it encompasses four distinct mechanisms
\begin{align}\label{eq:rates}
    \lambda_c(r) = \lambda_a + \lambda_f(r) + \lambda_s(r) + \lambda_v(r),
\end{align}
where $\lambda_a$ accounts for outdoor air exchange rate, $\lambda_f$ is the room filtration rate (filtration due to mechanical ventilation or people breathing in the room and absorbing infectious airborne particles), $\lambda_s$ is the net sedimentation rate, and $\lambda_v$ stands for the deactivation rate of the aerosolized pathogen (which depends on humidity and droplet size). 

Although the definition of air quality inside enclosed space varies over international standards \cite{KHOVALYG2020109819}, ASHRAE (American Society of Heating, Refrigerating and Air-Conditioning Engineers) described in the technical notes\footnote{ASHRAE 62.1 --- Ventilation for Acceptable Indoor Air Quality. } that the minimum recommended outdoor air exchange rate depends on the environment. Namely, for American homes $\lambda_a = 0.35~h^{-1}$ while for classrooms of 5 - 9 age children $\lambda_a = 0.8~h^{-1}$. Those are the minimal recommended values and as we will see correspond to the largest order of magnitude among all other terms in Equation \eqref{eq:rates}. 

For most air-conditioning systems in Brazil, a filtration system is absent and not coupled to the mechanical ventilation. However, in our model we assume aerosol consumption arises from people breathing in the classroom and filtrating air in their respiratory system. Therefore, we consider that the filtration rate can be estimated by $\lambda_f = N B/V $, where $N$ is the number of people in the room, $B$ is the average breathing rate, and $V$ is the classroom volume. We consider values of $B = 0.5~m^3/h$, $V=150~m^3$ (average volume of Maragogi classrooms) and $N=20$, which yields $\lambda_f =0.07~h^{-1}$.

The droplet size determines the sedimentation rate $\lambda_s$. For droplets larger than a critical radius $r > r_c$, the sedimentation rate due to gravity is high and contributes significantly to $\lambda_c$. Hereafter, we consider airborne transmission as that associated with droplets with radius $r < r_c$, since those droplets remain suspended in the air for long periods of time (typically a few hours in a closed classroom), and that contain viral loads capable of producing long-ranged airborne transmission. Realistic values for $r_c$ range from $1.3~ \mu m$ to $5.5~ \mu m$ \cite{Bazant_PNAS}. 

The sedimentation rate (drop settling rate) is given by $\lambda_s = \Bar{v}_s(\Bar{r})/H$, where $H$ is the height of the enclosed space. Fixing the velocity of sedimentation $\Bar{v}_s = 0.108 ~m/h$ \cite{Bazant_2021_co2} (the effective respiratory drop radius is $\overline{r} = 0.5 ~\mu m$), and the height $H$ of the classrooms in Maragogi being in the range of $2.57 - 2.85 ~m$, we estimate that $\lambda_s$ lies in the interval $0.038 - 0.042 ~h^{-1}$. Therefore, for biologically relevant droplets of submicron radius settling can be safely neglected \cite{Bazant_PNAS}.

In the following section, we will follow closely \cite{Bazant_PNAS,Bazant_2021_co2}, and assume a size dependent sedimentation rate $\lambda_s(r) = {v}_s(r)/H = \lambda_a (r/r_c)^2$ as the inverse of the time taken for a drop of radius $r$ to sediment
from ceiling to floor in a quiescent room. Hence, Bazant and co-authors propose that for the relevant droplet size range in consideration, one may write 
\begin{align}
    \lambda_c(r) = \lambda_a \Big[1+\Big(\frac{r}{r_c}\Big)^2\Big] +\lambda_v(r) +\lambda_f(r).
\end{align}


The viral deactivation (noninfectious) rate $\lambda_v(r)$ depends on the droplet radius and other quantities, such as tempetarute and humidity. So, aggregating data from influenza viruzes we can extrapolate a linear relationship between relative humidity in the environment $RH$ for SARS-CoV-2 \cite{Bazant_PNAS}. He we adopted $\lambda_v = 0.6 RH ~h^{-1}$ (since Maragogi is a coastal tropical city, $RH$ can be a significantly high factor). 

\subsubsection{Time-evolution of radius-resolved particle concentration}
\label{sec:radius}

We assume the air is well-mixed in the room to evaluate the time-dependent infectious airborne pathogen concentration suspended in a classroom of volume $V$ occupied by $N$ individuals, $I$ infected and $N-I$ susceptible individuals. Following Bazant and Bush (2021), we assume that the radius-resolved concentration of infectious
aerosol-borne pathogen in a classroom with well-mixed air conditions evolves according to
\begin{align}\label{eq:mass_balance1}
    V \frac{\partial c(r,t)}{\partial t} =  \sum_{j = 1}^I P_j(r,t) -V \lambda_c (r,t)  c(r,t),
\end{align}%
where $c(r,t)$ is the number-density of virion particles in the room carried by aerosol droplets with radius $r$ (given in virions per volume per radius), $P_j(r,t)$ is the pathogen production rate due to respiratory activity of a given infectious individual $j$ in the room, and $\lambda_c (r,t)$ is the pathogen concentration relaxation rate. 

The production term of a single infectious individual is given by
\begin{align}\label{eq:production}
P_j(r,t)= B_j(t) p_m^j(r) q_j(r,t),
\end{align}
where $B_j(t)$ is the individual breathing rate, $p_m^j(r)$ is the mask penetration factor of droplets of radius $r$, and $q_j(r,t)$ is an activity dependent concentration of exhaled virions in droplets of radius $r$ (number of virions per volume of air per radius of droplet). Moreover, we may specify that for each infectious individual $q_j(r,t)=n_d^j(r,t) V_d(r) c_v(r)$, where $n^j_d(r,t)$ is the size distribution of emitted droplets (number density of expelled droplets of radius $r$), $V_d(r)=4 \pi r^3/3$ is the droplet volume, and $c_v(r)$ is a microscopic viral concentration (concentration of virions per volume of the droplet). 

We point out that infected individuals emit virions in droplets with a given size distribution that quickly evolves (in a time scale shorter than one second) to a stationary profile $q(r)$ that can be suspended in the air for longer time (for minutes or hours). Therefore, for the relevant contagion time scale in a closed room (from minutes to hours), the production term $P$ in Equation \eqref{eq:production} is time-independent under a constant breathing rate $B$. Moreover, we also assume $\lambda_c (r,t)=\lambda_c (r)$ for steady ventilation conditions.

For simplicity, we assume that the average breathing rate for students and teachers is a constant value  $B$ regardless their activity. The mask penetration factor $p_m(r)$ lies in the unit interval $[0, 1]$ - so it might be associated to a probability of a particle to penetrate the mask tissue - and depends on the droplet size distribution. Based on experimental observations \cite{Chen_1992_penetration_mask}, from here on we assume the mask penetration factor is approximately constant over this submicrometer-size range, and evaluate $\overline{p_m}=p_m(\overline{r})$ at an effective aerosol radius $\overline{r}$ to be defined below in Equation \eqref{eq:lambda_eff}.

Consider that at $t=0$, $N$ individuals enter a room of volume $V$ and zero initial concentration of airborne viral particles, $c_0(r)=c(r,t=0)=0$. These individuals wear masks with equal penetration factor $\overline{p_m}$ and only one individual is infectious among them. They remain in the room for a given time period $\tau$, keeping constant respiratory activity (breathing and talking). The time evolution of the radius-resolved concentration is given by 
\begin{align}
    \frac{1}{\lambda_c(r)} \frac{\partial c(r,t)}{\partial t} =  \frac{P(r)}{V \lambda_c(r)} - c(r,t),
\end{align}
which can be integrated to
\begin{eqnarray}
\label{eq:solution1}
c(r,t)=c_0(r) e^{-\lambda_c(r) t}+\frac{P(r)}{V \lambda_c(r)} [1 -e^{-\lambda_c(r) t}], 
\end{eqnarray}
where $P(r)=B ~\overline{p_m} ~q(r)$ and $\lambda_c(r)>0$ for the relevant range of droplet size.

The probability of a susceptible person to be infected in the room depends not only on the total number of virions inhaled, but also on the power of a virion to cause an infection when it is carried a droplet of a given radius $r$. Therefore, we define  the infectious dose inhaled by an individual exposed to the room from $t=0$ to $t=\tau$ as
\begin{eqnarray}
\label{eq:dose1}
   D(\tau)=\int_{0}^{\tau}dt \int_{0}^{\infty} dr~ B~ p_m(r) c(r,t) ~i(r),
\end{eqnarray}
where $i(r)$ is the infectivity of the aerosolized pathogen in a droplet of radius $r$. $i(r)$ can be interpreted as being proportional to the probability of a single virion to cause an infection in a susceptible person when it is inhaled in a droplet of radius $r$ (in Refs. \cite{Bazant_PNAS,Bazant_2021_co2}, $i(r)$ is equivalent to $c_i(r)$).


The transient term in Equation \eqref{eq:solution1} vanishes for long exposition times $\tau \gg \lambda_c^{-1}$. In this condition we have the following linear dependence of the inhaled dose with $\tau$, 
\begin{eqnarray}
 D(\tau) \approx \frac{B^2}{V} \overline{p_m}^2 \tau \int_{0}^{\infty} dr~ \frac{q(r) i(r)}{\lambda_c(r)} = \frac{B^2}{V} \overline{p_m}^2 \tau \frac{C_q}{\overline{\lambda_c}}, 
\end{eqnarray}
where as in \cite{Bazant_PNAS} we have defined
\begin{align}
\label{eq:Cq1}
 C_q & \equiv \int_{0}^{\infty} dr~ q(r) i(r), \\
 \overline{\lambda_c}^{-1} &\equiv \frac{\int_{0}^{\infty} dr~ q(r) i(r) \lambda_c(r)^{-1}}{\int_{0}^{\infty} dr~ q(r) i(r)}.
 \label{eq:lambdabar}
\end{align}
Moreover, the effective infectious drop radius $\overline{r}$ can now be chosen such that
\begin{align}
\label{eq:lambda_eff}
    \lambda_c(r=\overline{r})=\overline{\lambda_c}.
\end{align}
Realistic physical parameters give us a range of $\overline{r}=0.3 - 5 ~ \mu m$. Bazant and co-authors \cite{Bazant_PNAS} have used $\overline{r}=2 ~ \mu m$ for fitting data from super-spreading events and the Wuhan outbreak; for monitoring air quality indoors Ref. \cite{Bazant_2021_co2} uses $\overline{r}=0.5 ~ \mu m$ for a closed space. 

We consider that the probability $p(\tau)$ of a susceptible individual to be infected when inhaling a given aerosolized pathogen dose $D(\tau)$ is given by the exponential distribution (Wells-Riley model)
\begin{align}
\label{eq:Wells}
    p(\tau) = 1 - e^{-s_r D(\tau)},
\end{align}
where $s_r$ is the age-dependent relative susceptibility of infection (an age-based measure \cite{Zhang_2020_susceptibility}) for a person. This expression follows from the simplest assumption that any infectious viral particle may trigger an infection by independent action of all inhaled viral particles, leading to a Poisson process \cite{Poydenot_2021_risk_assessment}.
For low dose inhalation, $D \ll 1$, the probability can be approximated by $p(\tau) \approx s_r D(\tau)$. This result is equivalent to the probability calculated for school safety guidelines in \cite{Bazant_PNAS}.

\subsubsection{Effective airborne transmission}
\label{sec:effective_airborne_transmission}
For our epidemiological model it is enough to estimate the mean infectious viral load concentration of exhaled air $C_q$ defined in Eq. (\ref{eq:Cq1}).
We will consider $q(r,t)=q(r)$ for any infectious individual in a room, thus $C_q$ is a time-independent constant that represents its average exhaled ``quanta'' concentration, depending on its respiratory activity. $C_q$ is typically expressed in units of quanta per volume of air and represents the important epidemiological parameter that can be numerically estimated based on real outbreak data.

The infectivity $i(r)$ (quanta RNA copies $^{-1}$) represents the probability of a pathogen surviving inside the host to initiate the infection, or we can interpret taking the inverse of the infectivity $i^{-1}$, which corresponds to the ``infectious dose'' of pathogens from inhaled aerosol droplets that cause infection with probability $1-(1/e) = 63\%$.  

To convert the infectious dose quantified in terms of RNA copies to infectious quanta (which is the measure we use in our model) two parameters must be known a priori: i) the number of infectious particles (RNA copies) needed to initiate the infection ($c_{RNA}$, RNA copies PFU$^{-1}$), and (ii) the quanta-to-plaque forming unit (PFU) conversion parameter ($c_{PFU}$, PFU quanta$^{-1}$). Hence, the expression for determining $i(r)$ is 
\begin{align*}
    i(r) = \frac{1}{c_{RNA}(r) ~c_{PFU}(r)}.
\end{align*}
Currently there are no $c_{PFU}$ values available for SARS-CoV-2 in the scientific literature for this value \cite{BUONANNO_2020_quantitative}, or the characterization of the size-dependent distributions $q(r)$, $n_d(r)$ and $c_v(r)$. So we estimate adopting values for SARS-CoV-1. On the other hand, the $c_{RNA}$ parameter has been estimated to be $1.3 \times 10^2~$RNA copies PFU$^{-1}$.


Equation \eqref{eq:Cq1} implies we should be able to characterize the concentration of virions suspended in the air on droplets of all sizes that are capable of causing an infection. Hence we define the total concentration of infectious aerosolized virions per volume of air as
\begin{align}\label{eq:total_concentration}
C(t)= \int_{0}^{\infty} c(r,t) i(r) dr,
\end{align}
where $C$ is given in units of quanta per volume of air. By multiplying Equation \eqref{eq:mass_balance1} by $i(r)$ and integrating for all $r$ one derives
\begin{align}\label{eq:mass_balance2}
    V \frac{d C}{d t} = -(\Lambda V + N B) C + B (C_s N_s p_{m}^s  + C_t N_t p_{m}^t),
\end{align}
where $C$ is the quanta per unit of volume of air in the room, $\Lambda + N B/V = \overline{\lambda_c}$ is the effective rate of relaxation of quanta concentration, $p_m=\overline{p_m}=p_m(\overline{r})$ is the effective mask penetration factor.
We consider that teachers' masks have $p_{m}^t$, and students' masks present $p_{m}^s$. The effective radius $\overline{r}$ for relevant infectious aerosol droplets are given by Eq. (\ref{eq:lambda_eff}), where we make the following approximation
\begin{align*}
\int_{0}^{\infty} \lambda_c(r) c(r,t) i(r) dr &\approx \overline{\lambda_c} C(t).    
\end{align*}

We consider in Equation \eqref{eq:mass_balance2} that the classroom of volume $V$ is occupied by $N$ individuals, in which $S$ are susceptible, $N_s$ are infected students and $N_t$ are infected teachers. Each person exchanges air masses with the environment at average breathing rate $B$, inhales a $C(t)$ quanta concentration and exhales a different concentration. We introduce heterogeneity in the concentration of quanta expelled by students and teachers, assuming they perform different breathing activities \cite{Bazant_PNAS}: $C_s = 40$ ($quanta/m^3$) is the concentration of quanta expelled from students such that $C_{q}^{students}=C_s$, and $C_t = 72$ ($quanta/m^3$) denotes the concentration expelled by teachers (corresponding to voiced counting \cite{MORAWSKA_2009_droplets_size}), such that $C_{q}^{teachers}=C_t$.   

The amount of quanta inhaled by a person inside the class over an exposition time $\tau$ is the inhaled dose in Eq. (\ref{eq:dose1}), which can be writen as
\begin{align*}
D(\tau) =  B ~p_m \int_0^{\tau} C(t) d t,
\end{align*}
where $t=0$ stands for the time the person enters the room and the total concentration of quanta $C$ ($quanta/m^3$) inside the classroom evolves according to Eq.~(\ref{eq:mass_balance2}). Finally, the probability $p(\tau)$ of a susceptible individual to be infected when inhaling a given aerosolized pathogen dose $D(\tau)$ is given by Eq. (\ref{eq:Wells}).

The infectivity is known to differ across distinct age groups and pathogen strains, a variability that is captured by the relative susceptibility $s_r$ in Eq. (\ref{eq:Wells}). For instance, based on the study of transmission in quarantined households in China \cite{Zhang_2020_susceptibility}, Bazant and Bush \cite{Bazant_PNAS} suggest assigning $s_r=1$ for the elderly (over 65 years old), $s_r=0.68$ for adults (aged 15-64) and $s_r=0.23$ for children (aged 0-14) for the original Wuhan strain of SARS-CoV-2, which we adopt here as well. 

\subsubsection{Characteristic parameter values}

\subsubsection{Outdoor air exchange rate}
\label{sec:outdoor_air_rate}
Although the definition of air quality inside enclosed space varies over international standards \cite{KHOVALYG2020109819}, we selected ASHRAE.
As described by the technical notes of ASHRAE 62.1 (Ventilation for Acceptable Indoor Air Quality), additional requirements for taking airborne transmission into account are not covered by the minimum ventilation rates used here. For ASHRAE 62.1 the minimum ventilation rate is calculated as 
\begin{align}
    \lambda_a = \Lambda_p  N + \Lambda_a A
\end{align}
where $\Lambda_p$ is the outdoor airflow rate required per person, $N$ is the number of people in the ventilation zone during use, $\Lambda_a$ is the outdoor air flow rate required per unit area and $A$ is the net occupiable floor area of the ventilation zone. Both $\Lambda_p$ and $\Lambda_a$ are reference ventilation rates determined by ASHRAE standard and depends on the type of enclosed space (we adopted values of Educational Facilities - Classrooms of ages 5 to 8 and age 9 plus). As aforementioned in the main text we adopted three reference values regarding distinct situations rather than any arbitrary values:
\begin{itemize}
\item \textit{Unoccupied}: it consists of the minimum ventilation rate letting $N = 0$. Take the mean area of the group of Maragogi classrooms in our database, we obtained the ventilation rate as $\Lambda_1 = 0.8 ~h^{-1}$.
\item \textit{Half occupied density}: it assumes half occupation density for the classrooms. So, the ventilation rate accounts for both factors, $N$ and $A$. Using the same mean area value than previously, we obtain $\Lambda_2 = 3.8~ h^{-1}$.
\item \textit{Full occupied density }: it consists of full occupation density, and repeating similar calculation we obtain $\Lambda_3 = 6.6 ~h^{-1}$.
\end{itemize}
Note all reference outdoor exchange air flow above are larger than sedimentation and inactivation rate in the model. 

\chapter{Generalization}

\section{Robustness of results for the capital Curitiba}
We show some results of our investigation on the effects of mitigation protocols in schools for the city Curitiba, the largest state capital in the south of Brazil with nearly 2 million inhabitants. This is a very well developed city, among the highest ranked in the country regarding HDI which is in the \textit{very high} range. 

The results presented consider potential interventions during the infection wave which occurred between June $14^{th}$ 2020 and October $12^{th}$ 2020. 

We look at main scenarios from Figure 4, namely scenarios I, III, V, VIII as well as the scenarios where schools remain open with no NPIs and the baseline where schools are closed. We observe that, while the city of Curitiba is less susceptible to the measures, with increase of cases showing smaller magnitude, \textbf{the results are structurally robust and present the same relative hierarchy of effectiveness as the one shown in our main study}.

\subsection{Inference of states from data of Curitiba}

The inference of states in the case of Curitiba is similar to the inference of states made for Maragogi in Section \ref{sec:data_processing}. The data used for this inference is available at OPENDATASUS \cite{SRAG_2020}. The structure of this data differs from the structure of the data collected for Maragogi mainly because we have no information about the brands of the tests used, meaning that we can not take in consideration false positives or false negatives.


The population of Curitiba is approximately 60 times bigger then the population of Maragogi. This enables us to avoid dealing with the attendance or hospitalization data, which is prone to bigger bias, and use the more robust death data to infer the states in a daily basis (as done in \cite{mellan2020report}).

To infer the states we use a negative binomial with the daily number of deaths and the overall probability of death (computed using the Table \ref{table:probs}), then we infer, using the distributions in \cite{kerr2020covasim}, the time each reconstructed individual spent in each state. As in the main study, this process is repeated 400 times to generate a distribution.

\subsection{Baseline scenario}

The baseline scenario we consider in this section is the one obtained from modelling the COVID-19 disease in the city during the period of June $14^{th}$ 2020 to October $12^{th}$ 2020. During the period considered, the city of CURITIBA-PR was also in lockdown, though interventions were softer when compared to those applied to the city of MARAGOGI-AL during the first wave of the disease. From the city's official instructions regarding the opening/closure of services during the first wave, we grouped the services allowed to open during the period in the following categories:
\begin{itemize}
    \item Hospitals: it comprises all type of health facilities in which possible COVID-19 infected patients were received, including campaign hospitals or not;
    \item Health Facilities: it includes all other type of health facilities not contained in the category above;
    \item Supermarkets: the set of all market facilities commercializing mainly food, of medium to large size according to \cite{SIDRA_services} (code $47.11-3$);
    \item Markets: the set of all market facilities commercializing mainly food, of small size according to \cite{SIDRA_services} (code $47.12-1$);
    \item Food stores: the set of small food stores commercializing essential products (meat, dairy, etc., codes $47.21-1$, $47.22-9$, $47.23-7$, $47.24-5$);
    \item Construction stores: the set of store facilities which sell construction equipment, sell vehicle fuel and provide maintenance to vehicle engines (codes $47.3$, $47.4$ and $45$);
    \item Drug stores: the set of pharmacies and similar stores (code $47.7$);
    \item Industry: the set of industries which depend on production lines to deliver its products (codes $10$ to $17$, and $19$ to $33$);
    \item Construction: the set of companies specialized in construction, which demand physical presence of many workers on site (codes $41$ to $43$);
    \item Non-essential: all type of services not included above, except schools.
\end{itemize}
Data for the total number of facilities and total number of employees, for most services, has been gathered from \cite{SIDRA_services}. Data for the total number of facilities and total number of employees of hospitals and health facilities was taken from \cite{cnes_services}. Schools were not opened during the period considered, but we have taken them into account in comparison scenarios (see Section \ref{sec:robustness_curitiba}). The data regarding students and teachers, as well as classes and schools, all have been taken from \cite{Inep_data_dados_abertos,INEP_students_per_class}.

During the period considered, according to the city's official instructions, almost all of the services mentioned above were opened, most with restrictions on the opening time and total number of people per square meter. In our simulations, we have considered that from July 1$^{st}$ 2020 to July 21$^{st}$ 2020, construction stores and non-essential services remained closed. These services were opened during the rest of the period considered in normal opening time. Other services were also opened in normal time during the period considered. The impact of restrictions to opening time and people capacity for services has been taken into account in the calibration of the average number of contacts in these services. See Section \ref{sec:cali_curitiba} for details.

The visitation period and contact network parameters for hospitals, health facilities, markets, supermarkets, food stores and construction stores have been assumed equal to those collected for Maragogi-AL (see Section \ref{sec:calibration_initialization}). The visitation period for drug stores was used as 4 times that of markets, and the network parameters for this service were chosen equal to that of markets as well. Services named construction, construction stores, industry and non-essential services did not receive clients, therefore their visitation period was, conceptually, infinite. However, the contact network parameters assumed for these services has been calibrated from the SEIR data (see Section \ref{sec:cali_curitiba} below). 

We have also considered that modelling the public transportation system was relevant for the spread of COVID-19 in Curitiba-PR (as opposed to what was assumed for Maragogi-AL). The contact network and general behavior of the transportation system has been described in Section \ref{sec:comorbuss_transport}.

\subsection{Calibration of the model}\label{sec:cali_curitiba}

The calibration process used in the city of Curitiba-AL is identical to that exposed in Section \ref{sec:optimization}, except that more parameters were optimized for in this case. We have calibrated $4$ parameters in total, which as listed in the following:
\begin{itemize}
    \item $p$: the infection probability parameter, the same type calibrated for the city of Maragogi-AL;
    \item $f_{viol}$: the fraction of non-essential services that violated city hall instructions regarding their opening during the period considered. This parameter was not assumed necessary at first, but it proved needed eventually during the calibration process;
    \item $c_{ne}$: the average number of $1$-hour contacts  between workers of industry, construction and non essential services. Notice that we have assumed the same parameter for the three types of service;
    \item $c_{transp}$: the average number of $1$-hour contacts between users of the public transportation system.
\end{itemize}
We have observed from a simple sensibility analysis that the first two of these parameters caused a much higher impact in the SEIR curves generated from COMORBUSS as an output. Since calibrating the four parameters simultaneously has been proved to be an intense and nearly impractical computational task, we have chosen to calibrate them in two steps. First, we optimize $c_{ne}$ and $c_{transp}$, keeping $p$ as in Section \ref{sec:optimization} and $f_{viol}=0$. This first calibration procedure gave us the following approximated values for these parameters: $c_{ne}=0.2$ and $c_{transp}=0.1$. By fixing $c_{ne}$ and $c_{transp}$ by these calibrated values, we optimized for $p$ and $f_{viol}$ in a second step. The final values for these last parameters were found to be $p=0.0434$ and $f_{viol}=0.879$, with an $L^1$-Wasserstein distance of $1.35\times 10^{-2}$ between the target and reference SEIR curve distributions. 

\subsection{Robustness of results}\label{sec:robustness_curitiba}
\begin{figure}
    \centering
    \includegraphics[width = 0.85\textwidth]{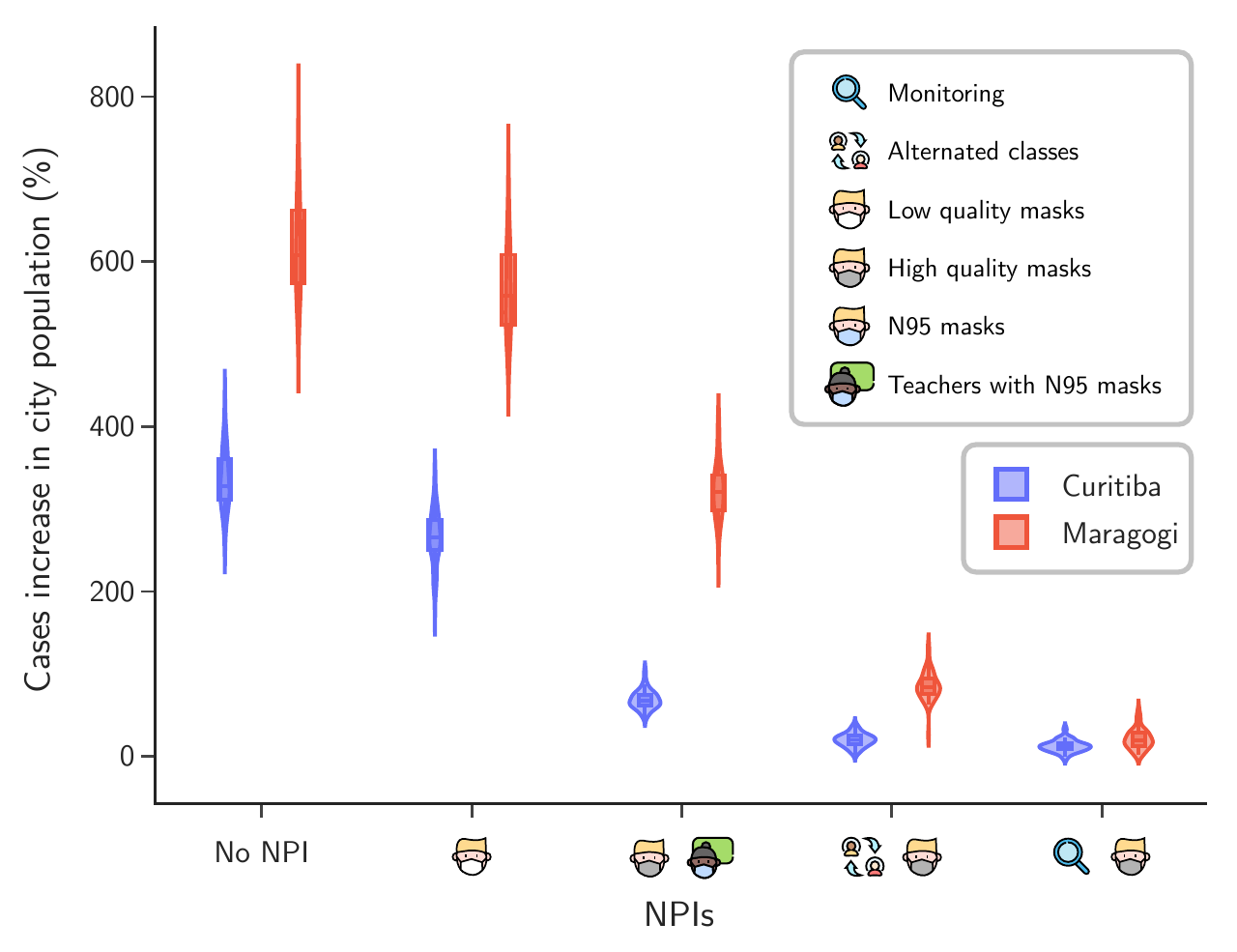}
    \caption{\textbf{Effectiveness comparison for different demographics.} Relative increase in cases for different scenarios for Curitiba-PR compared to Maragogi-AL.}
    \label{fig:curitiba_vs_maragogi_rel}
\end{figure}

After the modeling and calibration for the city of Curitiba, we carry out simulations with 60 seeds using 5 different policies scenarios which are compared again to the baseline where schools are kept closed. The case increase relative to this baseline is depicted for each scenario in Figure \ref{fig:curitiba_vs_maragogi_rel}. 

Most remarkably, the relative rank of protocol effectiveness is the same as observed for a city of small demography such as Maragogi. This showcases the robustness of the protocols across different demographics. Secondly, we note that cities of smaller demographics are susceptible to greater case increase due to bad choices of protocols. This highlights their greater vulnerability and, coupled to their larger representation in national and international demographic distributions, justifies our choice of focus for this study.


\bibliographystyle{plain}